\documentclass[traditabstract]{aa} 
\usepackage{amsmath}
\usepackage{graphicx}
\usepackage{xcolor}
\usepackage{siunitx}
\usepackage{txfonts}
\usepackage{graphicx}
\usepackage{caption}
\usepackage{natbib}
\usepackage{subfigure}
\usepackage{hyperref}
\usepackage[utf8]{inputenc}
\usepackage{placeins}
\usepackage{scrextend}
\usepackage{rotating}
\usepackage{multirow}
\usepackage{booktabs}
\usepackage{diagbox}

\newcommand{\zs}{$z_\mathrm{s}$}
\newcommand{\zd}{$z_\mathrm{d}$}

\begin{document}

   \title{HOLISMOKES}
   \subtitle{XVIII. Cosmology with strongly lensed type II supernovae: Effects of instrumental setups on $H_{0}$}
  \titlerunning{HOLISMOKES XVIII.}

   \author{J. Grupa\inst{1,2,3}
   			\and
             S. Taubenberger\inst{1,2}
                          \and    
          S. H. Suyu\inst{2,1}
             \and 
             D. Sluse\inst{4}
             \and 
   		 S. Huber \inst{1,2}  
                  \and 
             C. Vogl\inst{1,2,3}
          }

   \institute{Max-Planck-Institut f\"ur Astrophysik, Karl-Schwarzschild Str. 1, 85748 Garching, Germany\\
              \email{jana@MPA-Garching.MPG.DE}
         \and 
           Physik-Department, Technische Universit\"at M\"unchen, James-Franck-Stra\ss{}e~1, 85748 Garching, Germany
			\and 
           Exzellenzcluster ORIGINS, Boltzmannstr. 2, 85748 Garching, Germany
\and
           STAR Institute, Quartier Agora – Allée du six Août, 19c, 4000 Liège, Belgium}
 
  \abstract
  {The upcoming Rubin Observatory and subsequent follow-up observations should improve the determination of the Hubble constant ($H_0$) via time-delay cosmography of strongly lensed type II supernovae (LSNe II), by enabling the detection of many more such events.
  In our previous work, we developed a method for determining the supernova (SN) phase from spectral absorption features. Because obtaining spectra of faint targets such as distant SN II is expensive, we examined how low-resolution spectra influence the precision of time-delay retrieval, and consequently the precision on $H_0$. We considered spectral resolutions $R = \frac{\lambda}{\Delta \lambda}$ between 100 and 250, and we investigated three signal-to-noise ratio ($S/N$) values of 10, 15, and 20, for each resolution.  
  Furthermore, we forecast the precision on $H_{0}$ achievable with $S/N=10$ and compared the observing time required to reach it with ground-based and space-based facilities. 
  We find that the time delay can be determined without bias and with uncertainties as low as 1.3 days for the investigated resolutions and $S/N$ values when we combine time-delay measurements of multiple absorption lines. 
  For a typical LSN II system (absolute magnitude $\sim-$17 mag in the rest-frame V band, source redshift of \zs = 0.8), the required exposure times range from multiple hours for ground-based observations to a few minutes for space-based observations with the JWST.
  Our predictions on the precision of $H_0$ for a single lensed SN range from 14.2\% for $R = 100$ and $S/N$ = 10 to 7.5\% for $R = 250$ and $S/N$ = 20, enabling a 1\% determination of $H_0$ from $\sim$20 lensed SNe in the coming years.}


   \maketitle

\section{Introduction}

The ongoing tension in measurements of the Hubble constant, $H_{0}$, with discrepancies exceeding $5\,\sigma$ \citep[e.g.][]{Valentino2021,Verde2024}, began with measurements of the cosmic microwave background (CMB) \citep{Planck2020} that disagreed with local distance ladder inferences. These local inferences are based on Cepheid period-luminosity measurements and type Ia supernovae (SNe Ia) from the Supernova $H_{0}$ for the Equation of State (SH0ES) program \citep{Riess2022, Breuval2024, Riess2024} and pose a fundamental challenge to cosmology. Since then, additional methods have confirmed this tension by varying the anchors of the second rung of the distance ladder.
Among these are Mira variables \citep{Huang2020}, surface brightness fluctuations \citep{Blakeslee2021}, and Tully-Fisher distances \citep{Schombert2020,Tully2023}.
Notable exceptions that agree with both CMB and SH0ES measurements include works based on the tip of the red giant branch (TRGB) technique \citep{Freedman2019,Freedman2020, Freedman2021}, as well as the most  recent work by \cite{Freedman2024}, which incorporates Cepheids and the J-band asymptotic giant branch (JAGB) method \citep{Madore2020} based on James Webb Space Telescope (JWST) observations. However, other TRGB-based studies are consistent with the higher SH0ES measurement \citep{Anand2022,Scolnic2023,Uddin2023}.
The variations in $H_{0}$ values measured using different methods indicate a need for an independent approach to rule out possible underestimated systematics or to support approaches that hint at new physics. One such method is time-delay cosmography based on strong lensing.

This paper explores the application of strong gravitational lensing as an independent method for inferring cosmological parameters through time-delay cosmography. Originally proposed by \cite{Refsdal1964}, this technique has been applied to measure $H_{0}$ in the$H_{0}$ Lenses in COSMOGRAIL’s Wellspring (H0LiCOW) program \citep[e.g.,][]{Suyu2017, Chen2019, Wong2019, Rusu2019}, in collaboration with the COSmological MOnitoring of GRAvItational Lenses (COSMOGRAIL) \citep{Eigenbrod2005, Courbin2017, Bonvin2018}, and the Strong lensing at High Angular Resolution Program (SHARP) \citep{Chen2019}, combining multiple strongly lensed quasars.  The Time-Delay Cosmography (TDCOSMO) collaboration  \citep{Millon2020} is extending this method to a larger sample of lensed quasars \citep[e.g.,][]{Mozumdar2023, Dux2025} and developing new ways to further control systematic effects in $H_0$ measurements \citep[e.g.,][]{Birrer2020, Gomer2022, Yildirim2023, Wells2023, Knabel2025, Wang2025}. In this paper, we focus on lensed SNe (LSNe) as new sources for time-delay cosmography.

The resolved lensed type II supernova (LSN II) SN Refsdal, discovered in 2014 \citep{Kelly2015, Kelly2016a, Kelly2016b}, provided time-delay measurements that yield $H_{0} = 64.8^{+4.4}_{-4.3}\,\mathrm{km\,s^{-1}\,Mpc^{-1}}$ \citep{Kelly2023, Grillo2024, Liu2024}. The recently discovered lensed type Ia supernova SN H0pe at $z = 1.78$ \citep{Frye2023, Frye2024} yields $H_{0} = 75.4^{+8.1}_{-5.5}\,\mathrm{km\,s^{-1}\,Mpc^{-1}}$ \citep{Pascale2024, Pierel2024}. Both cases fall within $2\,\sigma$ of the CMB and SH0ES measurements, highlighting the potential of LSNe as a tool to resolve the $H_{0}$ tension once a larger sample of LSNe becomes available.
Such a large sample requires efficient and precise methods to build mass models and determine time delays.

In this paper, we focus on time delays through spectroscopy. 
Previous studies have applied spectroscopic methods to determine time delays for the LSN Ia iPTF16geu \citep{Johansson2020} and SN H0pe \citep{Chen2024}. In both instances, the authors used templates or models to determine the phase of each lensed image. This template-fitting method is effective for the two LSNe because both are type Ia, and it is a standard method for determining the age of SNe Ia \citep[e.g.,][]{Blondin2007}. However, because the spectra of SNe II are much more diverse,
template fitting is more difficult to apply to them. Therefore, we explored a method based solely on matching the evolution of spectral absorption wavelength in the multiple SN images, without the need for spectral templates.

The upcoming Rubin Observatory Legacy Survey of Space and Time \citep[LSST;][]{LSSTScienceCollaboration2009} is expected to detect hundreds of LSNe, 
with $\sim$70-80\% of them of type II (IIP, IIL, and IIn)
\citep{Wojtak2019, Goldstein2019, Goldstein2018, Goldstein2016}. 
This work builds on \citet[hereafter HOLISMOKES V]{Bayer2021}, which demonstrates the feasibility of determining the relative phase between LSN IIP images from spectral absorption lines as they shift to longer wavelengths during photospheric expansion. The earlier study used simulated mock spectra with a 3 \AA \ bin size generated using a modified version of the \textsc{tardis} code \citep{Kerzendorf2014,Vogl2019}. In the present work, we instead explore the impact of reduced spectral resolution and signal-to-noise ratio ($S/N$) on time-delay inference.
Specifically, we simulate low-resolution spectra at $R = 100$, $150$, $200$, and $250$, coupled with three S/N values: 10, 15, and 20. Using absorption lines of H$\mathrm{\alpha}$, H$\mathrm{\beta}$, and Fe\,\textsc{ii}, we analyze phase-shift retrieval under these conditions, incorporating microlensing effects caused by stars in the lensing galaxy. 

Additionally, we estimated the exposure time required to obtain sufficiently high-quality data for measuring a delay with ground-based instruments such as the FOcal Reducer and low-dispersion Spectrograph 2 (FORS2) and the Multi Unit Spectroscopic Explorer (MUSE) on the Very Large Telescope (VLT), as well as space-based instruments including the Space Telescope Imaging Spectrograph (STIS) on the Hubble Space Telescope (HST) and the Near Infrared Spectrograph (NIRSpec) on the JWST. For these spectrographs, we consider the wavelength range of the absorption lines for an assumed SN redshift of \zs $= 0.8$, the approximate mean redshift of LSNe expected from the LSST \citep{Oguri2010}.
Finally, we estimate the precision in $H_{0}$ achievable for a single lensed SN II based on these low-resolution spectra, incorporating uncertainties from lens modeling \citep{Suyu2020}. 
This work provides a crucial step toward enabling time-delay cosmography with LSNe II detected with the LSST.

We begin with the creation of the low-resolution mock spectra in Sect. \ref{sec: Type II Supernova models} and present the results of the phase retrievals of these spectra in Sect. \ref{sec: SN phase inference from spectra}. 
We then estimate the $H_0$ precision to expect for an individual lensed SN II for the different mock spectra in Sect. \ref{sec: H0 inference}. Additionally, we assess the number of LSNe expected to be detected in the LSST era in Sect. \ref{sec: SN_numbers}; these data will be useful for precise $H_0$ constraints based on the spectroscopic time-delay retrieval method. 
We calculate exposure times for ground- and space-based instruments in Sect. \ref{sec: Exposure time estimation}.
We provide a discussion and conclusions in Sect. \ref{sec: Discussion and Conclusion}.

\section{Type II supernova models}
\label{sec: Type II Supernova models}

We used supernova models to study cosmography with LSN IIP spectra, as  observed spectral data with the desired temporal cadence for LSNe II are currently unavailable, and we  included microlensing to study its influence on the precision of time-delay retrieval. In HOLISMOKES V, we describe the intricate modeling of simulated model spectra with the radiative-transfer code \textsc{tardis}. We provide a short summary in the following section before reducing the resolution and creating the dataset for this work.

\subsection{Microlensed \textsc{tardis} simulations}
\label{sec: tardis simulations}

We computed the model spectra with the extended version of the Monte Carlo radiative-transfer code \textsc{tardis} from \cite{Vogl2019}. We applied the code to observed spectra of one of the most well-observed prototypical SNe IIP, SN 1999em \citep[e.g.][]{Dessart2006, Leonard2002, Baron2004, Hamuy2001}.
  
We obtained five spectral snapshots, as \textsc{tardis} does not perform fully time-dependent radiative transfer computations, with a constant bin size of $\Delta \lambda_{\mathrm{bin, sim}} = 3 \ \text{\AA}$. We calculated the flux by placing the SN at a distance of 10 pc.

\begin{figure}[hbt!]
\centering
{\includegraphics[width=0.489\textwidth]{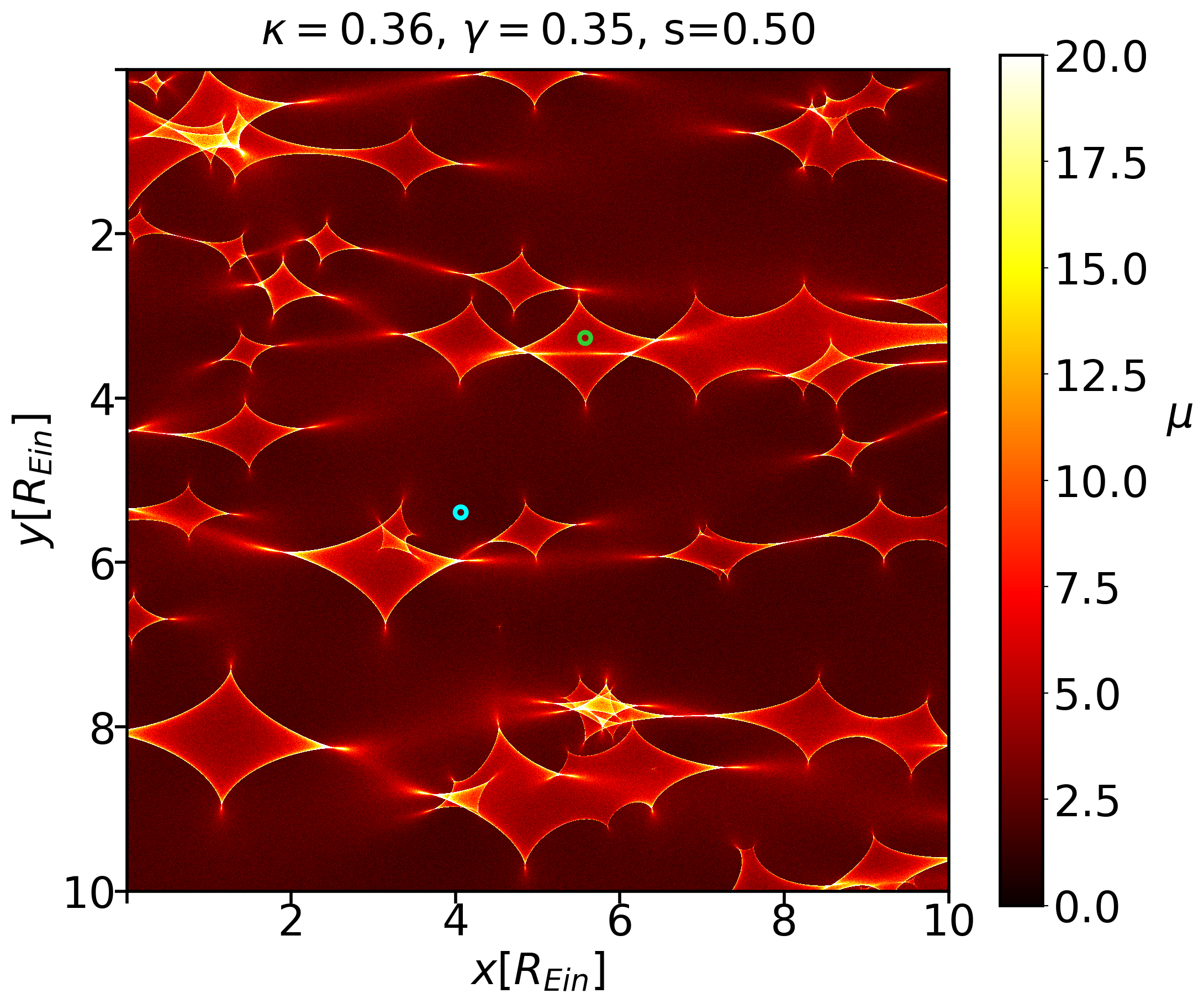}}\
\caption{\label{micromap}Microlensing magnification map for the first lensing image, SN-A, with convergence $\kappa$ = 0.36, shear $\gamma$ = 0.35, and smooth matter fraction $s$ = 0.50. The magnification $\mu(x,y)$ is shown by the color scale on the right. The two colored circles in the map mark the positions used for phase retrieval.}
\end{figure}

As shown in HOLISMOKES V, microlensing affects SN IIP spectra. Therefore, we modified the flux of the spectra according to \cite{Huber2019} to include microlensing.
We used two \textsc{gerlumph} microlensing magnification maps generated with GPU-D \citep{Thompson2010, Vernardos2015, Kayser1986, Wambsganss1992, Vernardos2013}. The maps have a resolution of 20000 × 20000 pixels, corresponding to a size of 10 $R_{\rm Ein}$ × 10 $R_{\rm Ein}$. We then calculated microlensing for 10000 random SN positions in each map. 
The magnification maps display the total magnification $\mu(x,y)$, including macro-magnification $\mu_{\mathrm{macro}}$ and micro-magnification $\mu_{\mathrm{micro}}$, in the source plane as a function of the Cartesian coordinates $x$ and $y$, in units of the Einstein radius defined as
\begin{equation} \label{R_Ein}
R_{\rm Ein} = \sqrt{\frac{4G \langle M \rangle}{c^{2}}\frac{D_{\rm s}D_{\rm ds}}{D_{\rm d}}},
\end{equation}
where $D_{\rm s}$ is the angular-diameter distance from the observer
to the source, $D_{\rm ds}$ is the angular-diameter distance from the
lens to the source, $D_{\rm d}$ is the angular-diameter distance from
the observer to the lens, and $c$ is the speed of light. We assumed a mean mass of the point-mass microlenses of $\langle M \rangle$ = 0.35 $\mathrm{M_{\odot}}$ for a Salpeter initial mass function \citep{Huber2021}. 
We used the same two microlensing maps already introduced in HOLISMOKES V, which correspond to two SN images within the same lens system with a lens redshift of \zd = 0.32. The first is a time-delay minimum (type I) with convergence $\kappa$ = 0.36, shear $\gamma$ = 0.35, and macro-magnification $\mu_{\mathrm{macro}}$ = 3.5, and the second is a saddle (type II) image with $\kappa$ = 0.7, $\gamma$ = 0.7, and macro-magnification $\mu_{\mathrm{macro}}$ = $-$2.5 \citep{Oguri2010, Huber2019, Suyu2020}. The smooth matter fraction $s$ is set to 0.5 in both maps.
Figure \ref{micromap} shows the microlensing map for the first image with two random positions marked in green and cyan, which are part of the set of positions used later in phase retrieval. The sizes of the circles correspond to the radii of the SN containing 99.9 \% of the total projected specific intensity \citep{Huber2019} for the fourth epoch, 22 rest-frame days after the explosion.

\begin{figure}[hbt!]
\centering
{\includegraphics[width=0.489\textwidth]{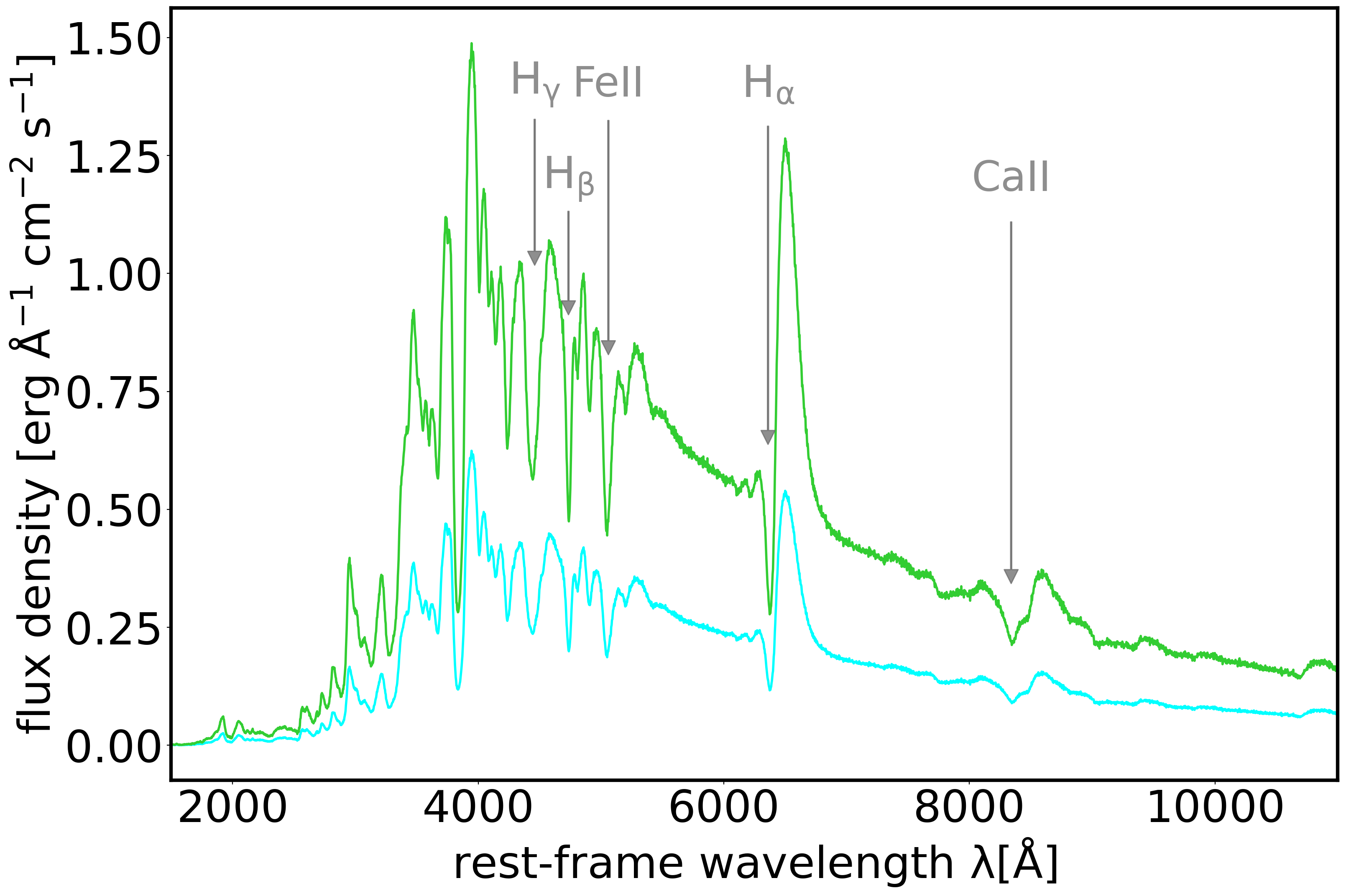}}\
\caption{\label{spectrum_micro}Two microlensing realizations of the rest-frame model spectrum of SN 1999em at day 22 (rest-frame) after the explosion, with a wavelength bin size of $\Delta \lambda_{\mathrm{bin}} =$ 3 \AA. Gray arrows indicate typical absorption features for this epoch \citep{Elmhamdi2003}.}
\end{figure}

We calculated the microlensed flux $F_{\lambda, \mathrm{o}}(t)$ by placing the SN model at a specific time bin into the microlensing map and calculating the change in flux due to the magnification at the specified map position, considering the size of the SN at that time, 
\begin{equation} \label{flux_eq}
F_{\lambda, \mathrm{o}}(t) = \frac{1}{D_{\mathrm{lum}}^{2}(1+z_{\mathrm{s}})} \int \mathrm{d}x \int \mathrm{d}y \ I_{\lambda, \mathrm{e}}(t, x, y) \ \mu(x, y),
\end{equation}
where $D_{\mathrm{lum}}$ is the luminosity distance to the source, $z_{\mathrm{s}}$ is the source redshift (in this case the SN), and
$I_{\lambda, \mathrm{e}}(t, x, y)$ is the specific intensity in the
source plane.
In Fig. \ref{spectrum_micro}, we show the two microlensing realizations marked in Fig. \ref{micromap} for a rest-frame model spectrum of SN 1999em at 22 days (rest-frame) after the explosion. The green position is magnified and therefore shifted to larger flux density values compared to the cyan position. We indicate typical absorption features for this epoch at the top with gray arrows \citep{Elmhamdi2003}.

\subsection{Redshift inclusion and spectral resolution reduction}
\label{sec: Reduction of spectral resolution}

At this stage, the simulated model spectra are currently in the rest-frame, as only the added microlensing effects account for the source and lens redshift, manifesting in the micro-magnification. In reality, we should emulate the LSN spectra in the observed frame to account for their redshift. We therefore added a redshift before reducing the spectral resolution to obtain more realistic spectra.
We redshifted each wavelength bin to \zs = 0.8 \citep{Oguri2010} and reduced the flux density by a factor of $1 + $ \zs.

\begin{figure}[hbt!]
\centering
{\includegraphics[width=0.489\textwidth]{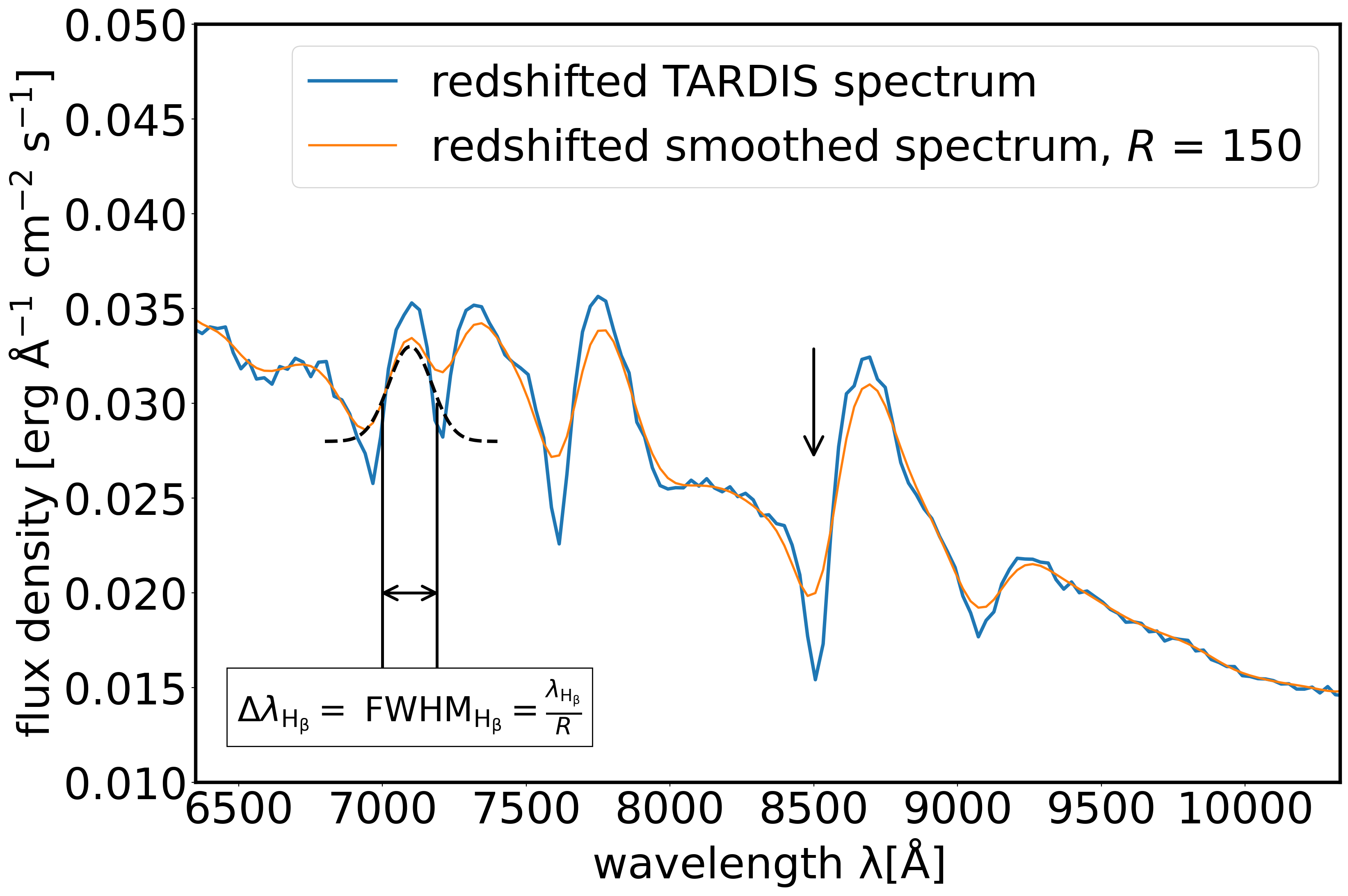}}\
\caption{\label{gaussian_smooth}Gaussian smoothing process used to emulate a spectral resolution $R = 150$. The blue line shows the redshifted \textsc{tardis} spectrum around the $\mathrm{H_{\beta}}$ line, which is indicated with a vertical black arrow at an observed wavelength of $\sim8525 \ \text{\AA}$. The dashed black line indicates the Gaussian kernel used for smoothing. The resulting smoothed spectral absorption line is shown in orange.}
\end{figure}

\begin{figure}[hbt!]
\centering
{\includegraphics[width=0.489\textwidth]{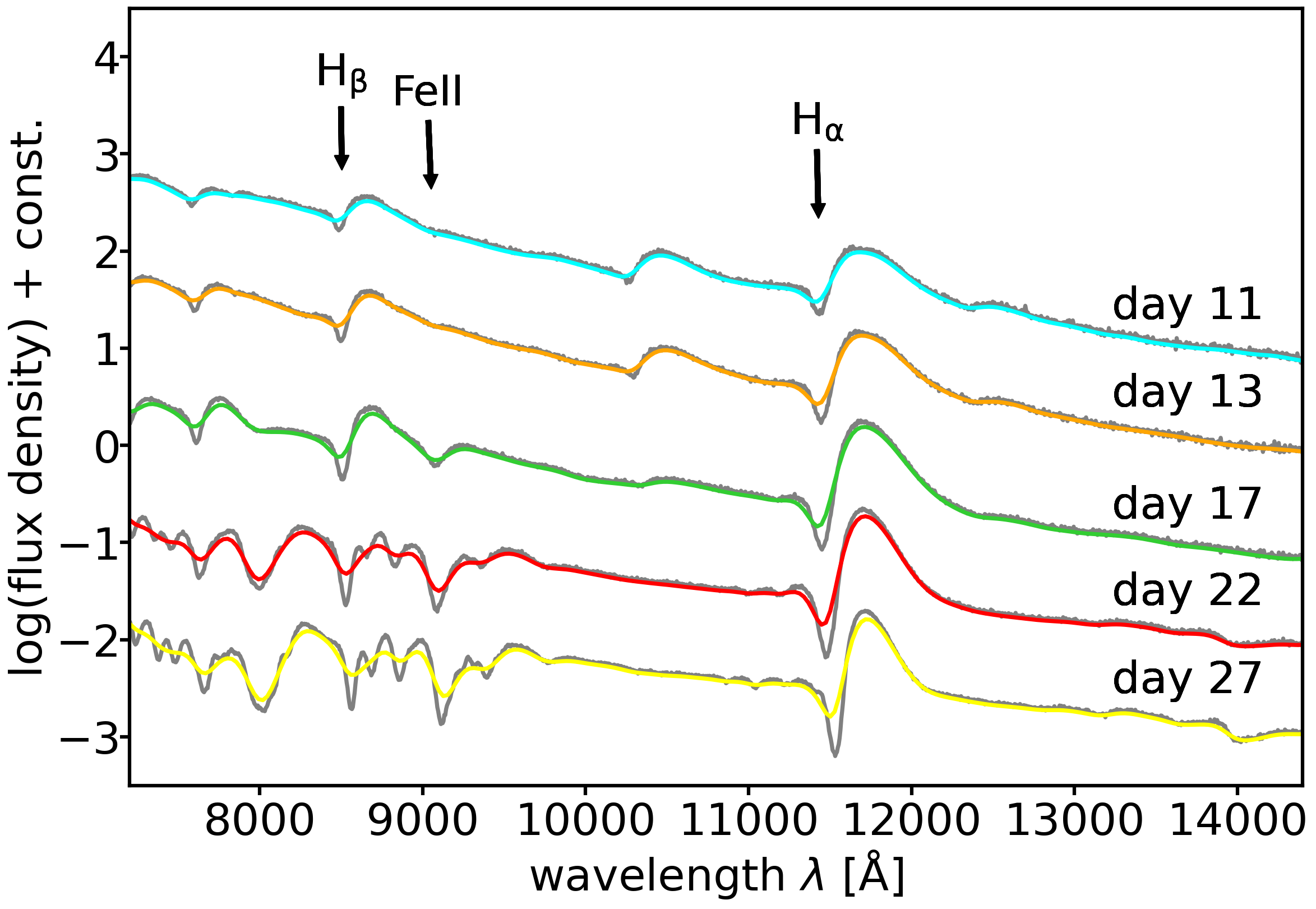}}\
\caption{\label{spectrum_smooth_all}
Model spectra of SN 1999em at redshift \zs = 0.8. The gray lines represent the rest-frame wavelength spectra with a bin size of 3 \AA, while the colored lines show the spectra with lower resolution $R = 150$ for the five investigated epochs (indicated in the legend in rest-frame days). Absorption lines used in the phase inference are indicated with arrows.}
\end{figure}

After redshifting the spectrum, we reduced the resolution because high-throughput, low-resolution spectrographs are among the most promising ways to obtain spectroscopic observations of the typically very faint LSNe.
Throughout the paper, we adopt the notation $\Delta\lambda_{\mathrm{bin}}$ to refer to the size of wavelength bins of the spectrum and $\Delta\lambda_{\mathrm{R}}$ to refer to the full width at half maximum (FWHM) of a resolvable spectral line.
The bin size of the redshifted spectrum is $\Delta \lambda_{\mathrm{bin,sim}} (1 +$\zs) = 3 \AA \ (1 +\zs) = 5.4 \AA. 
We implemented four lower resolution setups, $R =$ 100, 150, 200, and 250, to investigate how the use of degraded spectra impacts the time-delay inference through the time evolution of spectral features and, consequently, the $H_0$ uncertainties. We reduced the resolution in two steps:
\begin{enumerate}
    \item We smoothed the microlensed \textsc{tardis} spectrum with a Gaussian kernel. We determined the kernel size FWHM by the FWHM of a resolvable spectral line $\Delta \lambda_{\mathrm{R}} = \frac{\lambda_{\mathrm{element}}}{R(\lambda)}$, where the wavelength $\lambda_{\mathrm{element}}$ corresponds to the wavelength of the considered absorption-line minimum Fe\,\textsc{ii}, H$\mathrm{\beta}$, or H$\mathrm{\alpha}$\footnote{The wavelength dependence of $R(\lambda)$ is based on the specifications of the upcoming Roman telescope's low-resolution spectrograph \citep{Groff2021}. The dependency, which we adopt for our study, is approximately constant in the range of the investigated absorption lines.}.
   Figure \ref{gaussian_smooth} illustrates the smoothing process for the case of H$\mathrm{\beta}$.
    \item 

    To match observations, we downsampled the spectra to align with the sampling of a low-resolution spectrograph. We fixed the downsampling applied to the spectra to satisfy the Nyquist theorem \citep{Robertson2017}, sampling $\Delta \lambda_{\mathrm{R}}$ with $\sim$2.5 to $\sim$3 spectral bins.
     We set the exact $R$ value at the observed wavelength of $10800 \ \text{\AA}$, corresponding to a rest frame wavelength of $6000 \ \text{\AA}$ close to the investigated absorption lines in Sec. \ref{sec: SN phase inference from spectra}.    
    This leads to downsampling, grouping seven bins 
    for $R =$ 100, five bins for $R =$ 150, four bins for $R =$ 200, and three bins for $R =$ 250 into one bin, with corresponding sizes of $\Delta \lambda_{\mathrm{bin,mock}}$ of $\sim$38 \AA, $\sim$27 \AA, $\sim$22 \AA, and $\sim$16 \AA \ in the observed frame, respectively.
\end{enumerate}
We denote these noiseless, low-resolution spectra as those with $S/N$ = $\infty$. Figure \ref{spectrum_smooth_all} compares the \textsc{tardis} spectra (rest-frame wavelength bin size of 3 \AA \ in gray) compared to the lower-resolution spectra ($R = 150$) in color across all five epochs of the SN 1999em model. The smaller absorption features are lost, but the stronger features, such as the Fe\,\textsc{ii}, H$\mathrm{\beta}$, and H$\mathrm{\alpha}$ lines, remain visible.

To create mock spectra with noise, we followed a procedure similar to that in HOLISMOKES V.
We calculated the Gaussian-noise amplitude at an observed wavelength of 10800 \AA \ according to the desired $S/N$ value of 10, 15, and 20\footnote{In HOLISMOKES V, we selected the maximum flux value of the spectrum at around rest-frame 4000 \AA \ to calculate the Gaussian-noise amplitude, which resulted in lower $S/N$ values at the investigated absorption lines around rest-frame 6000 \AA.}.

\begin{figure}[hbt!]
\centering
{\includegraphics[width=0.489\textwidth]{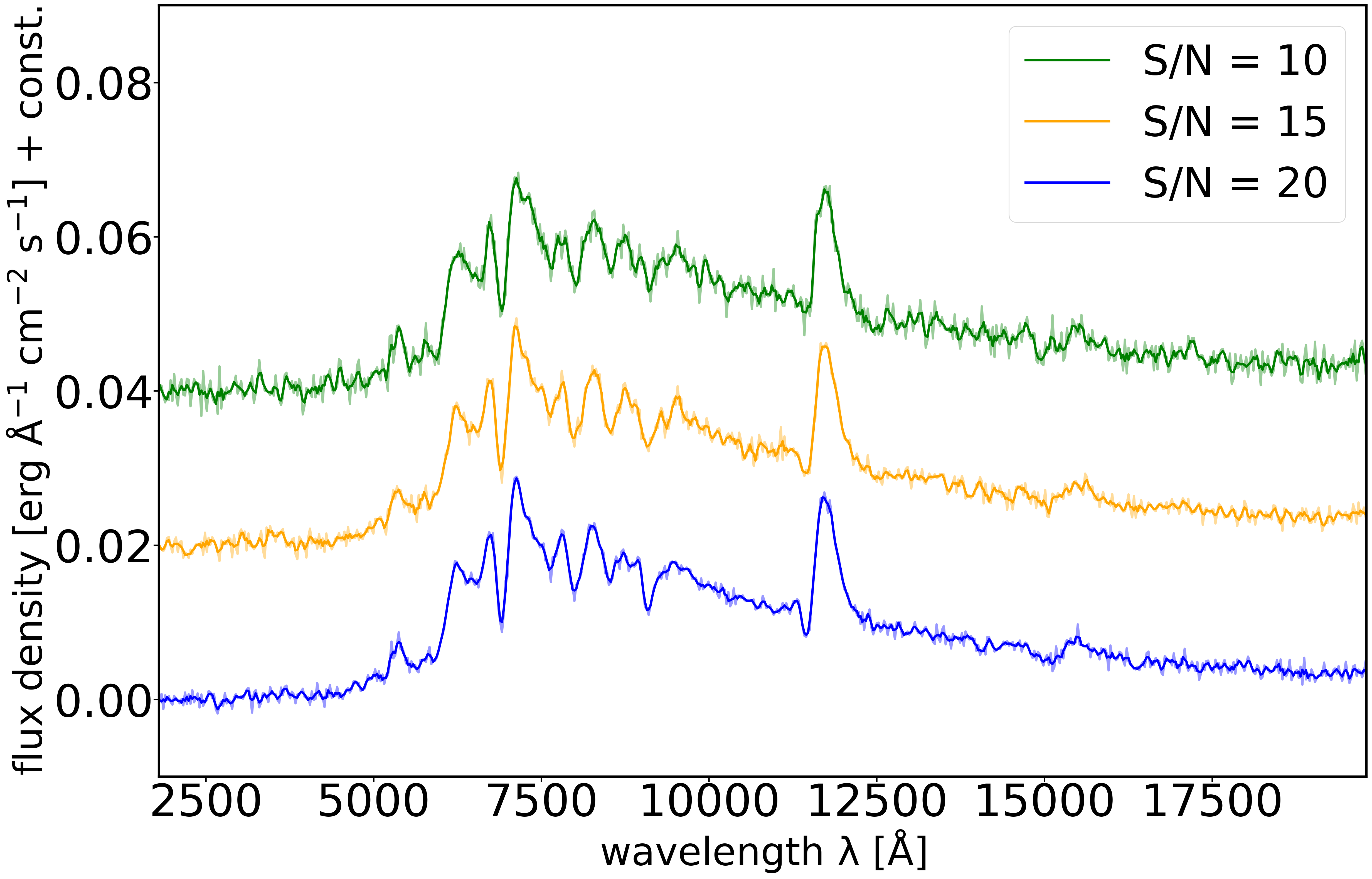}}\
\caption{\label{spectrum_noise}Mock spectra of SN 1999em at \zs = 0.8 on day 22 post-explosion, at resolution $R = 150$, for three different $S/N$ values: 10 (blue), 15 (orange), and 20 (green). The spectra are smoothed with the Savitzky-Golay filter. For comparison, the unsmoothed spectra are shown in lighter colors.}
\end{figure}

Subsequently, we applied a first-order Savitzky-Golay filter covering a width of three bins to smooth the spectra again to minimize the effect of the added noise on our phase-retrieval algorithm.
We selected the order and number of bins to ensure that the features remained sufficiently defined for the spectral time-delay method, as already applied in HOLISMOKES V. Figure \ref{spectrum_noise} shows the mock spectrum on day 22 after the explosion for the three $S/N$ at $R = 150$. It includes noise addition with and without the additional smoothing provided by the Savitzky-Golay filter.

\section{Lensed supernova phase inference from spectra}
\label{sec: SN phase inference from spectra}

For the determination of the phase, we used our spectral phase retrieval code developed in HOLISMOKES V and applied it to our redshifted low-resolution spectra with $R$ = 100, 150, 200, and 250 for the three $S/N$ values of 10, 15, and 20.
We first sampled 10000 random positions in the microlensing maps and considered five epochs $t_{i}$ with $i$ = 1, 2, ..., 5 for each of the two LSN images (SN-A and SN-B), corresponding to rest-frame days 11, 13, 17, 22, and 27 after the explosion. 
For the 10000 random positions in the microlensing maps, five epochs, and different noise realizations, we fit the absorption minima of H$\mathrm{\alpha}$, H$\mathrm{\beta}$, and Fe\,\textsc{ii} with a Gaussian function and determined the absorption line wavelength $\lambda(t_{i})$ from the minimum of the fit. We approximated the distribution of $\lambda(t_{i})$ with a Gaussian, which yields the continuous distribution mean $\lambda_{\mathrm{d}}(t)$ and the standard deviation $\sigma_{\mathrm{d}}(t)$, by linearly interpolating between $\lambda_{\mathrm{d}}(t_{i})$ and $\sigma_{\mathrm{d}}(t_{i})$.

\begin{figure*}[hbt!]
\centering
\subfigure{\label{temp_evol_legend}\includegraphics[width=\textwidth]{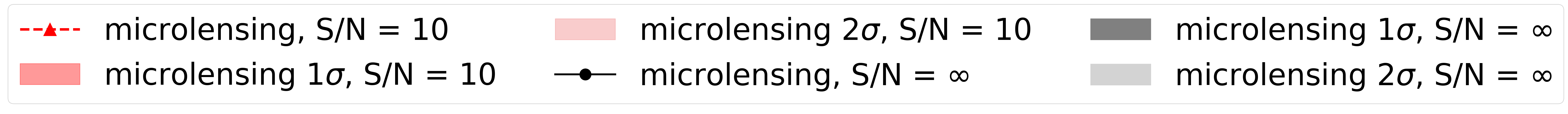}}\hfill
\subfigure{\label{temp_wave_min_H_alpha_1}\includegraphics[width=0.49\textwidth]{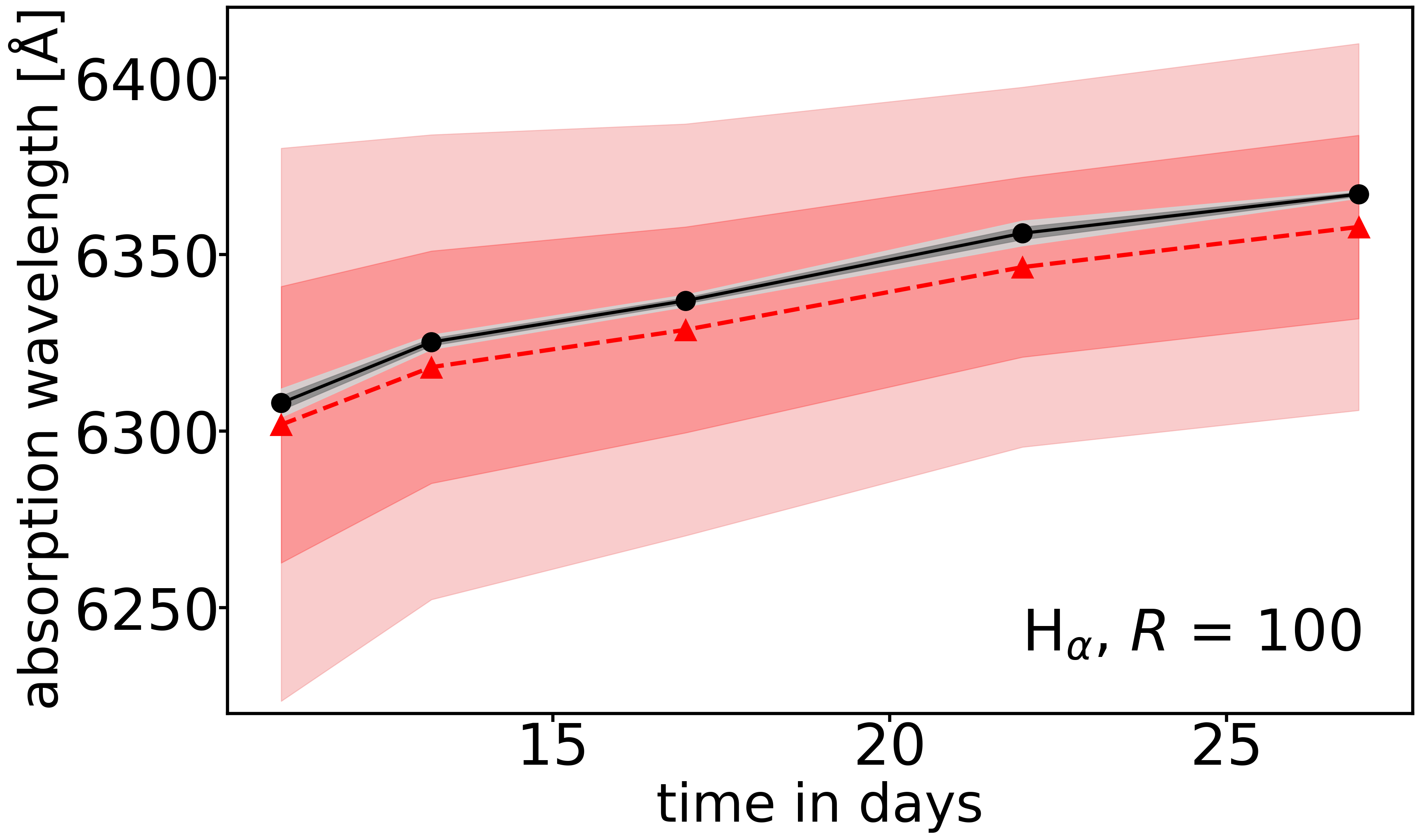}}\hfill
\subfigure{\label{temp_wave_min_H_alpha_2}\includegraphics[width=0.49\textwidth]{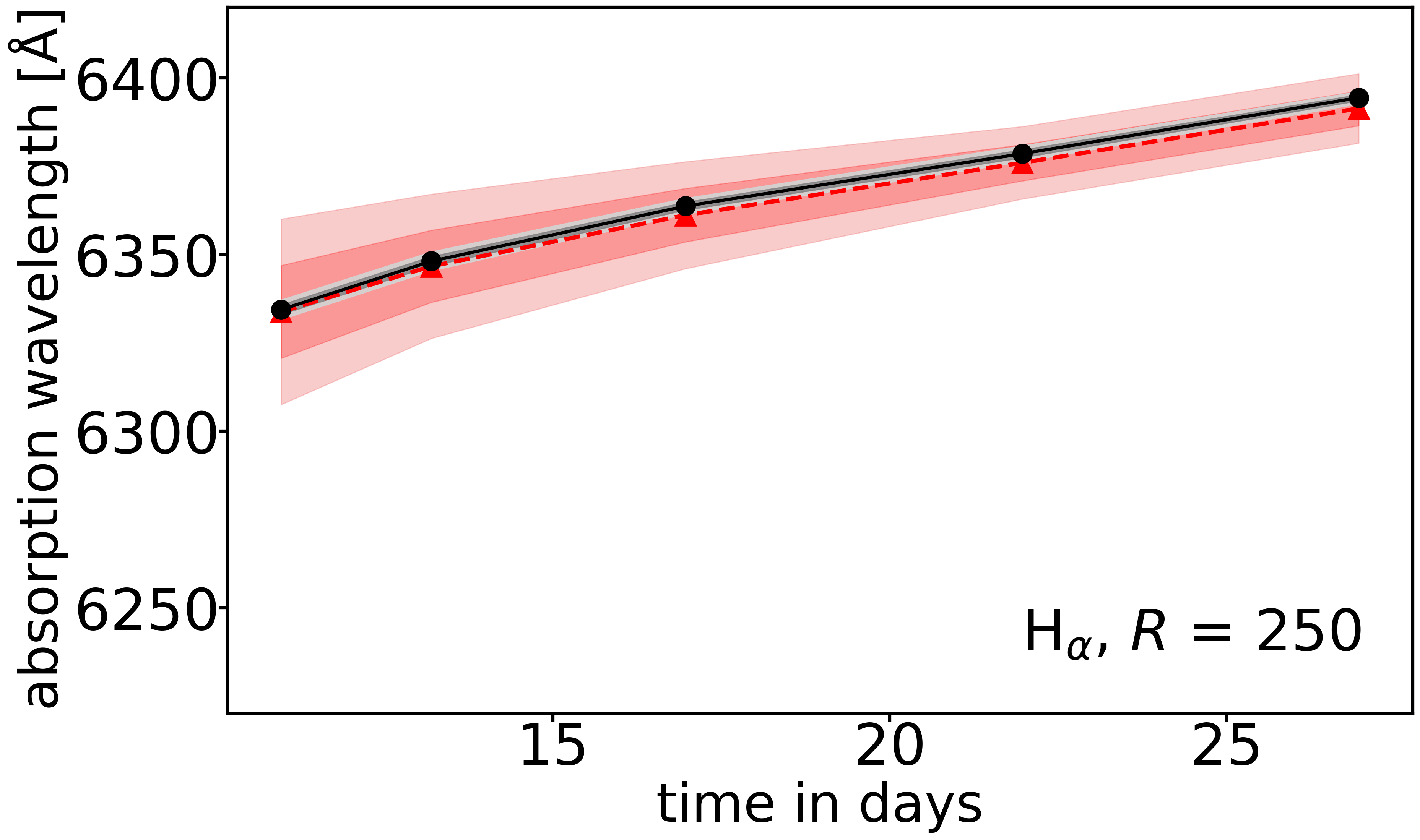}}\\
\subfigure{\label{temp_wave_min_H_beta_1}\includegraphics[width=0.49\textwidth]{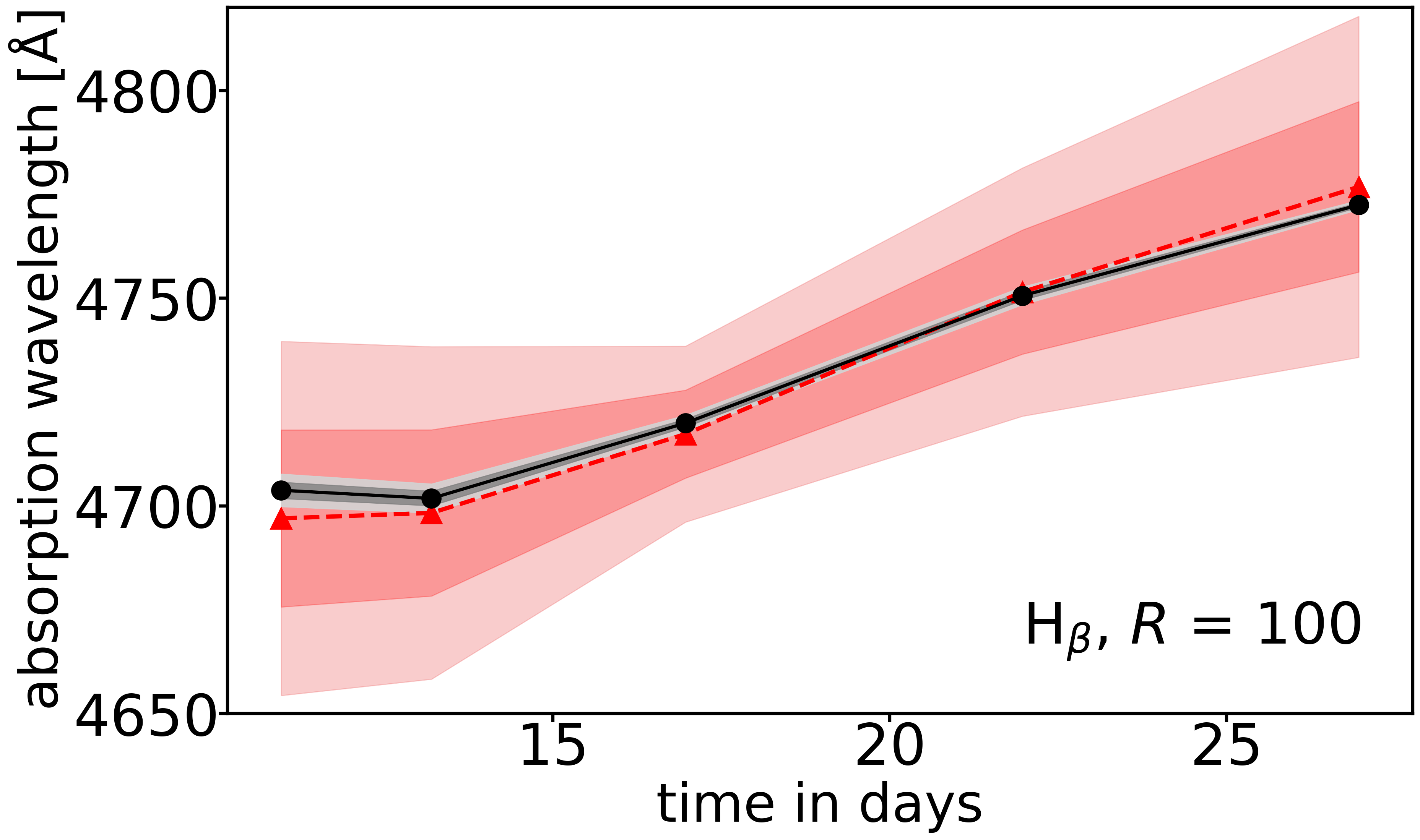}}\hfill
\subfigure{\label{temp_wave_min_H_beta_2}\includegraphics[width=0.49\textwidth]{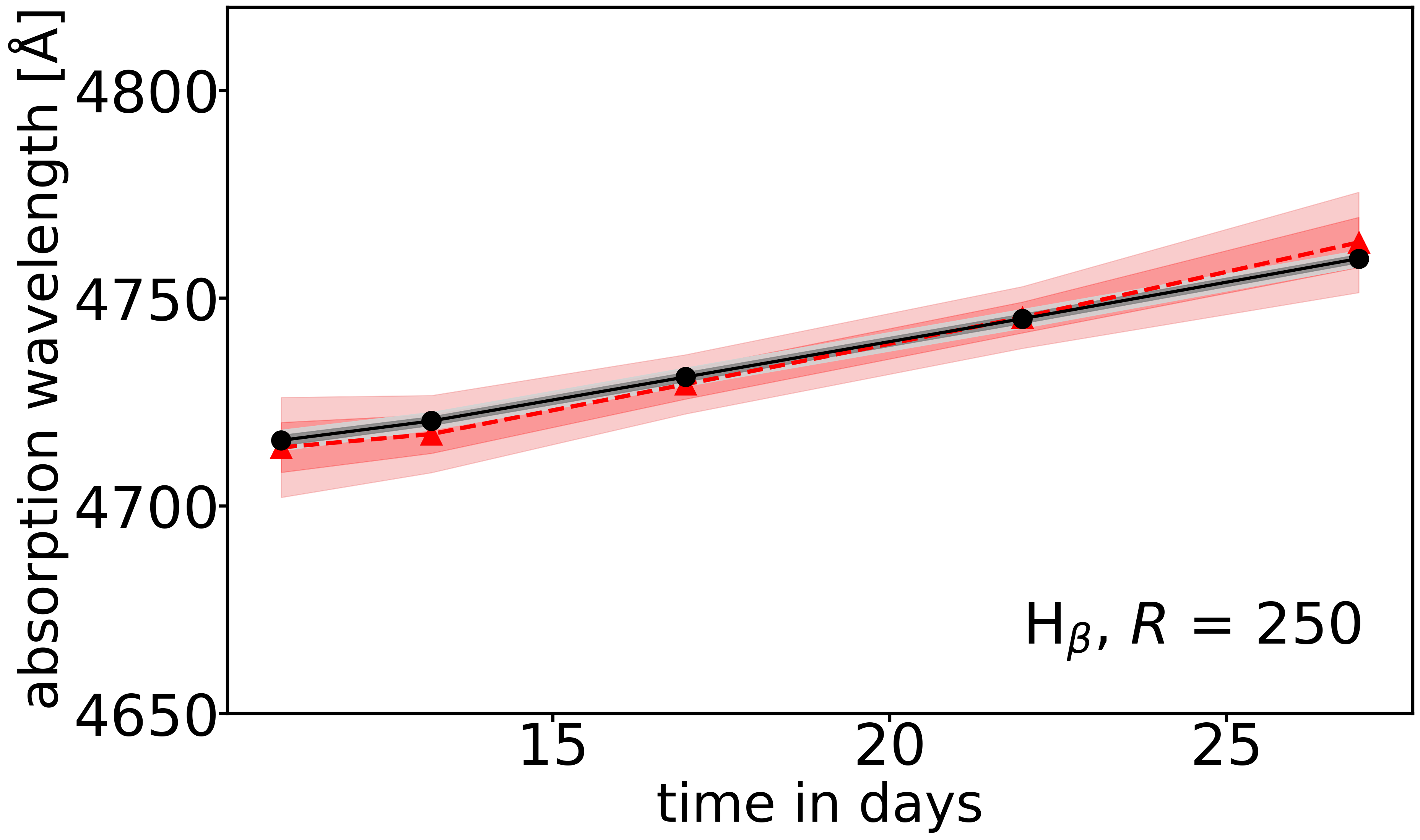}}\\
\subfigure{\label{temp_wave_min_FeII_1}\includegraphics[width=0.49\textwidth]{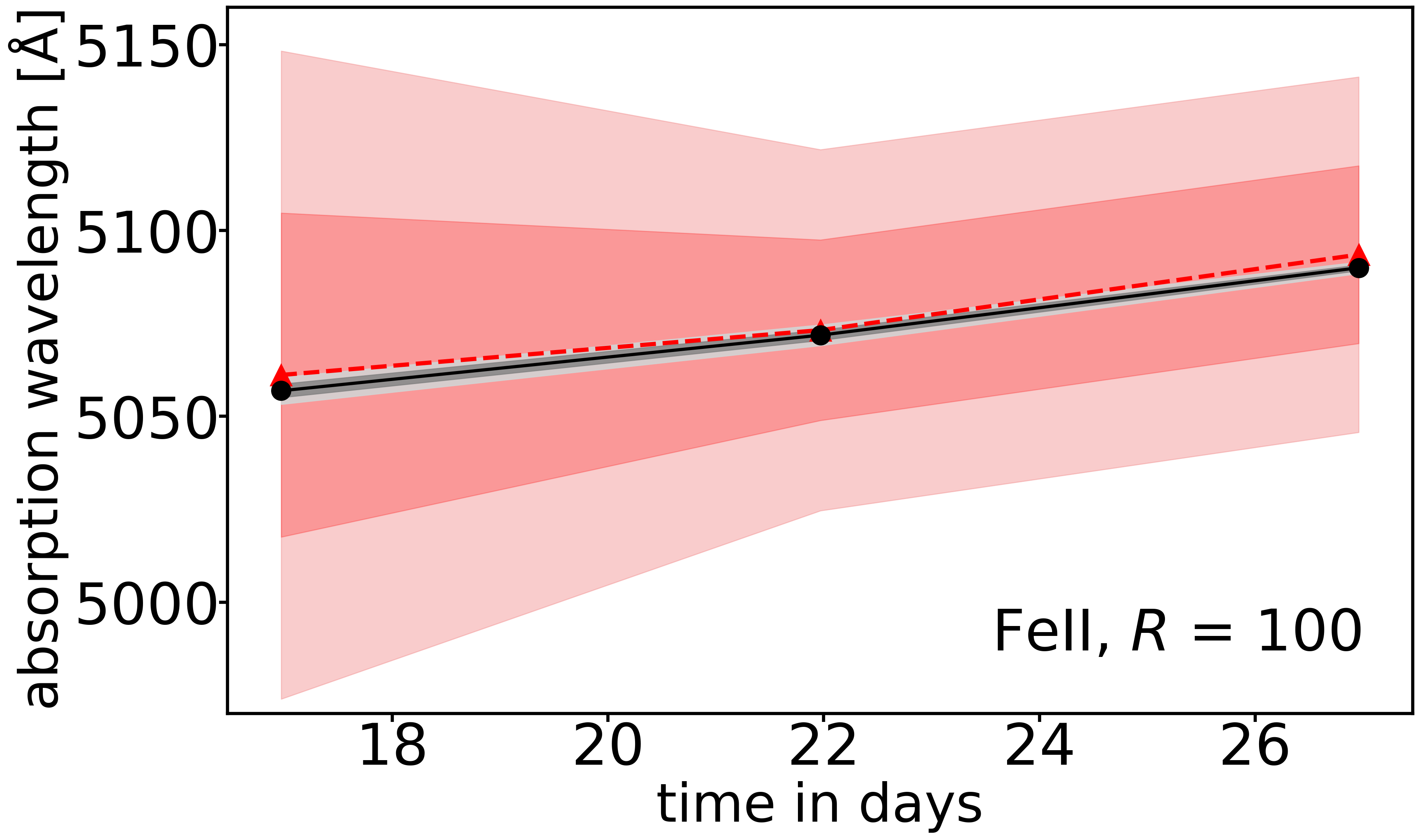}}\hfill
\subfigure{\label{temp_wave_min_FeII_2}\includegraphics[width=0.49\textwidth]{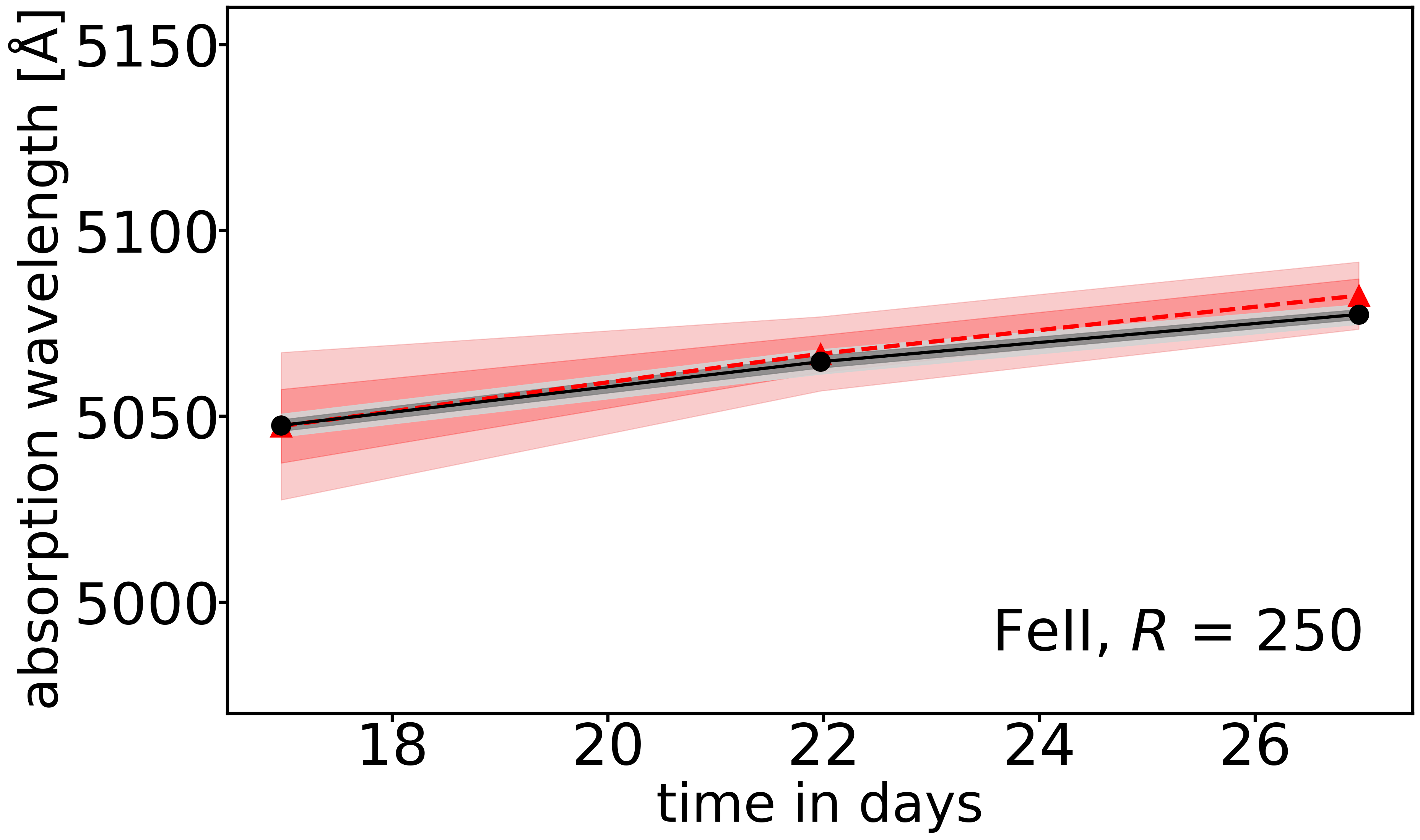}}\
\caption{\label{temp_wave_min} Temporal evolution of the rest-frame absorption
  minima of the H$\mathrm{\alpha}$, H$\mathrm{\beta}$, and Fe\,\textsc{ii}
  lines for microlensed spectra at $R$ = 100 (left) and $R$ = 250 (right) using the magnification map of the first image. The Fe\,\textsc{ii} absorption line is undetected before day 17 after explosion.
  For the noise-free microlensed spectra (labeled as $S/N$ = $\infty$), the median of the 10000 positions
  and the 1$\sigma$ and 2$\sigma$ ranges are shown in black and gray. The temporal evolution of the absorption lines with an added noise of $S/N$ = 10 is shown in red.}
\end{figure*}

Figure \ref{temp_wave_min} shows the temporal evolution for each absorption line for image SN-A, where we plot the absorption wavelengths retrieved for each epoch with 1$\sigma$ and 2$\sigma$ confidence intervals for the noisy case with $S/N$ = 10 and the noiseless case denoted as $S/N$ = $\infty$ for the two resolutions $R$ = 100 (left panels) and 250 (right panels).  We obtain the narrowest distributions of the measured wavelengths $\lambda(t_i)$ for $R$ = 250, as expected because this is the highest resolution investigated. Comparing the lowest resolution of $R$ = 100 to $R$ = 250, we see that the method starts to break, especially for the H$\mathrm{\alpha}$ absorption line, as the large binning associated with the low resolution washes out the spectral features and makes the determination of the minima difficult for the considered observational $S/N$ \footnote{Although the H$\mathrm{\beta}$ absorption line is shallower than the H$\mathrm{\alpha}$ line, it performs better because it is effectively observed with at higher resolution due to its shorter absorption wavelength.}. However, the noiseless case for $R$ = 100 still performs as well as $R$ = 250, but the evolution of the retrieved absorption minimum of H$\mathrm{\alpha}$ for $R$ = 100 does not increase monotonically. This suggests difficulties in fitting the minimum correctly at earlier epochs, where the line is less prominent than at later epochs and appears to be washed out by the binning applied for $R$ = 100.
 At a resolution of $R$ = 150, the wavelength determination of the absorption minima performs better and at a level similar to $R$ = 250, as seen in the phase-determination results presented below.

\begin{figure*}[hbt!]
\centering
\subfigure[]{\label{phase_FeII}\includegraphics[width=0.49\textwidth]{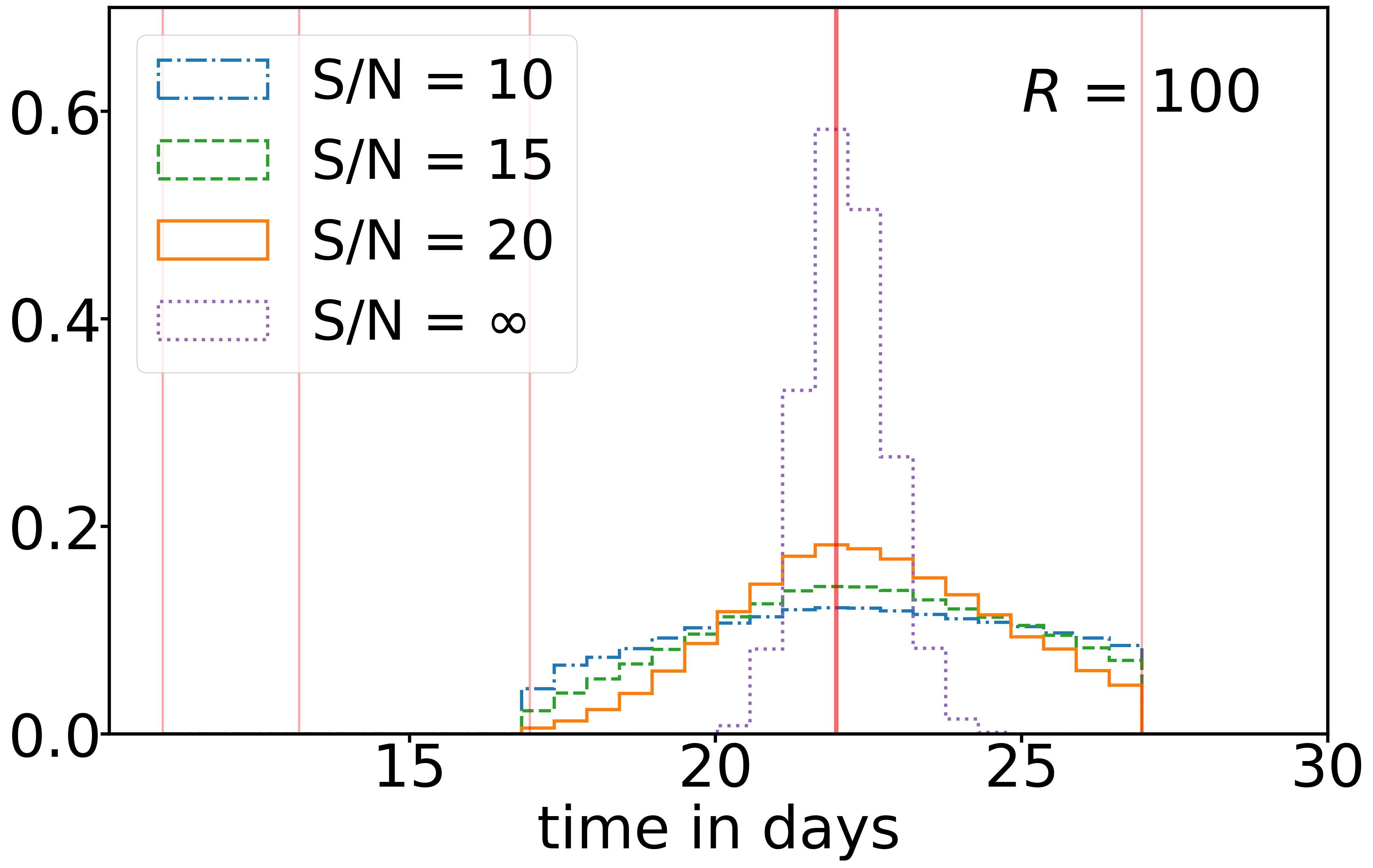}}\
\subfigure[]{\label{phase_H_alpha}\includegraphics[width=0.49\textwidth]{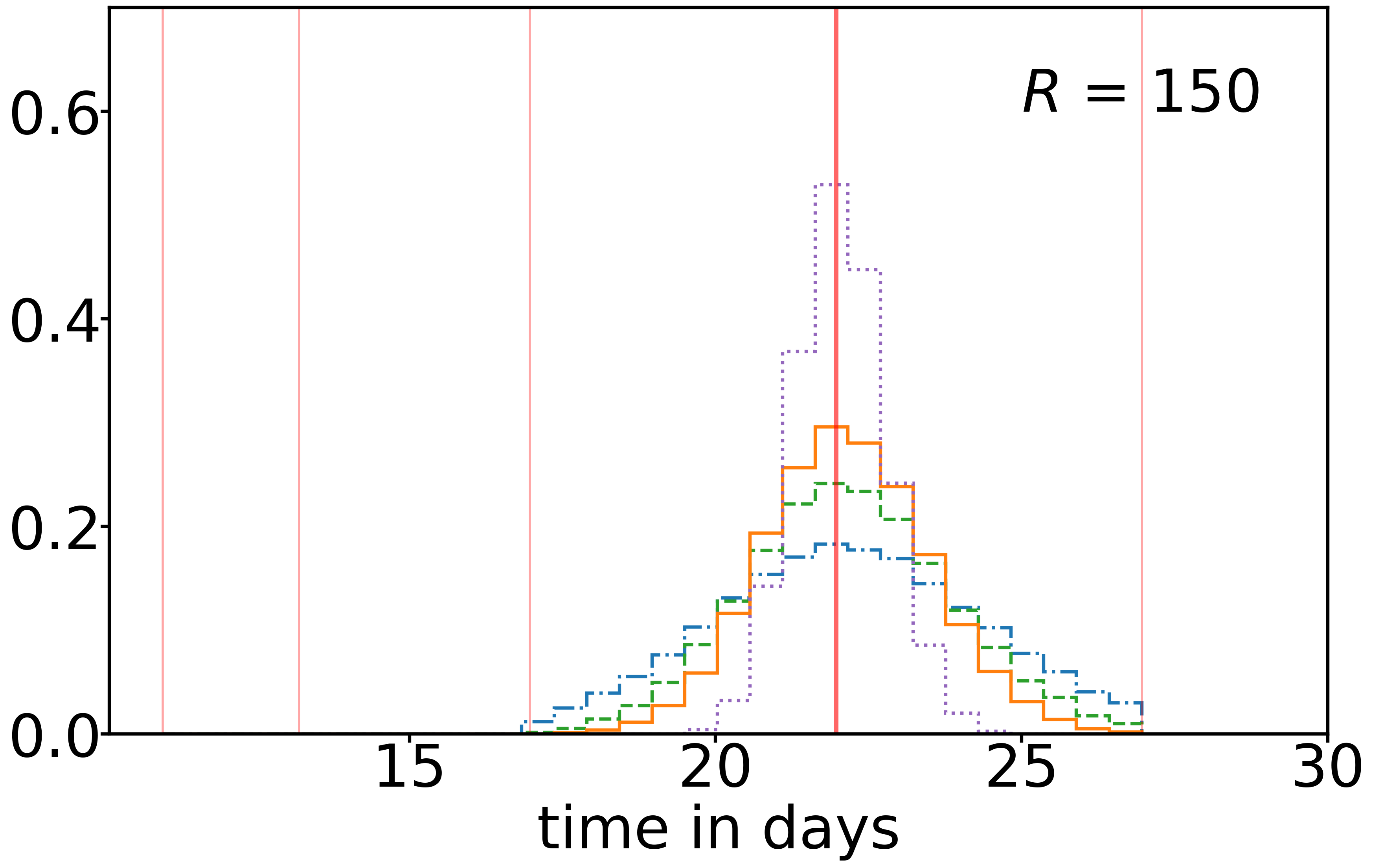}}\\
\subfigure[]{\label{phase_H_beta}\includegraphics[width=0.49\textwidth]{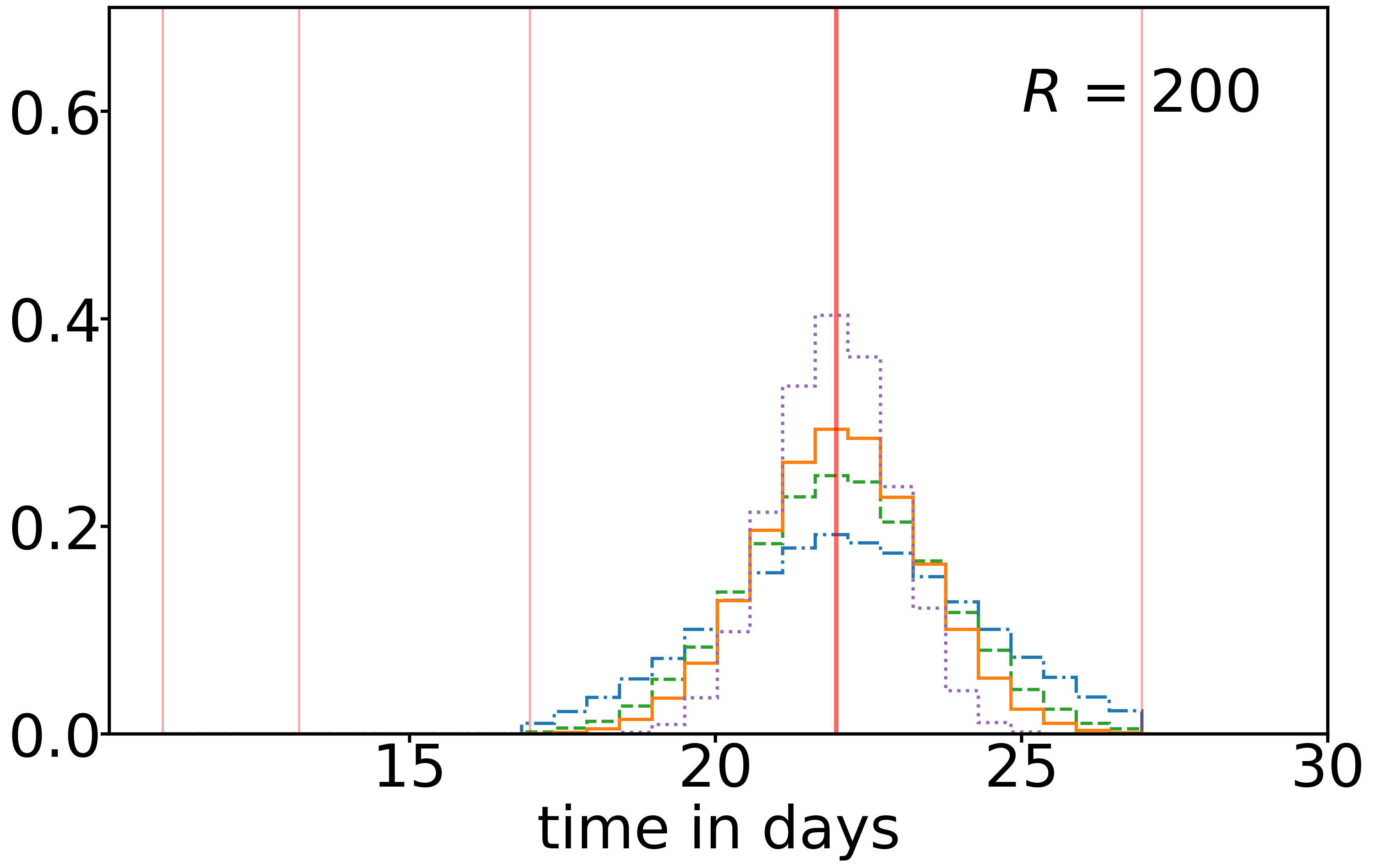}}\
\subfigure[]{\label{phase_combined}\includegraphics[width=0.49\textwidth]{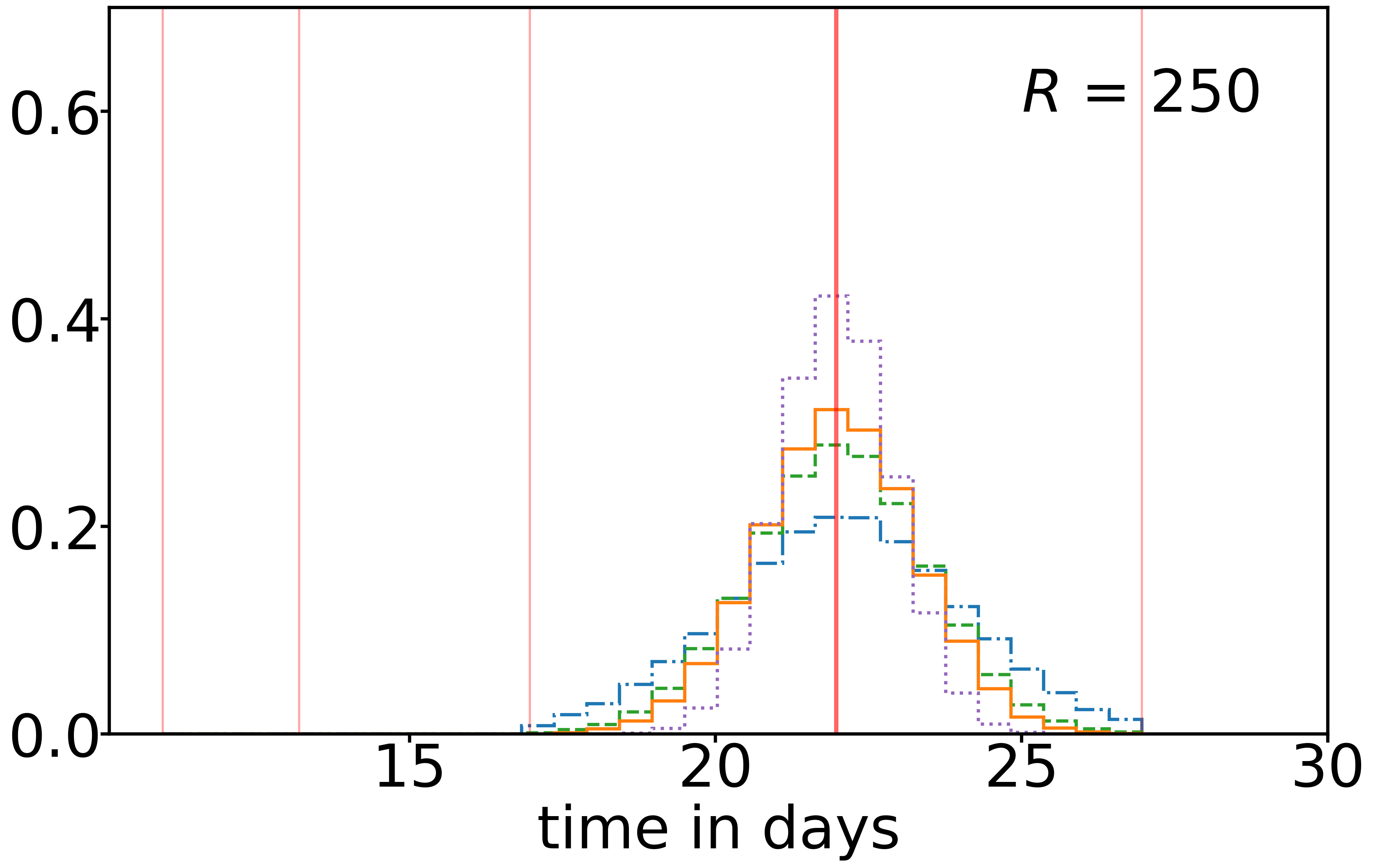}}\
\caption{\label{phases} Histograms of the retrieved phases of the
  second SN image using the Fe\,\textsc{ii},
  H$\mathrm{\alpha}$, and H$\mathrm{\beta}$ absorption lines. Phase inference
  is performed for $S/N$ = 10, 15, 20, and for noiseless spectra. The thin vertical red lines in the histograms indicate the epochs of the five
  available spectra of the first SN image. The thick vertical line
  marks the input value. The phase of the second SN image is correctly
  recovered.}
\end{figure*}

We considered a single wavelength measurement, $\lambda_{\mathrm{B}} \pm \sigma_{\mathrm{B}}$, of the SN-B image at one of the epochs to infer the time range that corresponds to the wavelength evolution in the SN-A image. Our goal was to obtain the probability distribution of $t$ in the observer frame of SN-A given the data of the first image $d_{\mathrm{A}}$ and $\lambda_{\mathrm{B}} \pm \sigma_{\mathrm{B}}$.
This can be written as $P(t\,|\, d_{\mathrm{A}}, d_{\mathrm{B}})$. From the available data, we retrieved $P(t\,|\,d_{\mathrm{A}}, d_{\mathrm{B}})$ using importance sampling \citep[e.g.][]{Lewis2002}, which is described in detail in HOLISMOKES V.
For the following results, we used the fourth epoch of SN-B as $\lambda_{\mathrm{B}} \pm \sigma_{\mathrm{B}}$ because it allowed us to use all three absorption lines of H$\mathrm{\alpha}$, H$\mathrm{\beta}$, and Fe\,\textsc{ii}.
Figure \ref{phases} shows the resulting phase retrievals as histograms combining all three absorption lines for $R$ = 100, 150, 200, and 250, considering all three noise cases with $S/N$ = 10, 15, and 20, as well as the limit set by the noiseless case ($S/N$ = $\infty$). We assumed that the noise and microlensing are uncorrelated when combining several absorption lines.
We list the retrieved values and 1$\sigma$ uncertainties in Table \ref{phase_uncertainties}.
Appendix \ref{app: additional_plots} shows the results for the individual absorption lines.

\begin{table*}[hbt!]
\caption{Retrieved phases and 1$\sigma$ uncertainties for the
  phase retrievals.}
\label{phase_uncertainties}
\centering
\begin{tabular}{l *{4}{c}}
\hline
\hline
\backslashbox{$R$}{$S/N$}
&\makebox[3em]{10}&\makebox[3em]{15}&\makebox[3em]{20}
&\makebox[3em]{$\infty$} \\
\hline  
100 & 22.2 $\pm$ 2.7 days & 22.4 $\pm$ 2.4 days & 22.5 $\pm$ 2.1 days & 22.1 $\pm$ 0.7 days \\ 
\hline 
150 & 22.1 $\pm$ 2.1 days & 22.2 $\pm$ 1.7 days & 22.1 $\pm$ 1.4 days & 22.0 $\pm$ 0.7 days \\ 
\hline 
200 & 22.1 $\pm$ 2.0 days & 22.1 $\pm$ 1.6 days & 22.0 $\pm$ 1.4 days & 22.0 $\pm$ 0.9 days \\ 
\hline 
250 & 22.1 $\pm$ 1.9 days & 22.0 $\pm$ 1.5 days & 22.0 $\pm$ 1.3 days & 22.0 $\pm$ 0.9 days \\ 
\hline 
\end{tabular}
\tablefoot{The values are computed by combining the Fe\,\textsc{ii},
  H$\mathrm{\alpha}$, and H$\mathrm{\beta}$ absorption lines. We assume that the noise and microlensing in these lines are
  uncorrelated.}
\end{table*}

Combining all three absorption lines, we find only little to no change in precision for the investigated resolutions above $R$ = 100, reaching uncertainties between $\sim$1-2 days, depending on $S/N$.  For the mock time-delay value of 22 days, this corresponds to $\lesssim$10\% uncertainty. Since systems with longer time delays still have the same uncertainty in days, a lens system with a time delay of $\gtrsim$30 days would allow a $\lesssim$5\% uncertainty for $S/N$=20. Phase-retrieval uncertainties generally depend more on the achievable $S/N$ than on the resolution. For $S/N$ = 10 to 20, any resolution discussed in this work above $R$ = 100 yields almost the same precision. For $S/N$ = 15 or greater, $R$ = 150 is already sufficient for precise time-delay measurements with less than 10\% uncertainty.
Compared to our previous study in HOLISMOKES V, we expected worse uncertainties for lower resolutions because of the loss of information in absorption-line features at lower spectral sampling.
The highest precision is achieved for $R$ = 250, almost matching that in our previous work on the \textsc{tardis} spectra in HOLISMOKES V. For $R = 100$, the binning is too sparse to sufficiently sample the absorption features, and the retrieval of the spectral delay fails. For $R \geq 150$, the method is applicable, but with less precision than the results shown in HOLISMOKES V, as expected.
In the limiting case of $S/N = \infty$, we achieve even better precision than previously predicted in HOLISMOKES V for all examined resolutions because we slightly improved the fitting procedure, making the initial line detection more robust and therefore facilitating the Gaussian fit procedure.
The results for the retrieved phase using the individual absorption lines, shown in Appendix \ref{app: additional_plots}, indicate the same trend: the precision of each line depends more strongly on $S/N$ than on $R$ for $R >$ 100.
Combining the phase retrieval of several absorption lines yields a less biased and more precise result than considering individual lines alone, in agreement with the analysis presented in HOLISMOKES V.
Appendix \ref{sec:app:background-contamination} further shows that spatial and spectral variations in the background from the host or lens galaxy are mild and do not affect our results.

\section{Forecast of $H_{0}$ precision}
\label{sec: H0 inference}

To estimate the $H_{0}$ precision for the different $S/N$ and spectral resolutions, we adopted the retrieved phase precisions as the time-delay measurement precisions. We then estimated the uncertainties of additional quantities required to forecast the $H_{0}$ precision,  following \cite{Suyu2020}.

We first determined the precision of the time-delay distance $D_{\Delta t}$ through error propagation, assuming that the measurements of the time delay $\Delta t$ and the Fermat potential $\Delta \phi$ are uncorrelated:
\begin{equation}
\label{eq:Ddeltat}
	\frac{\delta(D_{\Delta t})}{D_{\Delta t}} = \sqrt{\left(\frac{\delta(\Delta t)}{\Delta t}\right)^{2} + \left(\frac{\delta(\Delta \phi_{\mathrm{d, mod}})}{\Delta \phi_{\mathrm{d, mod}}}\right)^{2} + \left(\frac{\delta(\Delta \phi_{\mathrm{d,env}})}{\Delta \phi_{\mathrm{d,env}}}\right)^{2}}.
\end{equation}
This depends on the precisions of the time delay, $\frac{\delta(\Delta t)}{\Delta t}$, and the Fermat potential $\frac{\delta(\Delta \phi)}{\Delta \phi}$, for which we distinguish between the precision of lens modeling, $\frac{\delta(\Delta \phi_{\mathrm{d, mod}})}{\Delta \phi_{\mathrm{d, mod}}}$, and that of additional line-of-sight components and lens environment, $\frac{\delta(\Delta \phi_{\mathrm{d,env}})}{\Delta \phi_{\mathrm{d,env}}}$, which we also assume to be uncorrelated.
For the time-delay precision, we used the results from Sect. \ref{sec: SN phase inference from spectra} for low resolution spectra with $\Delta t = 22$ rest-frame days. We estimated the modeling uncertainty $\Delta \phi_{\mathrm{d, mod}}$ assuming imaging taken with a space telescope. For primary lens-galaxy mass modeling, we can reach $\frac{\delta(\Delta \phi_{\mathrm{d, mod}})}{\Delta \phi_{\mathrm{d, mod}}} \leq 3 \%$ \citep{Suyu2020} with spatially resolved kinematics of the foreground lens \citep{Yildirim2020, Yildirim2023, Wang2025}. 
For the remaining precision of the lens environment $\frac{\delta(\Delta \phi_{\mathrm{d,env}})}{\Delta \phi_{\mathrm{d,env}}}$, we similarly adopted a value $\leq 3 \%$ \citep{Greene2013}.
We assumed a radiation-free flat $\mathrm{\Lambda}$ cold dark matter ($\mathrm{\Lambda}$CDM) cosmology with $H_{0} = 72\, \mathrm{km s^{-1} Mpc^{-1}}$ \citep{Bonvin2017}, $\Omega_{\rm m}$ = 0.32, and $\Omega_{\rm \Lambda}$ = $1 - \Omega_{\rm m}$ = 0.68 \citep{Planck2020}.
To quantify the precision with which we can recover the input $H_0$ value, we followed the sampling procedure described in \cite{Suyu2020}, which is based on a Monte Carlo method. 
We set the lens redshift to $z_{\mathrm{d}} = 0.4$ and constructed a sample of 20 lensed SNe at a redshift $z_{\mathrm{s}} = 0.8$. For the various mock measurements of $D_{\Delta t}$, we assumed a Gaussian distribution with a standard deviation calculated from Eq. (\ref{eq:Ddeltat}) for the different resolutions considered. During the sampling process, we fixed the value of $\Omega_{\mathrm{m}}$ to 0.32 because we assumed no additional constraints on it.

\begin{figure}[hbt!]
\centering
{\includegraphics[width=0.49\textwidth]{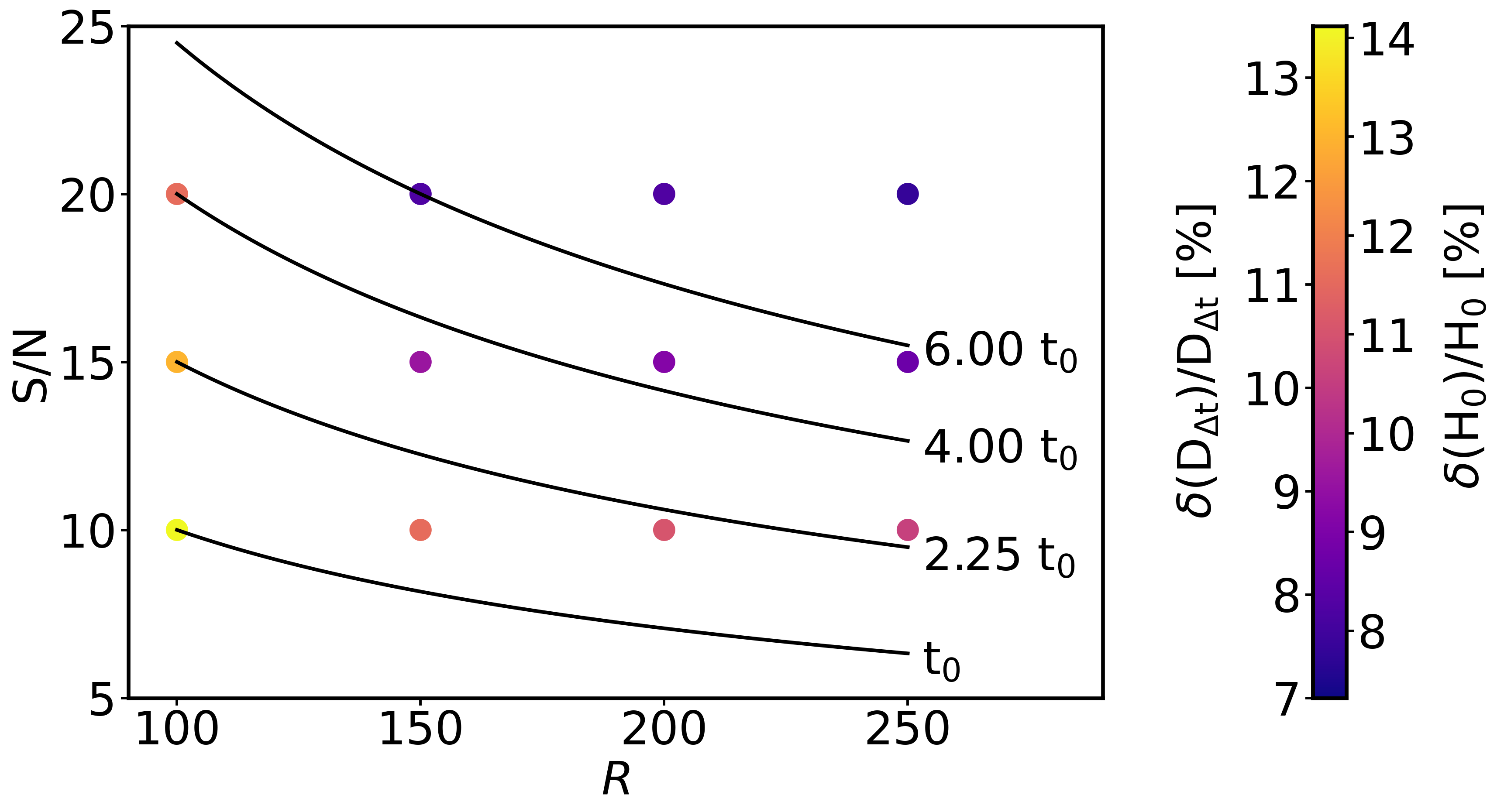}}\
\caption{\label{H0_prec}Precision on $D_{\Delta t}$ and $H_{0}$ for a single SN for each $R$ and $S/N$ case. The color gradient shows the precision values. We include lines of constant exposure time $t_{0}$, which scale with $R^{-1/2}$, in black.}
\end{figure}

Table \ref{H0_values} lists the resulting precision on $H_0$. These values are slightly worse than the precision of $D_{\Delta t}$ by a factor $\sim 1.05$.
Figure \ref{H0_prec} shows a summary of the precisions on $D_{\Delta t}$ and $H_0$ for each $R$ and $S/N$. We also include lines of constant arbitrary exposure times, denoted by $t_0$, in the plot.
To achieve twice the $S/N$, we require a factor of four in the exposure time. Doubling the resolution requires, on average, twice the exposure time. To improve precision on $H_0$ with a specific exposure time, it is beneficial to take lower-resolution spectra down to $R\sim150$, provided that the resolution does not fall below $R \sim$150.

\begin{table}[hbt!]
\caption{Predicted $H_{0}$ precision for a single lensed SN IIP.}
\label{H0_values}
\centering
\begin{tabular}{c c c c}
\hline
\hline
\backslashbox{$R$}{$S/N$}
&\makebox[3em]{10}&\makebox[3em]{15}&\makebox[3em]{20} \\
\hline 
100 & 14.2\% & 12.4\% & 11.1\% \\ 
\hline 
150 & 11.1\% & 9.1\% & 7.8\% \\ 
\hline 
200 & 10.6\% & 8.7\% & 7.8\% \\ 
\hline 
250 & 10.1\% & 8.3\% & 7.5\% \\ 
\hline 
\end{tabular}
\end{table}

\section{Rates of lensed SNe II in the era of LSST}
\label{sec: SN_numbers}

We used the works of \citet{Arendse2024} and \citet{Wojtak2019} to estimate the number of LSNe II expected over the ten years of the LSST for which sufficiently precise spectroscopic time delays can be measured with ground-based and space-based follow-up observations. Since lensed SNe IIP are rare, we extended the study to other SN types that are expected to perform equally well: SNe IIL and SNe Ia.
Because the method for determining the time delay from spectral features does not depend on any physical information of the SN mechanism, it can also be applied to other LSN types as long as sufficient temporal evolution of the spectra occurs and microlensing effects are negligible, as is, for example, the case for SNe Ia during the first approximately ten rest-frame days after explosion \citep{Suyu2020} as well as SNe IIL. Unfortunately, the most abundant LSN type, IIn, does not show broad P-Cygni features from which line velocities can be determined for use in our spectroscopic time-delay retrieval method.

We began with the catalog of simulated lensed SNe Ia provided by \citet{Arendse2024}, which contains the positions, time delays, and fluxes of 5000 LSN Ia doubly imaged systems (referred to as ``doubles'') and 5000 LSN Ia quadruply imaged systems (referred to as ``quads''). Applying the detection criteria of \citet{Wojtak2019}, 
the total detectable number of doubles is $n_{\mathrm{doubles}} = 4462$, and that of quads is $n_{\mathrm{quads}} = 4975$.
We anchored these numbers to the absolute yearly numbers of lensed SNe Ia in the LSST estimated by \citet{Wojtak2019}, as done by \citet{Arendse2024}.
These correspond to the actual expected annual detections of 64 LSN Ia double systems and 25 LSN Ia quads \citep{Wojtak2019}. Furthermore, taking into account the expected fraction of SN Ia of 29\% from \citet{Wojtak2019} relative to all other types of SN, the resulting normalization factor $C_{\mathrm{doubles}}$ for doubles (applied to detectable systems in the catalog of \citet{Arendse2024} to obtain the expected number of LSST detections over 10 years) is calculated as follows:
\begin{equation}
\label{norm_double}
    C_{\mathrm{doubles}} = \frac{64}{n_{\mathrm{doubles}}} \frac{10}{0.29} = 0.49,
\end{equation}
where the factor of ten corresponds to the 10-year duration of the LSST.
Similarly, we calculate the normalization factor $C_{\mathrm{quads}}$ for quads as
\begin{equation}
\label{norm_quad}
    C_{\mathrm{quads}} = \frac{25}{n_{\mathrm{quads}}} \frac{10}{0.29} = 0.17.
\end{equation}
From the full catalog provided by \citet{Arendse2024}, we note the following:
\begin{itemize}
    \item The last trailing image is usually the faintest, even after accounting for microlensing effects.
    \item  The second and third images in quad systems are often spatially close together and will be strongly blended in ground-based observations.
    \item Doubles often have much longer time delays $\Delta t$ than quads; however, when considering the first and last images of a quad system, the time delays are also sufficiently long, $\Delta t \gtrsim 15$ days, given the uncertainties predicted in Sect. \ref{sec: SN phase inference from spectra} and the desired time-delay precision of $\leq$10\% for precise measurements.
\end{itemize}
In addition to the normalization factors set from \citet{Wojtak2019}, we also applied certain selection criteria by imposing thresholds on the detectable peak magnitude, the minimum time delay, and the image separation to obtain accurate and precise spectroscopic time delays. We selected SNe with an $i$-band peak magnitude brighter than 23 mag to allow for a spectroscopic follow-up with sufficient $S/N$ (see Sect. \ref{sec: Exposure time estimation}). This corresponds to a typical SN II with an absolute magnitude of $\sim$$-17$ mag in the rest-frame V band \citep{Anderson2014} at redshift \zs = 0.8 and a lensing magnification $\mu \sim$ 2.
We also required a time delay of $\geq$15 days to ensure that the spectroscopic time delay could be determined with uncertainties $\leq$10\%. For ground-based observations, the image separation between each lensing images should be $\geq$1.5 arcsec, to allow deblending of the lensing images, as the typical seeing is around 1.0 arcsec. For space-based observations, we adopted an image separation limit of $\geq$0.2 arcsec, considering the spatial resolution of the JWST in the wavelength regime of the absorption lines in Sect. \ref{sec: SN phase inference from spectra}, as well as contamination of the host and lens galaxies
\footnote{
When the spatial resolution is high, the number of LSNe from space-based observations is limited by the time-delay constraint of $\geq 15$\, days rather than the image separation constraint of $\geq0.2 \ \mathrm{arcsec}$.  Even if the image separation constraint is more stringent with $\geq0.5 \ \mathrm{arcsec}$, the number of suitable LSN systems remains the same as that of the $\geq0.2 \ \mathrm{arcsec}$ constraint.}
We provide more details on the JWST spectrograph in Section \ref{sec: Exposure time estimation}

Using the normalization to \citet{Wojtak2019} as stated in Eq. \ref{norm_double} for double systems and in Eq. \ref{norm_quad} for quad systems, we expect $\sim$71 ground-based doubles and approximately nine ground-based quads or $\sim$133 space-based doubles and $\sim$49 space-based quads of all SN types within 10 years of the LSST. As \citet{Wojtak2019} expect about 8\% of all detected LSNe will be of type IIP and IIL, which we subsume here as type II, we expect approximately six LSN II doubles and $\sim$1 LSN II quad with ground-based observations and $\sim$11 LSN II doubles and approximately four LSN II quads with space-based observations in 10 years of the LSST to be useful for spectroscopic time-delay measurements. Applying the same approximations for SNe Ia, we expect $\sim$21 LSN Ia doubles and approximately three LSN Ia quads with ground-based observations and $\sim$38 LSN Ia doubles and $\sim$14 LSN Ia quads with space-based observations.
We summarize the numbers in Table \ref{LSN_numbers}.

\begin{table}[hbt!]
\caption{Predicted number of LSNe for precision cosmology based on spectroscopic time-delay retrieval in the era of LSST.}
\label{LSN_numbers}
\centering
\begin{tabular}{c c c c c}
\hline
\hline
 & &\makebox[3em]{LSNe II}&\makebox[3em]{LSNe Ia}&\makebox[3em]{total} \\
 & &\makebox[3em]{(IIP + IIL)} & & \\
\hline 
\multirow{2}{*}{ground-based} & doubles & 6 & 21 & 27 \\ \cline{2-5}
                   & quads & 1 & 3 & 4 \\ \hline
\multirow{2}{*}{space-based} & doubles & 11 & 38 & 49 \\ \cline{2-5}
                   & quads & 4 & 14 & 18 \\ \hline
\end{tabular}
\tablefoot{We distinguish between double and quad systems for ground-based and space-based observations. The sample of LSNe II include SNe of type IIP and type IIL.}
\end{table}

In summary, despite the low rate of LSNe IIP and LSNe IIL, we expect that the method will be applicable to $\sim$30 LSNe showing a temporal spectroscopic evolution in ground-based data. If space-based observations are available, the number of LSNe for spectroscopic time-delay retrieval more than doubles, as the criterion of the limited image separation can be neglected in this case. The limiting factor is the chosen time delay, as it also implicitly sets a limit on the image separation.

\section{Exposure time estimation}
\label{sec: Exposure time estimation}

We now explore the exposure times needed for the considered $S/N$ with ground- and space-based telescopes. 
Before discussing the individual instruments, we clarify that we used the resolutions $R$ of each instrument considered for our calculations. These resolutions are higher than those investigated up to this point. We adjusted them to an effective resolution $\tilde{R}$ by applying the spectral binning of $\Delta \lambda_{\mathrm{bin,mock}} = 16 \ \text{\AA}$,  used for the case with $R$ = 250, as described in Sect. \ref{sec: Exposure time calculations}.
For the ground-based case, we chose 8-meter class telescopes, and for space-based observations, we chose telescopes with two different mirror sizes. In all cases, we preferred long-slit spectroscopy as it provides higher throughput.
The two ground-based instruments considered are FORS2 \citep{Appenzeller1998}\footnote{FORS2 ETC \url{https://www.eso.org/observing/etc/bin/gen/form?INS.NAME=FORS+INS.MODE=spectro}} and MUSE \citep{Bacon2020}\footnote{MUSE ETC \url{https://www.eso.org/observing/etc/bin/gen/form?INS.NAME=MUSE+INS.MODE=swspectr}}, both mounted on the VLT. Although MUSE is an integral field spectrograph, it is not fiber-fed and therefore has high throughput.
For space-based facilities, we considered the STIS instrument on the HST \citep{Woodgate1998}\footnote{STIS ETC \url{https://etc.stsci.edu/etc/input/stis/spectroscopic/}} and NIRSpec on the JWST \citep{Jakobsen2022}\footnote{NIRSpec ETC \url{https://jwst.etc.stsci.edu/}}.
All considered instruments provide higher resolution than investigated in the sections above, but we compensate for this in Sec. \ref{sec: Exposure time calculations}.
To ensure comparable results, we individually set up the exposure time calculator (ETC) for each instrument to calculate $S/N_{\mathrm{1h}}$ for a one-hour exposure. This allowed us to determine the necessary exposure time $t_{\mathrm{exp}}$ required to achieve $S/N$ = 10.

\subsection{Instrumental setups}
\label{sec: Instrumental setups}

Each ETC has its own adjustability due to the specific features of each instrument. Before describing the individual setup, we summarize the parameters that were kept fixed across all ETCs.
We set the SN as a point source at the center of the field of view with a redshift of \zs = 0.8 with an apparent brightness of $m_{\mathrm{s}} = 23$ mag in AB magnitude in the $i$ band, as introduced in Sect. \ref{sec: SN_numbers}. The background includes the sky emission and an elliptical lensing galaxy based on the elliptical template spectrum from the Kinney-Calzetti Spectral Atlas of Galaxies \citep{Calzetti1994,Kinney1996}.
We assumed a scenario with a bright lensing galaxy at \zd$_{\mathrm{, bright}} = 0.17$ and a faint lensing galaxy at \zd$_{\mathrm{, faint}} = 0.5$. These redshifts correspond to the 16th and 84th percentiles of the distribution of lensing galaxies in the \citet{Oguri2010} lens catalog (OM10). This corresponds to a median apparent $i$-band surface brightness of $m_{\mathrm{d, bright}} = 20.77 \ \mathrm{mag \ arcsec^{-2}}$ for the bright lens and $m_{\mathrm{d, faint}} = 24.32 \ \mathrm{mag \ arcsec^{-2}}$ for the faint lens, assuming the SN image is located at the effective radius $r_{\mathrm{eff}}$.
In reality, the lensed images are distributed around the mean image separation of $\sim$1.5 arcsec taken from OM10, which is comparable to $r_{\mathrm{eff}}$, ranging from 1.2 to 2.6 arcsec for the faint and bright lens galaxy scenarios, respectively, assuming a flat surface-brightness profile due to the lack of selection options in some of the ETCs.
We modeled the lensing galaxy as a uniform and infinitely extended source in the ETC. This approach ensured that regardless of the lensing galaxy's position relative to the source, we evaluated it consistently at the effective surface brightness $r_{\mathrm{eff}}$ in our calculations.

We neglected the SN host galaxy and estimated its apparent surface brightness similarly to the lensing galaxies based on the OM10 lensing catalog, but at a redshift of \zs = 0.8, which corresponds to values below 24 $\mathrm{mag \ arcsec^{-2}}$.
For the ground-based sky conditions, we adopted five moon phases, from new to full moon, and two airmass values of 1.5 and 1.3. Furthermore, we assumed conservative values for the precipitable water vapor of 30.0 mm and 1.0 arcsec FWHM for the image quality for FORS2 and 0.5 arcsec FWHM for MUSE, which uses ground-layer adaptive optics (AO) to improve image quality. We set the space-based FWHM for STIS  to $\sim$0.1 arcsec, based on the diffraction limit of HST at 10000 \AA , and for NIRSpec  we adopt a value of $\sim$0.04 arcsec.
For every setup, we aimed for a spectral bin size matching $\Delta \lambda_{\mathrm{bin,mock}} = 16 \ \text{\AA}$, corresponding to the smallest bin size of our low-resolution mock spectra, and a spatial binning matched to the spatial FWHM to remain as consistent as possible with the mock spectra.

We differentiated between the different spectrographs for the individual instrumental setups while keeping the individual parameters as comparable as possible. We selected grisms and gratings that cover the wavelength range of the redshifted Fe\,\textsc{ii} and H$\mathrm{\beta}$ absorption lines for FORS2, MUSE, and STIS. For NIRSpec, we also included H$\mathrm{\alpha}$ ,as it is the only instrument capable of covering the infrared.

For FORS2 spectroscopy, we used the grism $GRIS\_300I+11$ with a 1.0 arcsec slit.  We set the source magnitude in the Bessell $I$ filter. We adopted the standard-resolution setup with the MIT red-optimized CCD detector, using 2 $\times$ 2 high-gain readout mode with a frequency of 100 kHz, and no polarimetry. In the spatial direction, we binned 8 pixels of size $\Delta x_{\mathrm{bin,spa}}$ = 0.25 arcsec/pixel, since the ETC returns all values per 8 pixels in spatial direction, and we aimed to emulate the SN flux integrated over 2 $\times$ 1.0 arcsec FWHM. Because the FORS2 ETC only returns values integrated over 2 $\times$ FWHM, we adopted the same integration for the other instruments.
\footnote{FORS2 manual \url{https://www.eso.org/sci/facilities/paranal/instruments/fors/doc/VLT-MAN-ESO-13100-1543_P01.pdf}}
    
Since MUSE is not capable of slit spectroscopy but  instead employs integral-field units (IFU), we adopted a setup as close as possible to the FORS2 setup. For MUSE, we applied the wide field mode with AO, the nominal wavelength range, and the Sloan $i$ filter to set the source magnitude. We coadded 5 $\times$ 5 spatial pixels of size $\Delta x_{\mathrm{bin,spa}}$ = 0.2 arcsec/pixel, which approximately matches the spatial coverage of the expected 2 $\times$ FWHM of 2 $\times$ 0.5 arcsec with AO, assuming an uncorrected seeing of 1 arcsec.\footnote{MUSE manual \url{https://www.eso.org/sci/facilities/paranal/instruments/muse/doc/ESO-261650_MUSE_User_Manual_P115.pdf}}

For STIS, we used the G750L grating with the 52 arcsec $\times$ 0.2 arcsec slit and the Sloan $i$ filter to set the source magnitude. We applied the 1 $\times$ 1 pixel binning with a gain of 1 e$^{-}$/ADU in one frame and set the detector dark current at its medium value. The extraction region height in the spatial direction is 5 pixels, corresponding to 0.25 arcsec. For the background, we used the average standard zodiacal light and average earth shine light normalizations.\footnote{STIS manual \url{https://hst-docs.stsci.edu/stisihb}}
    
For NIRSpec, we used the grating G140M with the F070LP filter and the S200A1 slit (0.2 arcsec $\times$ 3.3 arcsec). We adopted slit spectroscopy for the JWST to maintain comparability with HST and FORS2, although we expected similar exposure times for NIRSpec IFU. The source magnitude is given in the SDSS $i$ band, and we assumed a medium background. To achieve a one-hour exposure, we used 578 groups per integration, one integration, and one dither per exposure with the SUBS200A1 subarray and the near-infrared spectrograph (NRS) readout pattern. To cover 2 $\times$ FWHM, we set the aperture full height to 0.2 arcsec. For the background subtraction, we used the background region annulus with an inner edge at $\pm$ 0.15 arcsec and an outer edge at $\pm$ 0.25 arcsec measured from the slit center to match the slit width and ensure correct background extraction. The placement of the background region relative to the center of the extraction region does not affect the calculations, as the NIRSpec ETC subtracts the source light when determining the background.\footnote{NIRSpec manual \url{https://jwst-docs.stsci.edu/jwst-near-infrared-spectrograph\#gsc.tab=0}}

We list the setup for each instrument in Table \ref{set_up}.

\begin{table*}[htb!]
\caption{Parameters used in each exposure time calculator setup.}
\label{set_up}
\centering
\begin{tabular}{l *{2}{c}}
\hline
\hline
Instrument
&\makebox[3em]{Grism/grating; slit; filter; instrument specific adjustments} \\
\hline 
& GRIS\_300I+11; 1.0 arcsec; Bessell I;\\ 
VLT FORS2 & standard resolution; MIT red-optimized CCD; 100 kHz, 2x2, high readout mode; no polarimetry; \\
& $n_{\mathrm{spa}} = 8$; $\Delta x_{\mathrm{bin,spa}} = 0.25 \ \mathrm{arcsec \ pixel^{-1}}$; $\Delta \lambda_{\mathrm{bin}} = 1.6 \ \text{\AA}$; $f_{\mathrm{rebin}} = 2.2$ \\ 
\hline 
& -; -; Sloan i; \\
VLT MUSE & Wide Field Mode with AO; nominal wavelength range; coadded spatial pixels: 5x5; \\ 
& coadded spectral pixels: 1; $n_{\mathrm{spa}} = 25$; $\Delta x_{\mathrm{bin,spa}} = 0.2 \ \mathrm{arcsec \ pixel^{-1}}$; $\Delta \lambda_{\mathrm{bin}} = 1.3 \ \text{\AA}$; $f_{\mathrm{rebin}} = 3.6$ \\ 
\hline 
& G750L; 52 arcsec $\times$ 0.2 arcsec; Sloan i;\\
HST STIS & 1 pixel x 1 pixel binning; 1 e$^{-}$/ADU gain; 1 Frame; medium dark; \\
& average standard zodiacal light normalization; average standard earth shine light normalizations; \\
& $n_{\mathrm{spa}} = 5$; $\Delta x_{\mathrm{bin,spa}} = 0.05 \ \mathrm{arcsec \ pixel^{-1}}$; $\Delta \lambda_{\mathrm{bin}} = 4.9 \ \text{\AA}$; $f_{\mathrm{rebin}} = 1.8$ \\ 
\hline 
& G140M/F070LP; S200A1 (0.2 arcsec x 3.3 arcsec); SDSS i;\\
& medium background; 578 groups per integration $\hat{\approx}$ 3600 s; 1 integration per exposure; 1 dither;\\ 
JWST NIRSpec & Subarray SUBS200A1; Readout pattern NRS; 0.2 arcsec aperture full-height; \\
& background subtraction using background region; sky background sample region inner edge $\pm$ 0.15 arcsec; \\
& sky background sample region outer edge $\pm$ 0.25 arcsec; $n_{\mathrm{spa}} = 2$; $\Delta x_{\mathrm{bin,spa}} = 0.1 \ \mathrm{arcsec \ pixel^{-1}}$; \\
& $\Delta \lambda_{\mathrm{bin}} = 1.4 \ \text{\AA}$; $f_{\mathrm{rebin}} = 3.4$ \\ 
\hline 
\end{tabular}
\end{table*}

\subsection{Exposure time calculations}
\label{sec: Exposure time calculations}

To compare the exposure time of the different instruments, we adjusted the data to a similar reference, as defined in Sect. \ref{sec: Type II Supernova models}. This was not directly possible in the ETCs. Specifically, we evaluated the $S/N$ over the same spatial aperture in terms of the FWHM for the different instruments, as well as the same spectral resolutions. We therefore corrected the $S/N$ values calculated with the ETC output following the procedure below.
After setting up the ETCs, we first calculated the $S/N_{\mathrm{1h}}$ for a one-hour exposure and used this to calculate $t_{\mathrm{exp}}$ for $S/N$ = 10. We followed the same approach for each facility, calculating the $S/N$ with $t = 3600$ s from the source counts, $S_{\mathrm{s}}$, and background counts, $S_{\mathrm{n}}$, at the absorption line wavelengths of interest at \zs = 0.8, $\lambda_{\mathrm{H\beta}} = 8525 \ \text{\AA}$, $\lambda_{\mathrm{Fe\,\textsc{ii}}} = 9081 \ \text{\AA}$, and $\lambda_{\mathrm{H\alpha}} = 11473 \ \text{\AA}$. We took the values of $\lambda_{\mathrm{H\beta}}$ and $\lambda_{\mathrm{H\alpha}}$  from the spectrum of day 17 and $\lambda_{\mathrm{Fe\,\textsc{ii}}}$ from day 22, as these are the central epochs of the available spectral data for each of the three absorption lines. The $S_{\mathrm{n}}$ values include the average sky background $S_{\mathrm{sky}}$, lens light $S_{\mathrm{l}}$, dark current $S_{\mathrm{DC}}$, and read-out noise $N_{\mathrm{RON}}$ for one spectral bin.
To account for the spatial extent of the source on the detector, we multiplied $S_{\mathrm{DC}}$ and $N_{\mathrm{RON}}^{2}$ by the number of extracted spatial bins for each instrument, denoted as $n_{\mathrm{spa}}$. This ensured integration over all meaningful flux based on the spatial bin size specified for each instrument, which is measured in arcsec per pixel. We set the number of spectral bins to one and therefore did not include it as an additional factor besides $n_{\mathrm{spa}}$ in the equation for $S/N$. We already accounted for $S_{\mathrm{s}}, \ S_{\mathrm{l}},  \text{and} \ S_{\mathrm{sky}}$  in all spatial pixels in the ETCs and were therefore not additionally multiplied by $n_{\mathrm{spa}}$. We based $S_{\mathrm{DC}}$ and $N_{\mathrm{RON}}$ on values reported in the user manuals, where they are given per spatial bin. We determined $n_{\mathrm{spa}}$ from each instrument's desired spatial flux integration region, as stated in Sect. \ref{sec: Instrumental setups}. We list the values of $n_{\mathrm{spa}}$ for each instrument in Table \ref{set_up}.
We calculated the $S/N$ for a one-hour exposure as
\begin{equation}
\begin{array}{l}
    S/N_{\mathrm{1h}} = \frac{S_{\mathrm{s}}}{\sqrt{S_{\mathrm{s}} + S_{\mathrm{n}}}} \\ 
    \noindent\hspace*{10mm}
    = \frac{S_{\mathrm{s}}}{\sqrt{S_{\mathrm{s}} + S_{\mathrm{l}} +  S_{\mathrm{sky}} + n_{\mathrm{spa}} \times S_{\mathrm{DC}} + n_{\mathrm{spa}} \times N_{\mathrm{RON}}^{2}}},
\end{array}
\end{equation}
where all noise contributions follow a Poisson distribution except $N_{\mathrm{RON}}$, which represents the standard deviation of a number of electrons from the conversion process of a CCD pixel to a measurable signal rather than a counting process and is therefore squared in the equation.

To ensure comparability in terms of spectral binning, we did not modify the source and background counts in the equation of $S/N_{\mathrm{1h}}$ in the ETC itself, because the spectral binning is not a free parameter for each ETC. We scaled $S/N_{\mathrm{1h}}$ with a factor $f_{\mathrm{rebin}}$ computed from comparing the observed frame spectral binning of $\Delta \lambda_{\mathrm{bin,mock}} = 16 \ \text{\AA}$ with a spectral pixel-size $\Delta \lambda_{\mathrm{bin}}$ from the instrument,
\begin{equation}
\label{eq: rebin}
    f_{\mathrm{rebin}} = \sqrt{\frac{\Delta \lambda_{\mathrm{bin,mock}}}{\Delta \lambda_{\mathrm{bin}}}}.
\end{equation}
We list the values for $\Delta \lambda_{\mathrm{bin}}$ and $f_{\mathrm{rebin}}$ in Table \ref{set_up}.

To estimate the exposure time to achieve $S/N$ = 10, we assumed that exposures are shot-noise-dominated. 
As $S/N \propto \sqrt{t_{\mathrm{exp}}}$, the exposure time is given by
\begin{equation}
    t_{\mathrm{exp}} = \left( \frac{10}{S/N_{\mathrm{1h}} \times f_{\mathrm{rebin}}} \right) ^{2} \times 3600 \ \mathrm{s},
\end{equation}
including the rebinning factor $f_{\mathrm{rebin}}$ from equation (\ref{eq: rebin}).
 This does not apply to HST STIS as it is dominated by readout noise (RON) and dark current (DC). As $S/N_{\mathrm{1h}}$ is very low, we reduced the exposure time estimate for this instrument.
 The inefficiency of STIS in our scenario arises because the grating at the wavelengths of interest is only about 2-3\%, resulting in very small $S_{\mathrm{s}}$ compared to $S_{\mathrm{DC}}$ and $N_{\mathrm{RON}}$.

The spectral rebinning described above effectively corresponds to a rescaling of the instrumental spectral resolution. Therefore, we define an effective instrumental spectral resolution $\tilde{R}$ as
\begin{equation}
\label{equ:eff_resol}
    \tilde{R}= \frac{R}{f_{\mathrm{corr}}},
\end{equation}
where $f_{\mathrm{corr}}$ is the correction factor arising from the coverage of the FWHM $\Delta \lambda_{\mathrm{R}}$ of a spectral line with the new bin size $\Delta \lambda_{\mathrm{bin,mock}}$\footnote{see RoentDek detectors Fig. 10 \url{https://www.roentdek.com/info/OnResolution.pdf}; for bin sizes in units of the FWHM larger than 1.6 we assumed roughly linear extrapolation}.
We list the resolutions and effective resolutions for each instrument at each wavelength of the considered absorption lines in Table \ref{resol}, together with $f_{\mathrm{corr}}$ determined from the bin size $\Delta \lambda_{\mathrm{bin,mock}}$ in units of $\Delta \lambda_{\mathrm{R}}$.
\begin{table*}[htb!]
\caption{Resolutions $R$ and effective resolutions $\tilde{R}$ for each instrument.}
\label{resol}
\centering
\begin{tabular}{c c c c c c c c c c}
\hline 
\noalign{\vskip 1pt}
Instrument & $R_{\mathrm{H\mathrm{\beta}}}$ & $R_{\mathrm{Fe\,\textsc{ii}}}$ & $R_{\mathrm{H\mathrm{\alpha}}}$ & $f_{\mathrm{corr, H\mathrm{\beta}}}$ & 
$f_{\mathrm{corr, Fe\,\textsc{ii}}}$ & 
$f_{\mathrm{corr, H\mathrm{\alpha}}}$ & $\tilde{R}_{\mathrm{H\mathrm{\beta}}}$ & $\tilde{R}_{\mathrm{Fe\,\textsc{ii}}}$ & $\tilde{R}_{\mathrm{H\mathrm{\alpha}}}$ \\
\hline 
\hline
FORS2 & 660 & 660 & - & 1.30 & 1.30 & - & 508 & 508 & - \\ 
\hline 
MUSE & 3350 & 3465 & - & 3.85 & 3.75 & - & 870 & 924 & - \\ 
\hline 
STIS & 880 & 922 & - & 1.55 & 1.50 & - & 568 & 615 & - \\ 
\hline 
NIRSpec & 450 & 600 & 750 & 1.16 & 1.27 & 1.23 & 388 & 472 & 610 \\ 
\hline 
\end{tabular}
\tablefoot{We differentiate between the wavelengths of $\mathrm{H\mathrm{\beta}}$, $\mathrm{Fe\,\textsc{ii}}$, and $\mathrm{H\mathrm{\alpha}}$ absorption lines. We also list the correction factor $f_{\mathrm{corr}}$ for each instrument and absorption line.}
\end{table*}

\begin{figure}[hbt!]
\centering
{\includegraphics[width=0.489\textwidth]{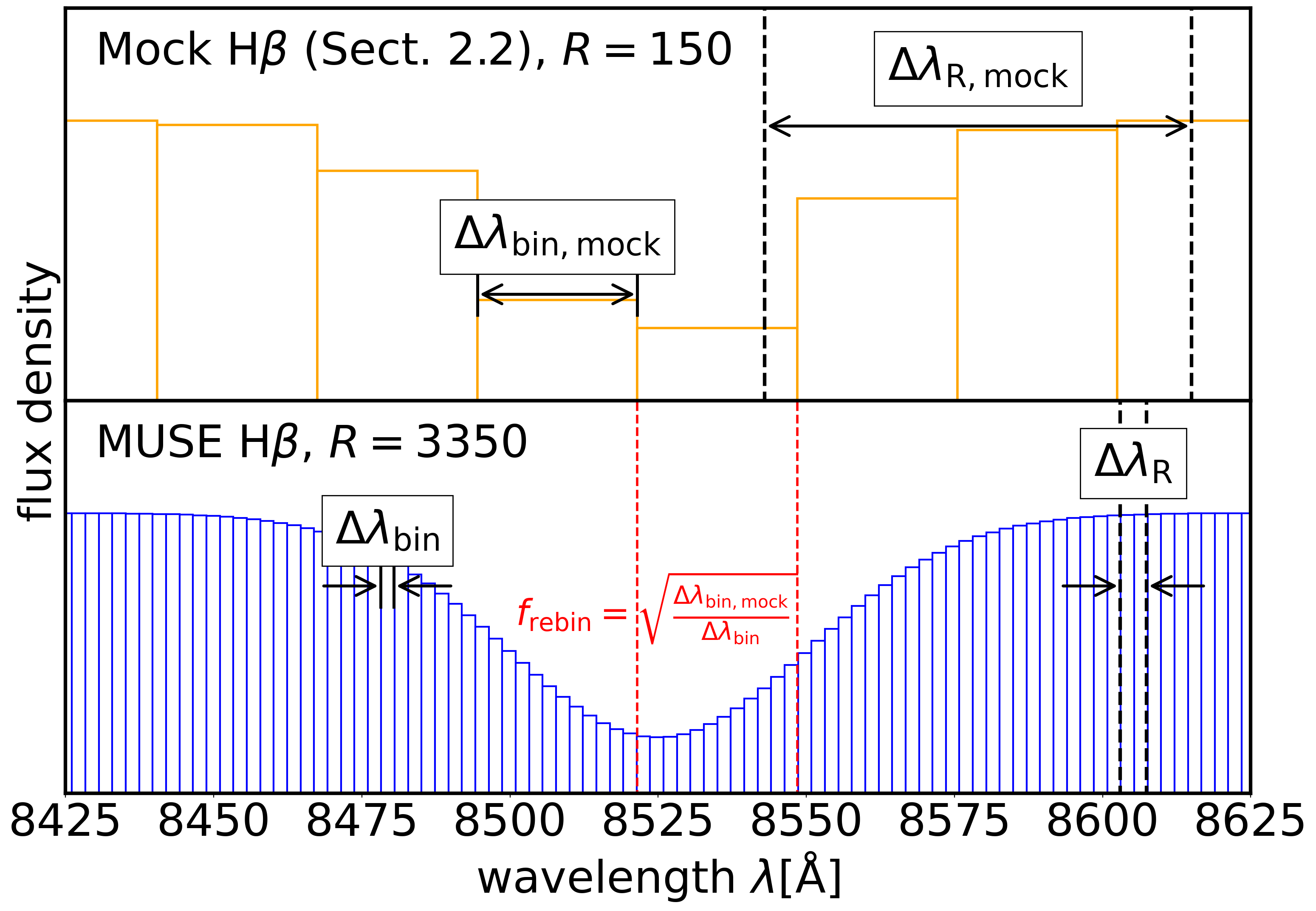}}\
\caption{\label{resol_scale}Conversion between $\Delta \lambda_{\mathrm{bin}}$ to $\Delta \lambda_{\mathrm{bin,mock}}$ and between $\Delta \lambda_{\mathrm{R}}$ to $\Delta \lambda_{\mathrm{R,mock}}$. The bottom panel shows an artificial $\mathrm{H \beta}$ absorption line observed with MUSE at resolution $R = 3350$, sampled with a spectral binning $\Delta \lambda_{\mathrm{bin}}$, (blue). This spectrum is rebinned to $\Delta \lambda_{\mathrm{bin,mock}}$ in the top panel (orange) using the conversion factor $f_{\mathrm{rebin}}$. This procedure defines a new $\Delta \lambda_{\mathrm{R,mock}}$ and enables conversion to a lower resolution of $R = 150$.}
\end{figure}

Figure \ref{resol_scale} further illustrates the difference between the desired $\Delta \lambda_{\mathrm{bin,mock}}$ and the instrument's $\Delta \lambda_{\mathrm{bin}}$, along with their conversion using $f_{\mathrm{rebin}}$. We also show $\Delta \lambda_{\mathrm{R,mock}}$ and $\Delta \lambda_{\mathrm{R}}$, which result from the resolution $R$ of the mock data (in orange, top) and of MUSE (in blue, bottom).
In the top panel, $\Delta \lambda_{\mathrm{R,mock}} = 72$ \AA \ is defined by $\frac{\lambda}{R_{\mathrm{mock}}}$ using the wavelength $\lambda$ = 10800 \AA \ as an example. In the bottom plot, we adopt the MUSE binning to illustrate $\Delta \lambda_{\mathrm{bin}}$, the difference between $\Delta \lambda_{\mathrm{R}}$ and $\Delta \lambda_{\mathrm{R,mock}}$, and the conversion with $f_{\mathrm{rebin}}$.

We list the resulting exposure times $t_{\mathrm{exp}}$ for the remaining three facilities in Table \ref{exp_times}. For the ground-based facilities, we list the results for an airmass of 1.3 and differentiate between the bright- and faint lensing galaxy scenarios, indicated by the subscripts bright and faint. The exposure times $t_{\mathrm{exp}}$ for an airmass of 1.5 are up to 10\% longer than for an airmass of 1.3.
\begin{table*}[htb!]
\caption{Estimated exposure times for FORS2, MUSE, and NIRSpec.}
\label{exp_times}
\centering
\begin{tabular}{c c c c c c c c c}
\hline
\hline
Instrument & Airmass & Moon Phase & $t_{\mathrm{exp}; \mathrm{H\beta; bright}}$ & $t_{\mathrm{exp}; \mathrm{Fe\,\textsc{ii}; bright}}$ & $t_{\mathrm{exp}; \mathrm{H\alpha; bright}}$ & $t_{\mathrm{exp}; \mathrm{H\beta; faint}}$ & $t_{\mathrm{exp}; \mathrm{Fe\,\textsc{ii}; faint}}$ & $t_{\mathrm{exp}; \mathrm{H\alpha; faint}}$ \\
\hline  
 &  & 0.0 & 6.2 h & 12.1 h &  & 5.9 h & 11.5 h &  \\ 
 &  & 0.25 & 6.6 h & 12.8 h &  & 6.3 h & 12.2 h &  \\ 
VLT FORS2 & 1.3 & 0.5 & 7.6 h & 14.6 h & - & 7.4 h & 13.9 h & - \\ 
 &  & 0.75 & 9.6 h & 18.0 h &  & 9.3 h & 17.4 h &  \\ 
 &  & 1.0 & 23.7 h & 42.6 h &  & 23.4 h & 42.0 h &  \\ 
\hline
 &  & 0.0 & 3.2 h & 5.4 h &  & 2.0 h & 3.6 h &  \\ 
 &  & 0.25 & 3.4 h & 6.0 h &  & 2.2 h & 4.1 h &  \\ 
VLT MUSE & 1.3 & 0.5 & 3.6 h & 6.6 h & - & 2.4 h & 4.8 h & - \\ 
 &  & 0.75 & 4.7 h & 7.9 h &  & 3.5 h & 6.0 h &  \\ 
 &  & 1.0 & 11.7 h & 15.3 h &  & 10.5 h & 13.4 h &  \\ 
\hline 
JWST NIRSpec & - & - & 8.9 min & 5.8 min & 3.6 min & 8.9 min & 5.8 min & 3.6 min \\ 
\hline 
\end{tabular}
\tablefoot{We assume $S/N$ = 10 for the mock  $\mathrm{H\mathrm{\beta}}$, $\mathrm{Fe\,\textsc{ii}}$, and $\mathrm{H\mathrm{\alpha}}$ spectral absorption lines. The subscripts ``bright'' and ``faint'' in $t_{\rm exp}$ denote  the respective lens galaxy brightness scenarios. For ground-based cases, we also differentiate between moon phases and airmasses.}
\end{table*}

For both ground-based facilities, we estimated exposure times of several hours, with values for MUSE roughly a factor of two lower than for FORS2, for which we estimate exposure times exceeding one day at full moon. 
Observing with FORS2 under the considered conditions is not viable, as it would typically require more than two full nights of observations.
The MUSE instrument provides better performance because AO enhances the seeing, and most observations require approximately two to seven hours of exposure time. For the space-based NIRSpec, exposure times are on the order of a few minutes.
We identify several trends from the investigated cases:
\begin{itemize}
    \item The airmass dependence is small compared to the dependence on the moon phase, with high exposure times around full moon.
    \item The exposure time strongly depends on the wavelength or the wavelength range of interest.
    \item Adaptive optics (AO) strongly reduces exposure time, making ground-based observations feasible for our setup.
    \item For the values considered in our work, the brightness of the lens galaxy is relevant for ground-based observations, increasing the exposure time in the scenario with the bright-lens by up to one hour compared to the faint-lens case. Space-based observations with NIRSpec do not depend on the brightness of the deflector in the investigated cases.
\end{itemize}

\subsection{Takeaways for future facilities}

Following our investigation of the different instruments and their required exposure times, one can envisage a hypothetical design for a future instrument that would be ideally suited to spectroscopic observations of lensed SNe for retrieving time delays.

The most efficient spectroscopic measurements can be achieved with a space-based telescope. Given that the JWST requires only minutes to achieve the required data quality, even a space telescope with a smaller main mirror would suffice. A factor-of-two difference in mirror size corresponds to a factor-of-four change in exposure time if the spatial resolution remains the same, as the $S/N$ scales linearly with the mirror diameter, and the required exposure time scales as the square of the desired $S/N$. Explicitly, halving the diameter of JWST's main mirror would increase the exposure time by a factor of four, resulting in exposure times of up to about 20  minutes for the same $S/N$.
Ideally, the resolution should be at least 150-200 to provide sufficient detail in the absorption-line features and precisely determine the wavelength of their minimum with our method. Furthermore, our investigation of HST exposure times shows that the throughput must be sufficiently high to ensure that the observations are shot-noise-dominated. Otherwise, readout noise drastically limits the possible exposure time.
The recently announced Lazuli Space Observatory \citep{Roy2026} meets these criteria well, featuring a 3 m primary mirror and an integral-field spectrograph with a resolution up to $R$ = 500 in the infrared. Designed for time-domain observations and cosmology, it complements existing observational facilities for exploring transient sources. Such facilities will enable the precise determination of time delays for strongly lensed SNe II during the LSST era, thus aiding in the precise measurement of $H_{0}$ from lensed SNe.

\section{Discussion and conclusions}
\label{sec: Discussion and Conclusion}

This work focuses on determining time delays of trailing images of LSNe IIP from the temporal evolution of absorption lines in low-resolution spectra. We used the precision achieved from the phase extraction results to forecast the precision of $H_{0}$ for a single LSN II. We also predicted the exposure times required to obtain these spectra with  $S/N=10$ for the ground-based instruments FORS2 and MUSE at the VLT and for the space-based instrument NIRSpec onboard the JWST, to assess the feasibility of the method. 

We used the temporal evolution of the absorption minima of the H$\mathrm{\alpha}$, H$\mathrm{\beta}$, and Fe\,\textsc{ii} lines from model spectra of SN 1999em. We simulated various low resolutions ($R$ = 100, 150, 200, and 250) and $S/N$ values (10, 15, and 20). Using the spectral phase-retrieval framework from HOLISMOKES V, we demonstrate that phase information can be retrieved with cosmology-grade precision even with low-resolution data.

As expected, increasing the resolution reduces the uncertainties in the retrieved absorption wavelengths. However, the comparison between $R$ = 150 and $R$ = 250 reveals only marginal differences in the uncertainties, highlighting the robustness of the retrieval process even at relatively low resolutions. Only at $R$ = 100 does the precision start to deteriorate significantly. In general, the uncertainties depend more strongly on the achievable $S/N$ than on the resolution. 
Combining the phase retrieval from all three absorption lines, assuming uncorrelated microlensing and noise, further enhances the precision and reduces bias, as shown in HOLISMOKES V. The consistency of the results with noiseless cases, where uncertainties are below one day, underscores the reliability of our method in extracting phase information even from low-resolution data.

Based on the predicted LSN IIP spectral time-delay precisions, we forecast the Hubble constant $H_0$.
Our findings demonstrate that the precision of $H_{0}$ depends more strongly on spectral $S/N$ than the resolution $R$, which results from the precision of the time-delay inference. The results shown in Fig. \ref{H0_prec} highlight the interplay between these parameters and show the stronger dependence on $S/N$. Notably, the precision on $H_{0}$ from a single LSN ranges between 7.5-14.2\% across the examined cases, with the upper value likely underestimated due to the artificial truncation of possible time delays from the restricted phase range available for the first image. We achieve better precisions for higher $S/N$ and $R$. 
This reinforces the importance of optimizing observational strategies, particularly since achieving twice the $S/N$ requires approximately $4 \times t_{\mathrm{exp}}$, while doubling the resolution entails roughly $2 \times t_{\mathrm{exp}}$.
The modeling assumptions and adopted uncertainty values for $\Delta \phi_{\mathrm{d, mod}}$ and $\Delta \phi_{\mathrm{d, env}}$, play a critical role in achieving the stated precision levels. Our assumption of $\leq 3\%$ uncertainty in both these components aligns with the benchmarks set by previous studies \citep{Suyu2020, Yildirim2020, Yildirim2023, Wang2025}. However, achieving these precision levels in practice requires spatially resolved kinematics for accurate lens mass modeling.

Since the spectroscopic time-delay retrieval method is not dependent on the detailed explosion physics of the SN and is relatively robust against microlensing during early phases, it can also be applied to other lensed SN classes beyond SNe IIP, such as SNe Ia and SNe IIL.
To evaluate the potential of using spectroscopic time-delay measurements for LSNe II (IIP and IIL) and LSNe Ia in the 10-year LSST survey, we estimated the number of suitable systems based on the simulated catalog of \citet{Arendse2024}, normalized to the LSST predictions of \citet{Wojtak2019}. After applying observational selection criteria on the peak magnitude in the $i$ band ($< 23$ mag), minimum time delay ($\geq 15$ days), and image separation ($\geq 1.5$ arcsec), we find that $\sim$31 LSNe are expected to be suitable for precise spectroscopic time-delay measurements over the full survey duration, assuming ground-based follow-up observations. This number more than doubles if space-based observations are available for a lensing image separation of $\geq 0.2$ arcsec.

The analysis of exposure-time requirements across various ground- and space-based telescopes highlights the instrumental capabilities and limitations when observing LSNe II at \zs = 0.8. The results provide insights into the efficiency of different setups, particularly in achieving the required $S/N$ = 10 for the investigated spectral resolution of $R =$ 150.
From our calculations, we observe significant differences in $t_{\mathrm{exp}}$ between the instruments studied:
\begin{itemize}
    \item Ground-based follow-up observations of LSNe II with existing facilities are very expensive because the transient nature of SNe offers little flexibility to optimize the seeing; AO is essentially the only option to reduce seeing effects. Adaptive optics (AO) on MUSE therefore decreases the exposure time to 2-3 hours for moon phases around the new moon, making this an option for future spectroscopic follow-up. 
    Additionally, the high sky brightness of around 9000 \AA \ and the low quantum efficiency of the optical CCDs further reduce the achievable $S/N$.
    \item The JWST NIRSpec calculations, yielding exposure times of minutes per spectrum, demonstrate that space-based spectral observations are a good option for spectroscopic follow-up observations.
    \item The investigation of HST STIS shows that a space-based instrument also requires sufficient throughput above 8000 \AA \ to be useful. 
    \item The derived effective resolutions $\tilde{R}$ confirm that all instruments are capable of meeting the spectral requirements of the mock models. Future facilities with lower resolutions could further reduce the exposure-time demands.
\end{itemize}

From these conclusions, one can conceptualize an instrument that is ideal for spectroscopy of lensed SNe.
The telescope should be space-based and the resolution should ideally exceed $R=150$.
Such a facility would enable precise time-delay measurements between strongly lensed SNe II images, and thereby precise $H_{0}$ determinations from lensed SNe.

In conclusion, our study extends the applicability of the HOLISMOKES V spectral phase-retrieval framework to low-resolution, microlensed spectra, confirming its efficacy in extracting reliable phase information for future precise $H_{0}$ measurements during the LSST era across various noise levels and resolutions. To efficiently collect the data required for precise time delays retrieved with this method, a space-based facility such as the JWST or a future facility capable of collecting the data in a short period of time would be ideal. Meanwhile, ground-based instruments such as MUSE remain competitive with AO, but with exposure times on the order of hours.

\begin{acknowledgements}
We thank W.~E.~Kerzendorf for discussions about the usage and adjustments of the exposure time calculators. We also thank S.~Perlmutter and M.~Rigault for feedback on our exposure time estimations. Further, we thank B.~Brooks of the STScI help desk for providing us with detailed information on the JWST NIRSpec ETC. We also thank the anonymous referee for their constructive feedback, which enhanced the quality of this work. 
We thank the Max Planck Society for support through the Max Planck Research Fellowship for SHS.
D.~Sluse acknowledges the support of the Fonds de la Recherche Scientifique-FNRS, Belgium, under grant No. 4.4503.1.
This project has received funding from the
European Research Council (ERC) under the European Union’s Horizon
2020 research and innovation programme (LENSNOVA: grant agreement
No. 771776; COSMICLENS: grant agreement No. 787886).
This research is supported in part by the Excellence Cluster ORIGINS, which is funded by the Deutsche Forschungsgemeinschaft (DFG, German Research Foundation) under Germany's Excellence Strategy -- EXC-2094 -- 390783311.
\end{acknowledgements}

\bibliographystyle{aa}
\bibliography{low_resol_spectra_H0.bib}

\FloatBarrier
\clearpage
\appendix

\section{Additional phase inference plots and data}
\label{app: additional_plots}

\begin{figure*}[hbt!]
\centering
\subfigure{\label{}\includegraphics[width=0.31\textwidth]{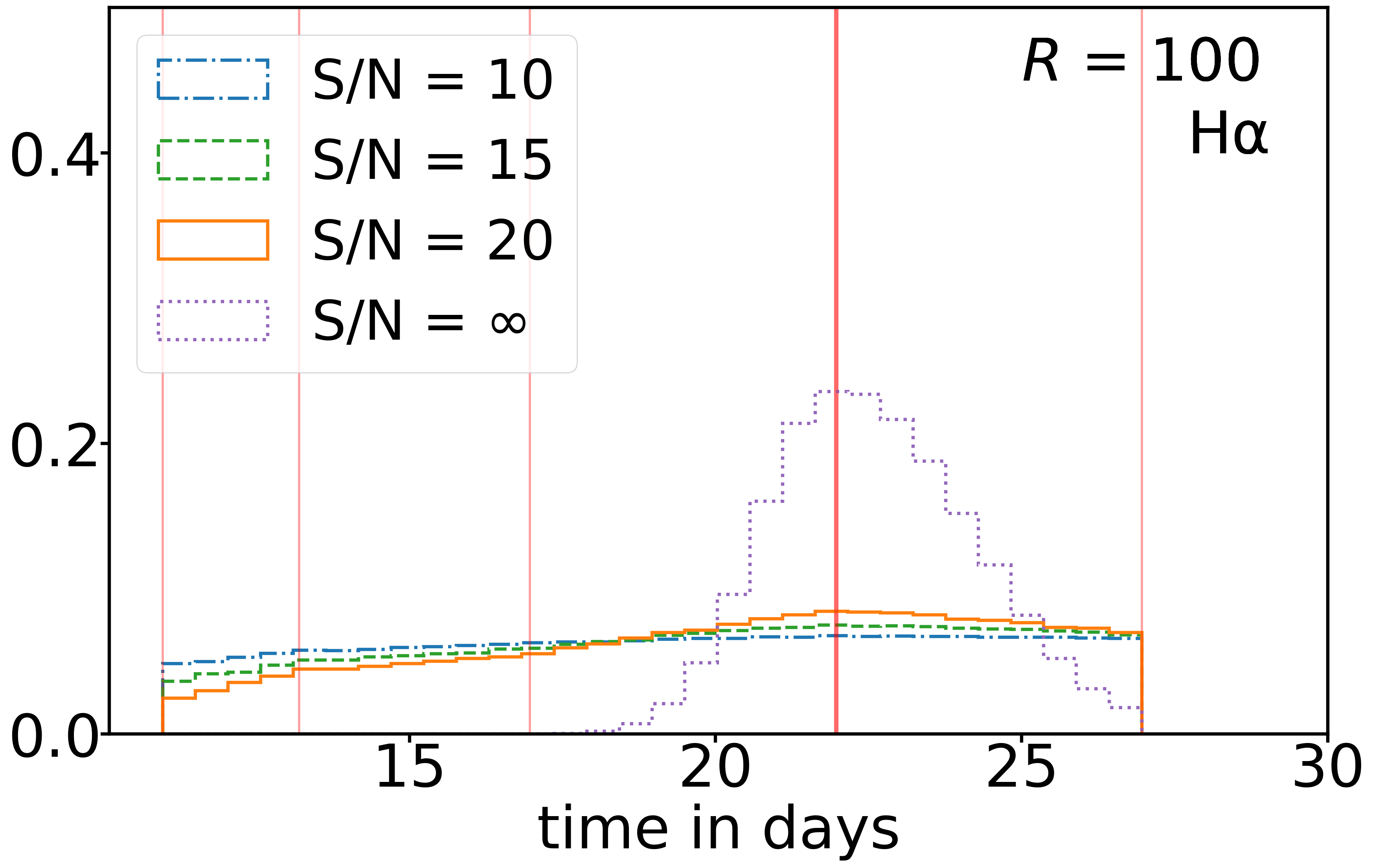}}\
\subfigure{\label{}\includegraphics[width=0.31\textwidth]{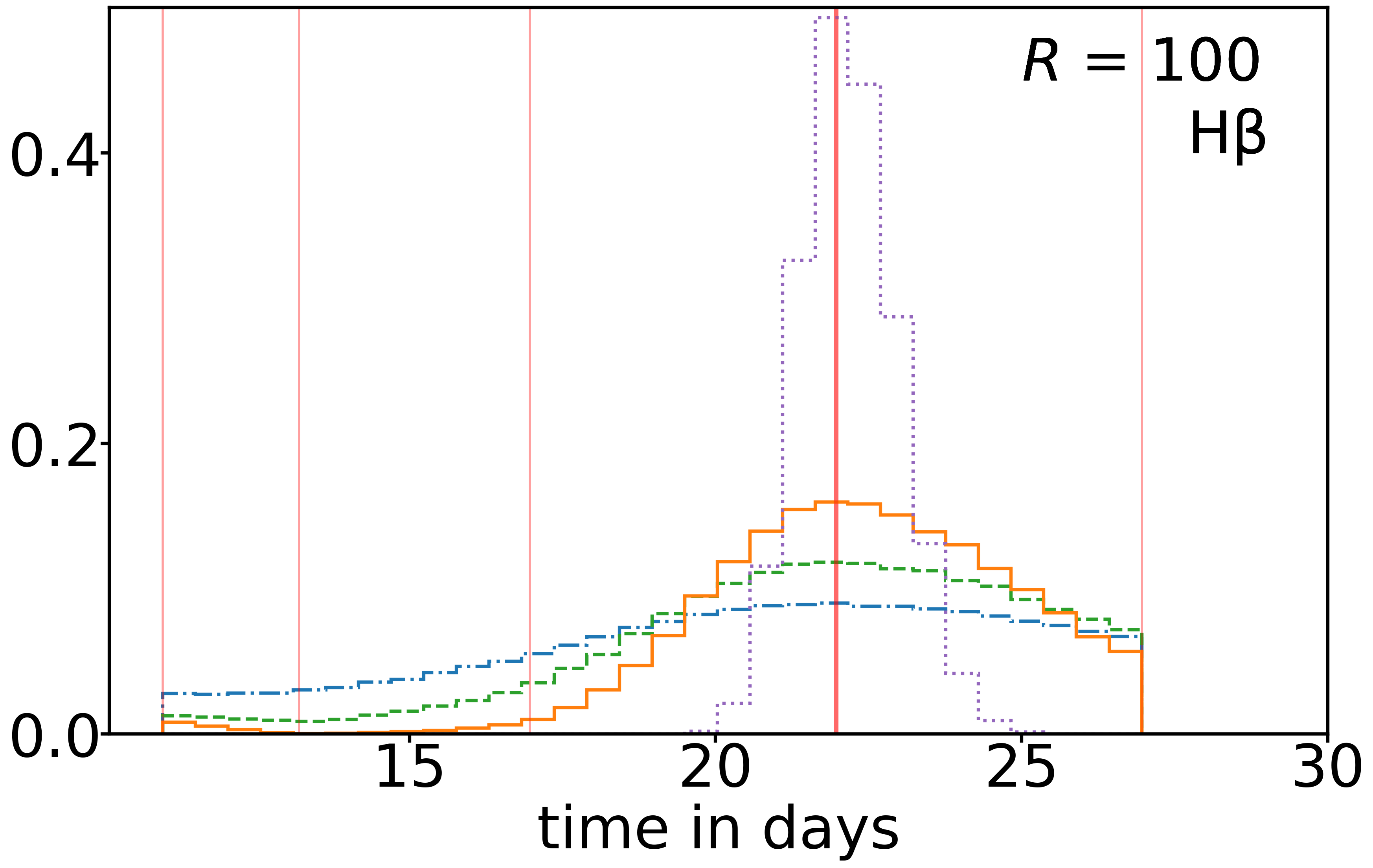}}\
\subfigure{\label{}\includegraphics[width=0.31\textwidth]{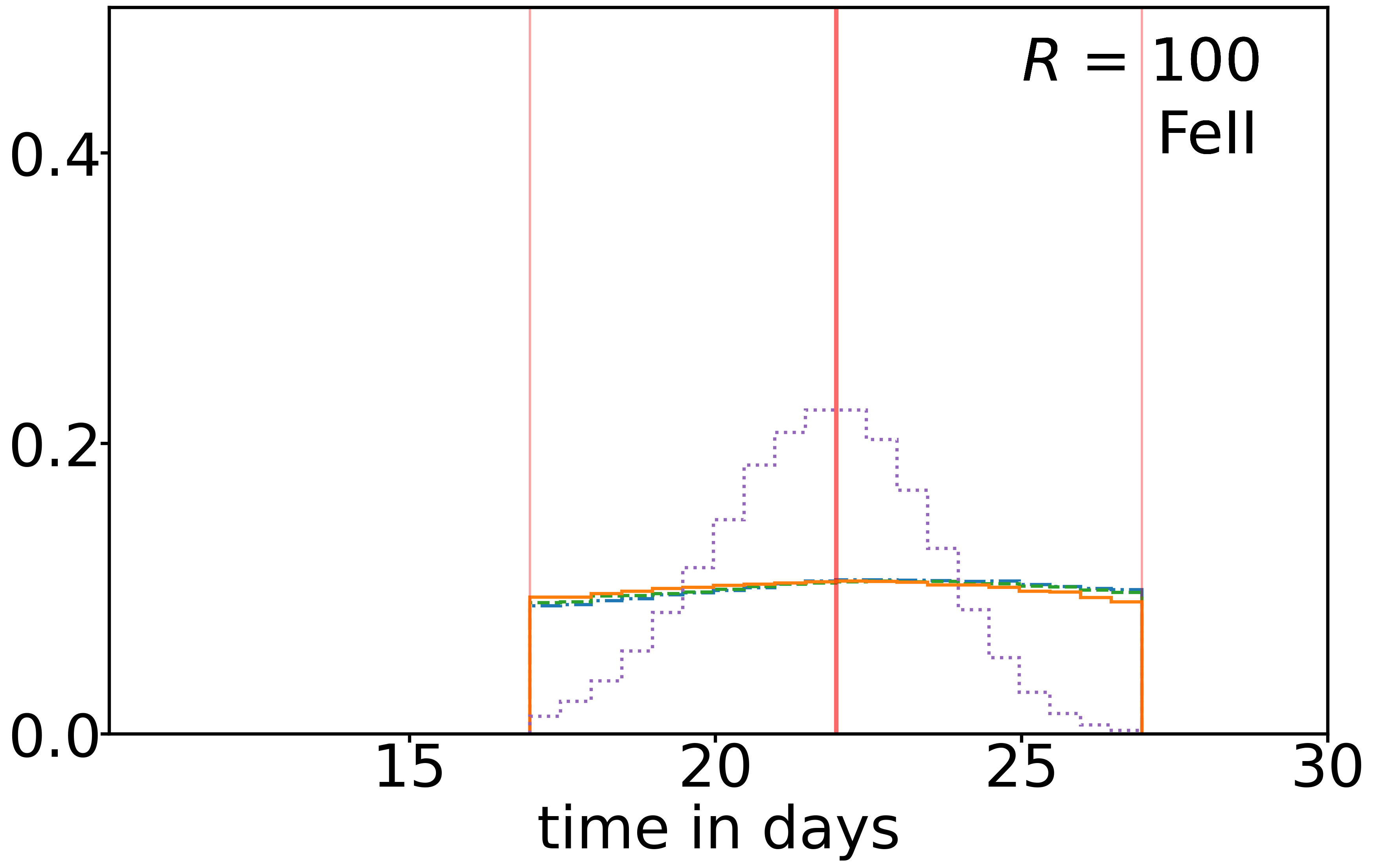}}\\
\subfigure{\label{}\includegraphics[width=0.31\textwidth]{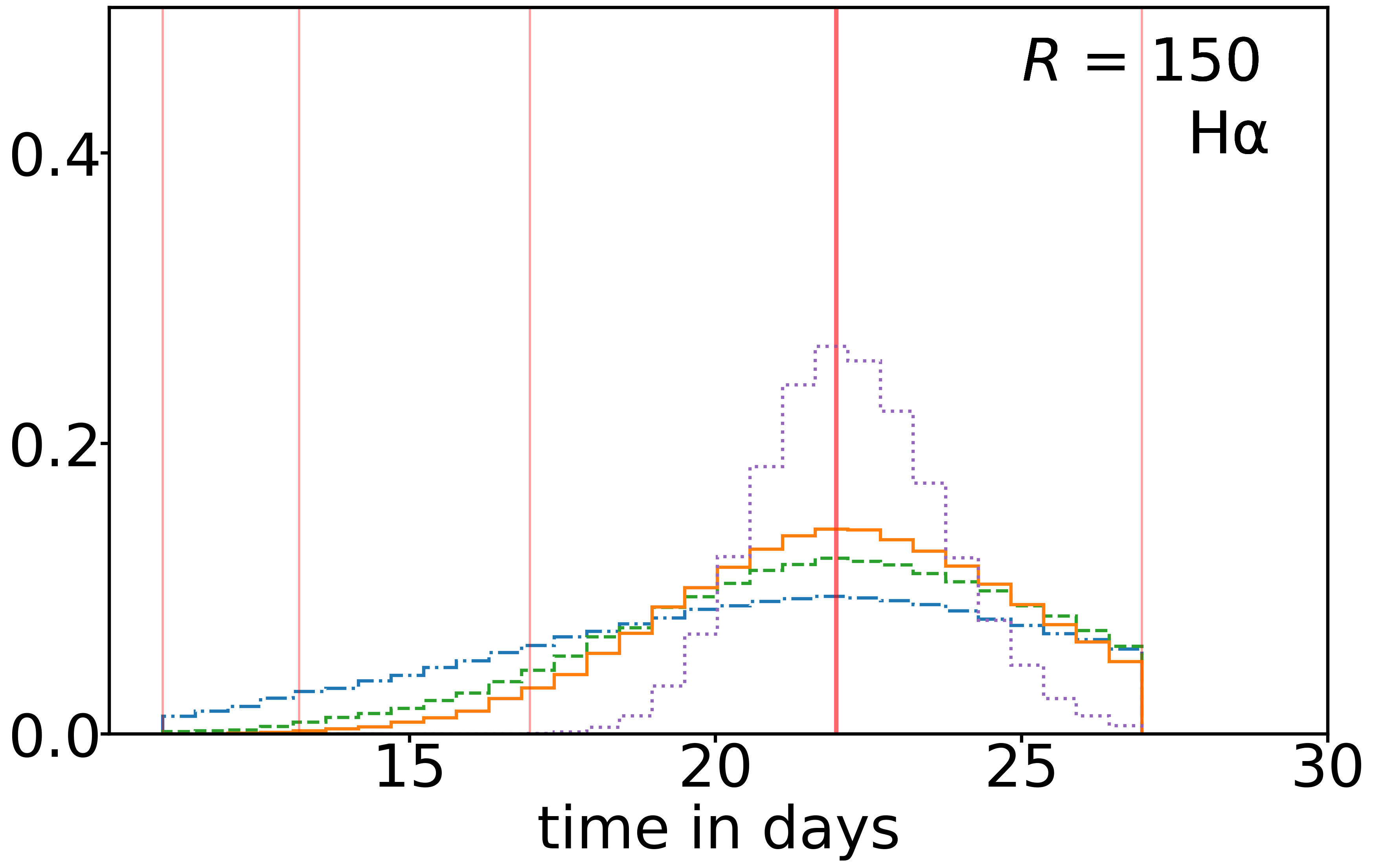}}\
\subfigure{\label{}\includegraphics[width=0.31\textwidth]{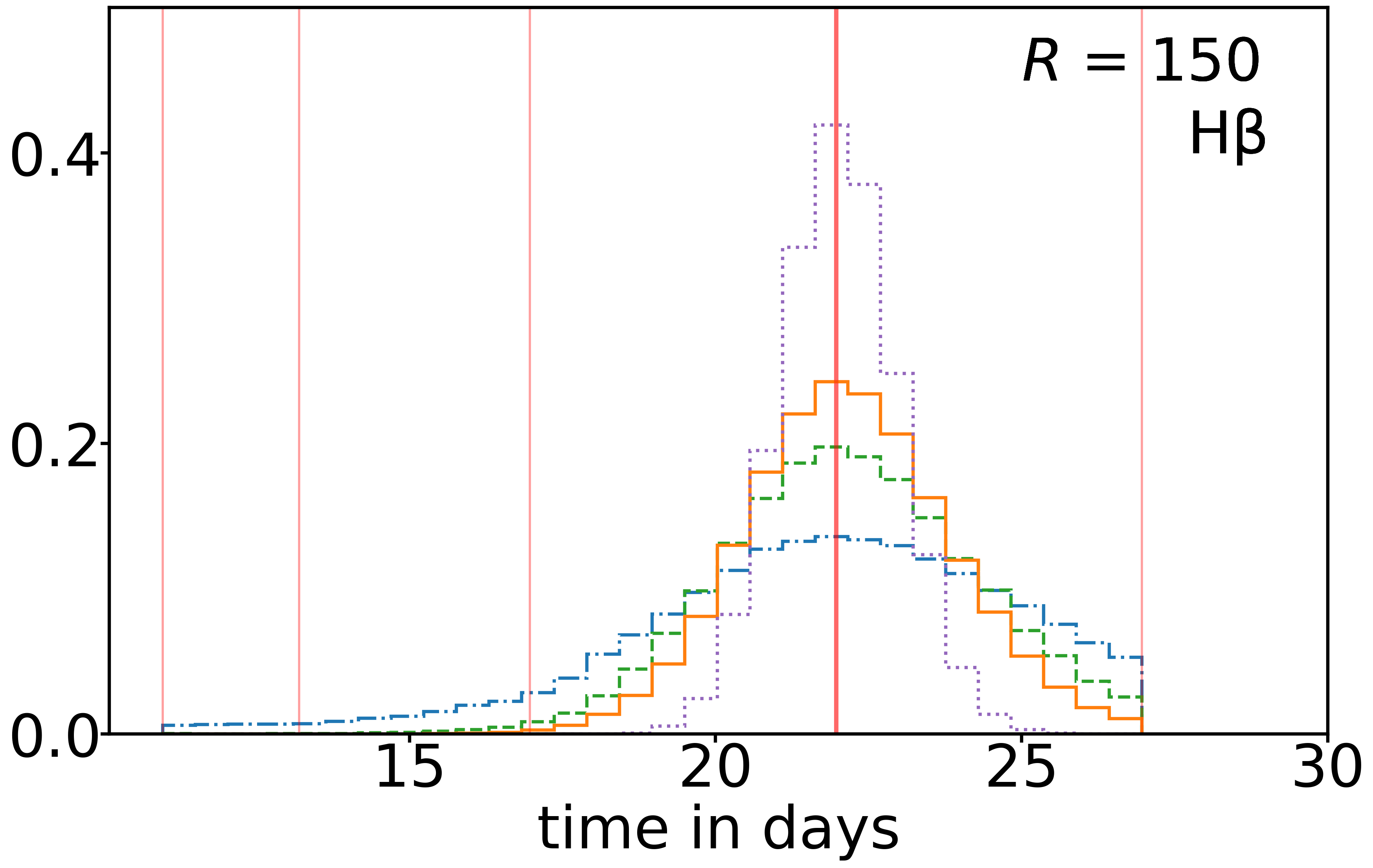}}\
\subfigure{\label{}\includegraphics[width=0.31\textwidth]{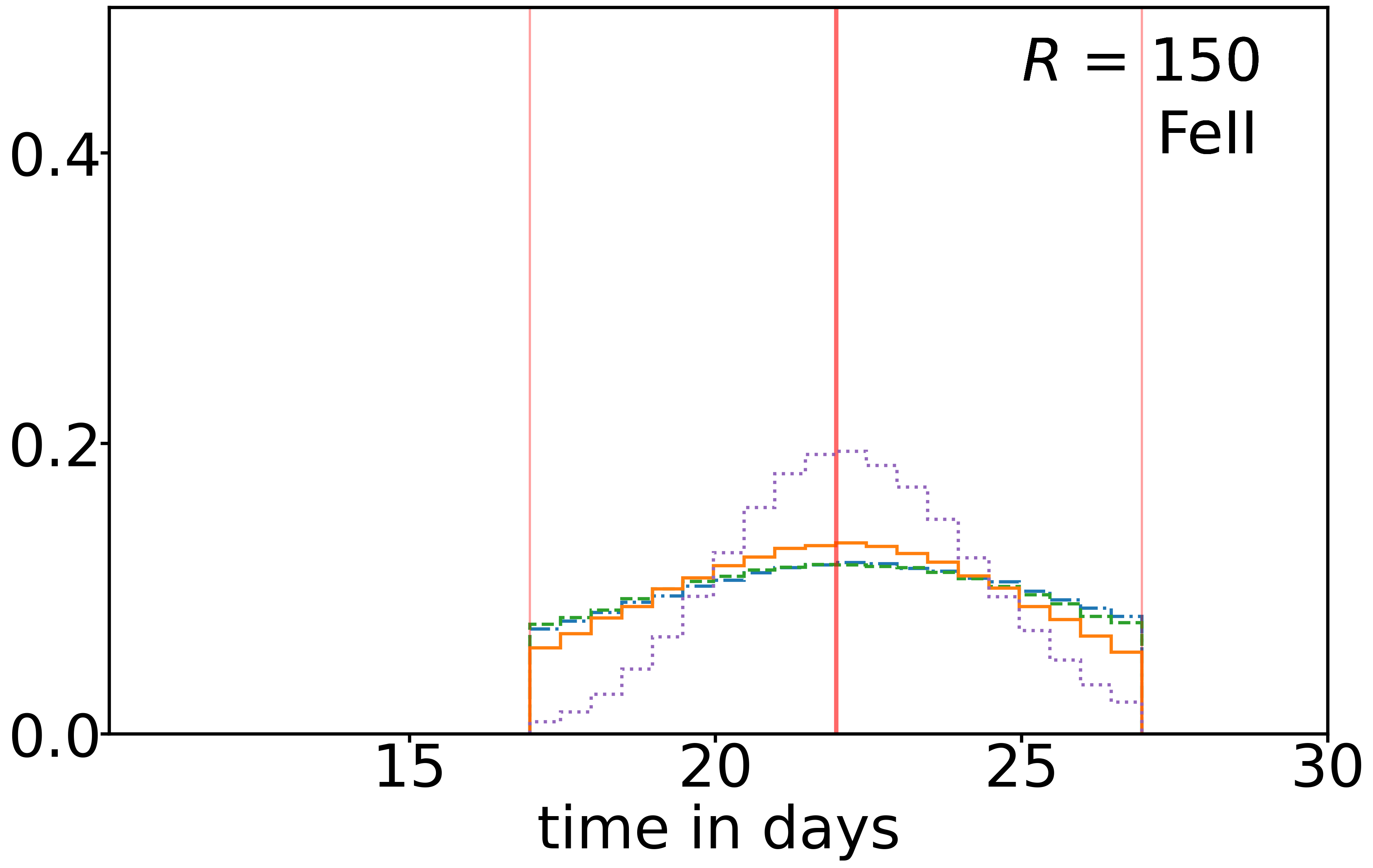}}\\
\subfigure{\label{}\includegraphics[width=0.31\textwidth]{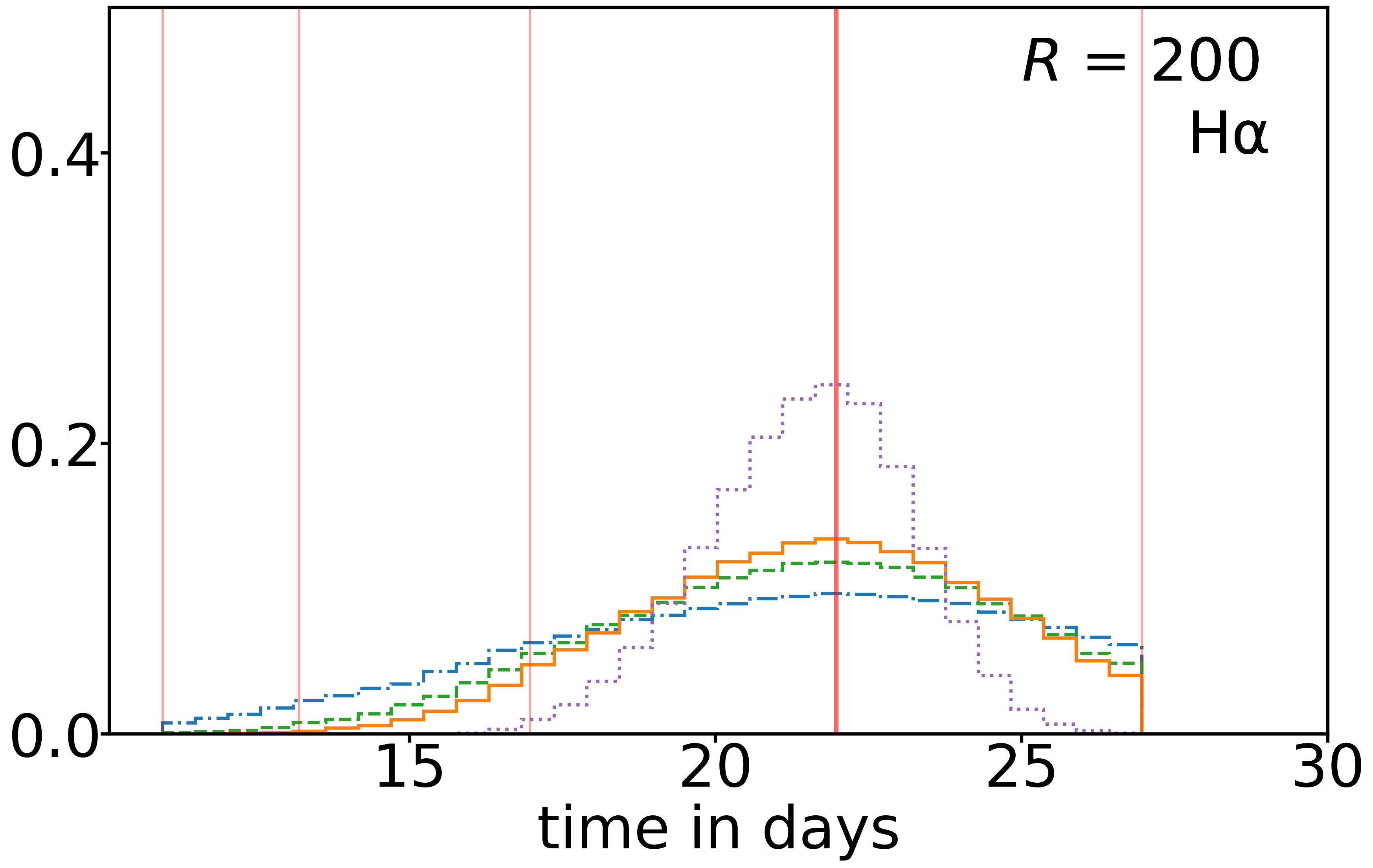}}\
\subfigure{\label{}\includegraphics[width=0.31\textwidth]{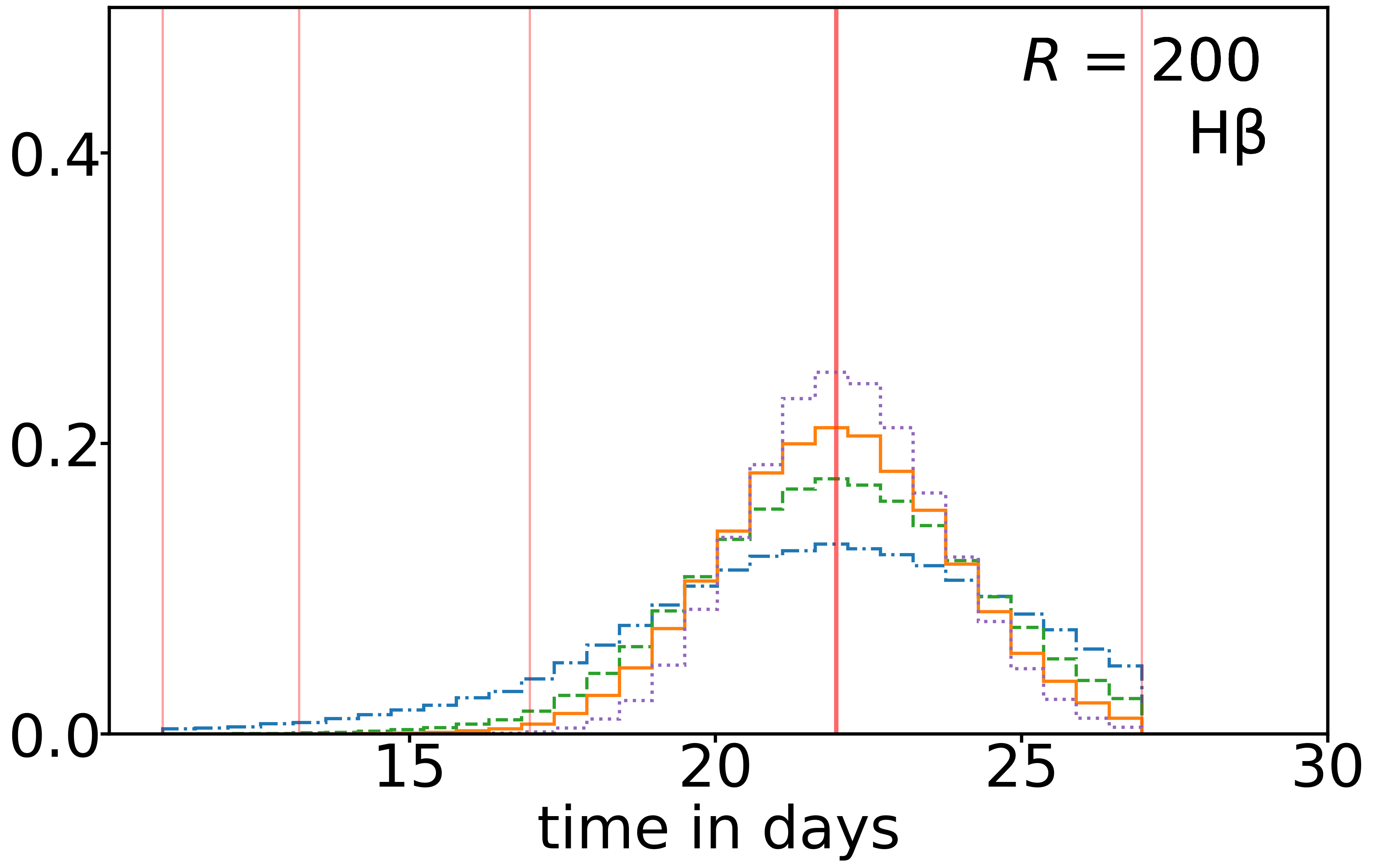}}\
\subfigure{\label{}\includegraphics[width=0.31\textwidth]{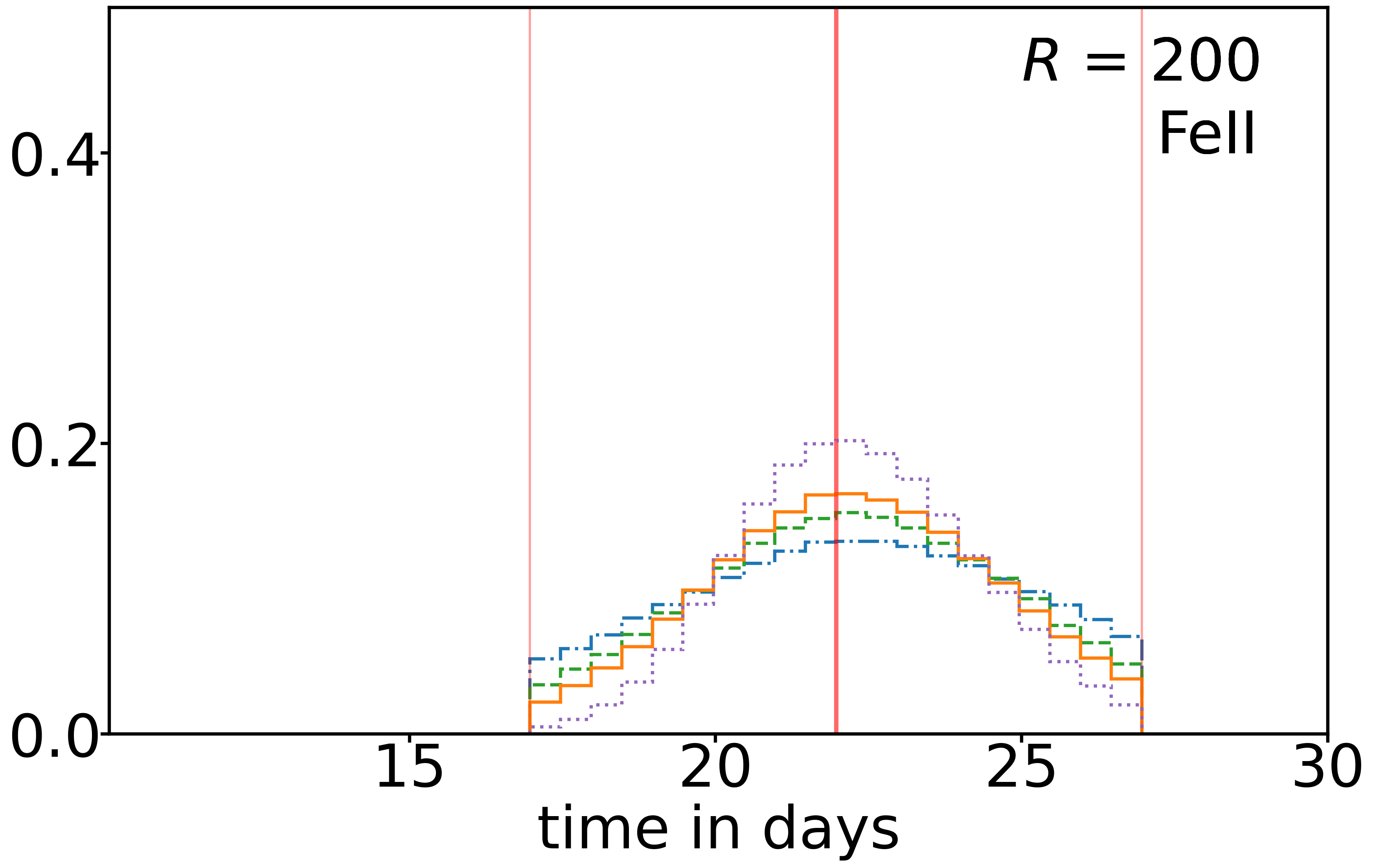}}\\
\subfigure{\label{}\includegraphics[width=0.31\textwidth]{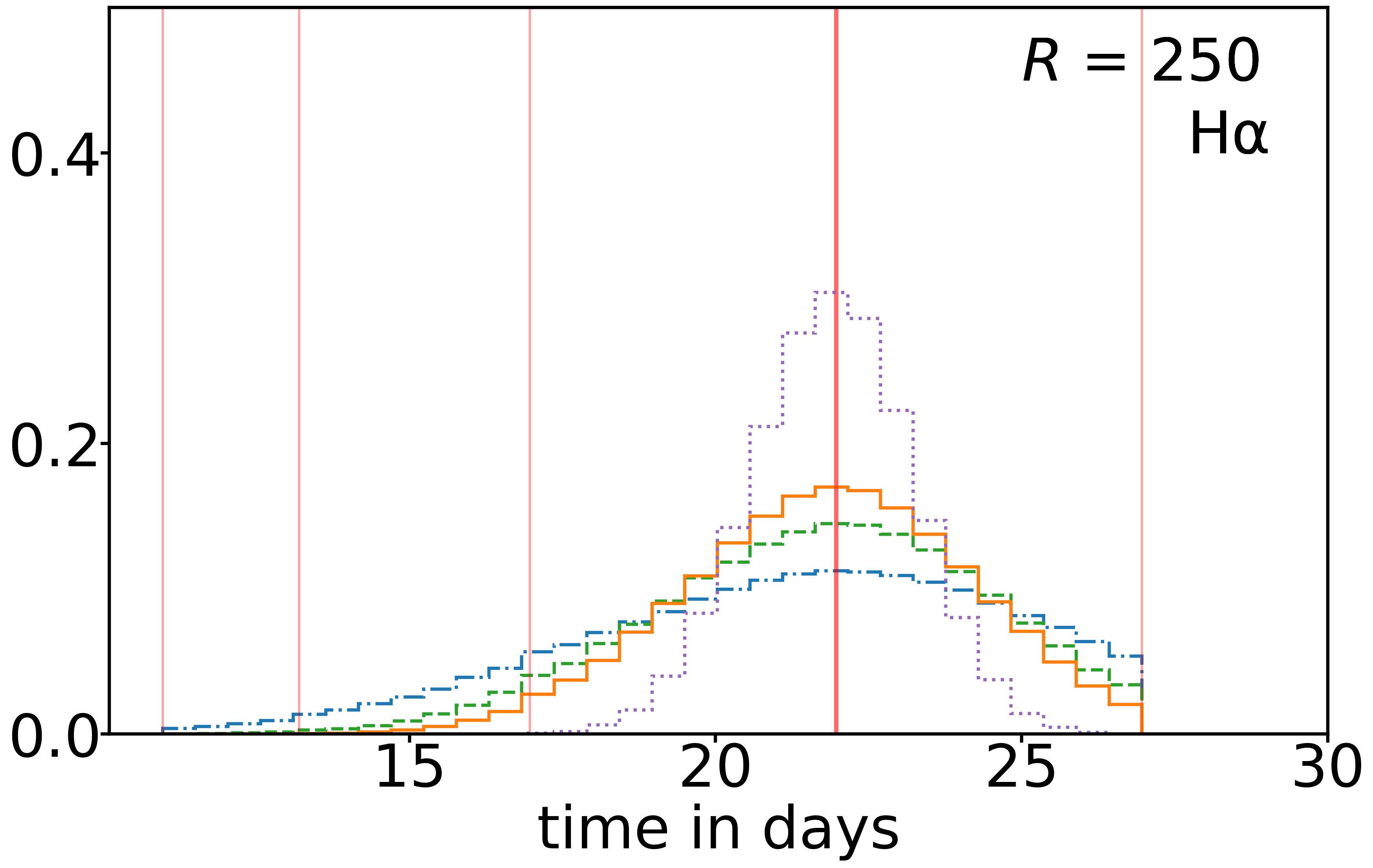}}\
\subfigure{\label{}\includegraphics[width=0.31\textwidth]{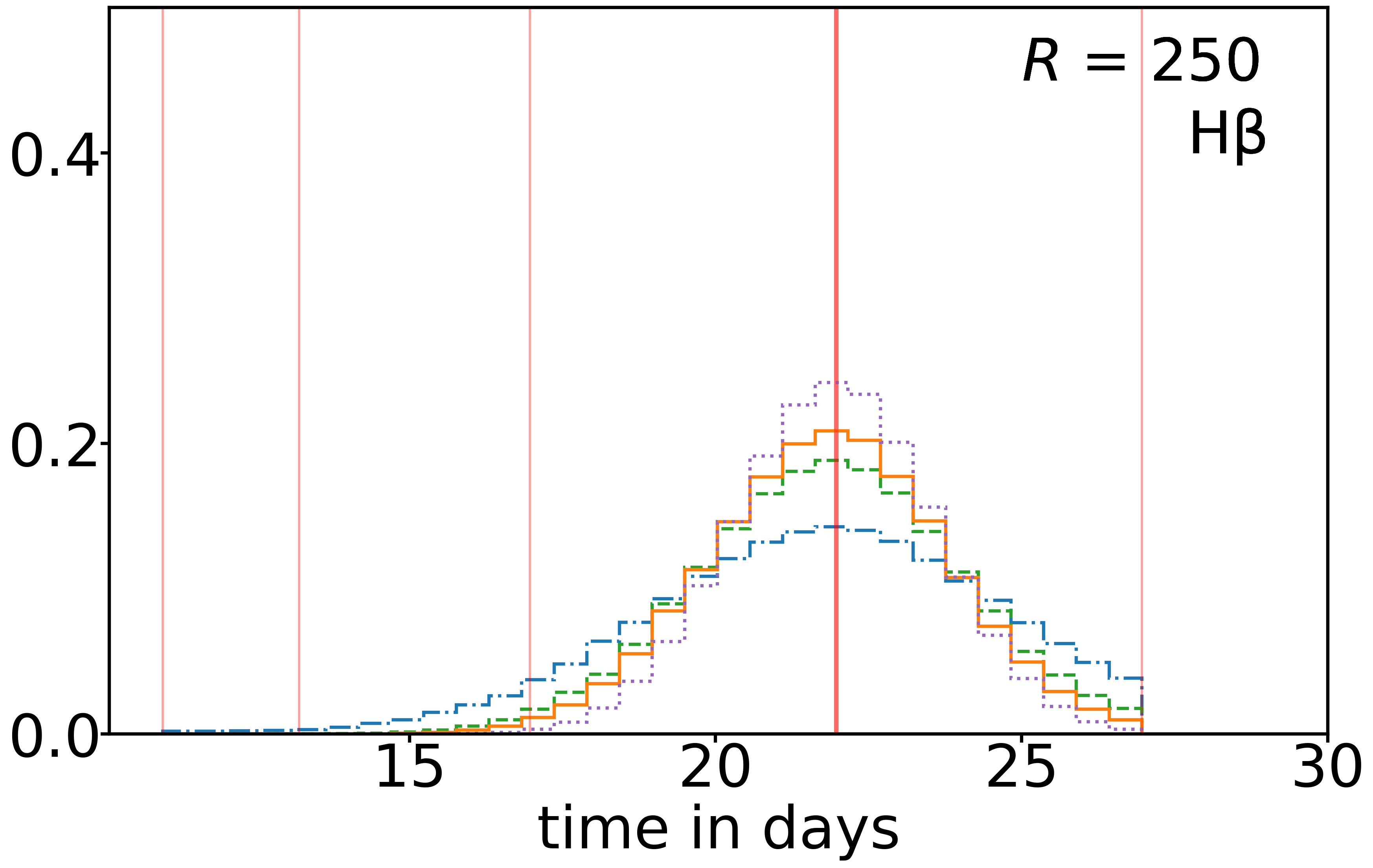}}\
\subfigure{\label{}\includegraphics[width=0.31\textwidth]{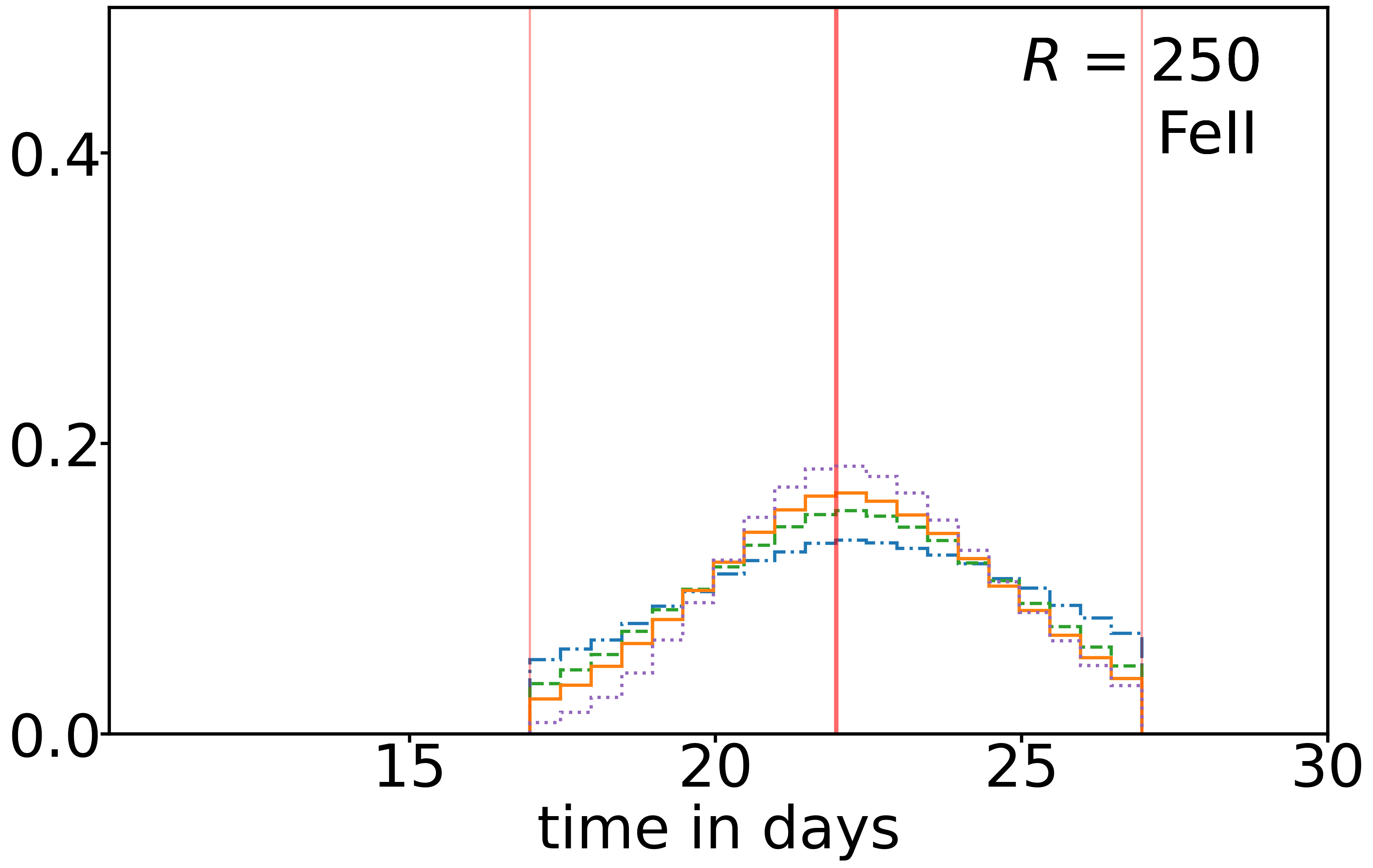}}\\
\caption{\label{phases_2} Histograms of the retrieved phases of the second SN image for the individual absorption lines 
  H$\mathrm{\alpha}$, H$\mathrm{\beta}$, and Fe\,\textsc{ii} as indicated in each panel. The phase inference
  is done for $S/N$ = 10, 15, 20, and noiseless spectra. The thin red vertical lines in the histograms indicate the epochs of the 5
  available spectra of the first SN image. 
  }
\end{figure*}

In this appendix, we present the time-delay inference results of the individual absorption lines H$\mathrm{\alpha}$, H$\mathrm{\beta}$, and Fe\,\textsc{ii} as histograms in Fig. \ref{phases_2}. The determined time-delay values are shown in Table \ref{fig:ap_alpha} for H$\mathrm{\alpha}$, in Table \ref{fig:ap_beta} for H$\mathrm{\beta}$, and in Table \ref{fig:ap_fe} for Fe\,\textsc{ii}.

We would like to point out that for some combinations of $R$ and $S/N$ for a certain absorption line, the uncertainty is limited to lower values by the available epochs of the first image.
In particular for H$\mathrm{\alpha}$ with $R = 100$ with noisy spectra and for almost all Fe\,\textsc{ii} cases, the width of the distribution is very broad, reaching the limit of the third epoch and fifth epoch in the case of Fe\,\textsc{ii} and the first epoch and fifth epochs, causing a truncation of the distribution at these borders. Therefore, the uncertainty will never be estimated larger than $\sim$5 days, as the full range of available epochs in this work covers 16 days, corresponding to a 68\% confidence interval of $\sim$11 days if one would like to infer the central epoch for the second image.
For real observations, we recommend having as many epochs as possible and ensuring to cover the epoch of the second image as well as possible in order to avoid artificial truncation.

\begin{table*}[hbt!]
\caption{
Retrieved phase values and 1$\sigma$ uncertainties of the phase retrievals for the absorption line H$\mathrm{\alpha}$.}
\label{fig:ap_alpha}
\centering
\begin{tabular}{l *{4}{c}}
\hline
\hline
\backslashbox{$R$}{$S/N$}
&\makebox[3em]{10}&\makebox[3em]{15}&\makebox[3em]{20}
&\makebox[3em]{$\infty$} \\
\hline 
100 & 19.3 $\pm$ 4.5 days & 19.7 $\pm$ 4.4 days & 20.1 $\pm$ 4.2 days & 22.6 $\pm$ 1.6 days \\ 
\hline 
150 & 20.5 $\pm$ 3.8 days & 21.5 $\pm$ 3.1 days & 21.9 $\pm$ 2.7 days & 22.2 $\pm$ 1.5 days \\ 
\hline 
200 & 20.7 $\pm$ 3.7 days & 21.3 $\pm$ 3.1 days & 21.5 $\pm$ 2.8 days & 21.5 $\pm$ 1.6 days \\ 
\hline 
250 & 21.1 $\pm$ 3.3 days & 21.6 $\pm$ 2.7 days & 21.8 $\pm$ 2.3 days & 21.9 $\pm$ 1.3 days \\ 
\hline 
\end{tabular}
\tablefoot{Three
  different $S/N$ values (10, 15, and 20) and a noiseless case, indicated by
  $S/N$ = $\infty$, are listed.}
\end{table*}

\begin{table*}[hbt!]
\caption{Retrieved phase values and 1$\sigma$ uncertainties of the phase retrievals for the absorption line H$\mathrm{\beta}$.}
\label{fig:ap_beta}
\centering
\begin{tabular}{l *{4}{c}}
\hline
\hline
\backslashbox{$R$}{$S/N$}
&\makebox[3em]{10}&\makebox[3em]{15}&\makebox[3em]{20}
&\makebox[3em]{$\infty$} \\
\hline  
100 & 20.4 $\pm$ 4.1 days & 21.5 $\pm$ 3.6 days & 22.3 $\pm$ 2.5 days & 22.2 $\pm$ 0.8 days \\ 
\hline 
150 & 21.6 $\pm$ 3.0 days & 22.1 $\pm$ 2.0 days & 22.2 $\pm$ 1.7 days & 22.0 $\pm$ 0.9 days \\ 
\hline 
200 & 21.5 $\pm$ 3.1 days & 21.9 $\pm$ 2.2 days & 22.0 $\pm$ 1.9 days & 22.1 $\pm$ 1.6 days \\ 
\hline 
250 & 21.5 $\pm$ 2.3 days & 21.8 $\pm$ 2.1 days & 21.8 $\pm$ 1.9 days & 21.9 $\pm$ 1.6 days \\ 
\hline 
\end{tabular}
\tablefoot{Three
  different $S/N$ values (10, 15, and 20) and a noiseless case, indicated by
  $S/N$ = $\infty$, are listed.}
\end{table*}

\begin{table*}[hbt!]
\caption{Retrieved phase values and 1$\sigma$ uncertainties of the phase retrievals for the absorption line Fe\,\textsc{ii}.}
\label{fig:ap_fe}
\centering
\begin{tabular}{l *{4}{c}}
\hline
\hline
\backslashbox{$R$}{$S/N$}
&\makebox[3em]{10}&\makebox[3em]{15}&\makebox[3em]{20}
&\makebox[3em]{$\infty$} \\
\hline
100 & 22.1 $\pm$ 2.8 days & 22.0 $\pm$ 2.8 days & 22.0 $\pm$ 2.8 days & 21.7 $\pm$ 1.7 days \\ 
\hline 
150 & 22.1 $\pm$ 2.7 days & 22.0 $\pm$ 2.7 days & 22.0 $\pm$ 2.6 days & 22.2 $\pm$ 2.0 days \\ 
\hline 
200 & 22.2 $\pm$ 2.5 days & 22.2 $\pm$ 2.4 days & 22.2 $\pm$ 2.2 days & 22.3 $\pm$ 1.9 days \\ 
\hline 
250 & 22.2 $\pm$ 2.5 days & 22.2 $\pm$ 2.4 days & 22.2 $\pm$ 2.2 days & 22.4 $\pm$ 2.0 days \\ 
\hline 
\end{tabular}
\tablefoot{Three
  different $S/N$ values (10, 15, and 20) and a noiseless case, indicated by
  $S/N$ = $\infty$, are listed.}
\end{table*}

\section{
Impact of Lens and host galaxy background on spectral phase inference}
\label{sec:app:background-contamination}

We consider a contamination of the SN spectrum by light from the SN host galaxy and the lens galaxy, and assess the impact on our method. The addition of the spectra of the SN, its host galaxy, and lens galaxy results in a composite spectrum from all three objects, which alters the overall shape of the lensed SN's spectrum and may distort individual absorption lines or reduce their strength. The magnitude of this effect will depend on the brightness of the lens and host galaxies and the alignment of the lensing system.
To quantify the significance, we investigate two scenarios. In scenario one, we consider a bright host galaxy, while in scenario two, the lensing galaxy produces a high nuisance. In both cases, we use an early-type galaxy spectrum taken from the SDSS spectral cross-correlation templates \citep{York2000, Adelman2007}\footnote{https://classic.sdss.org/dr5/algorithms/spectemplates/}, shifted either to the host-galaxy redshift \zs\ $= 0.8$ or to a typical lens redshift \zd\ $= 0.4$ \citep{Oguri2010}. 
The superposition of these galaxy spectra forms an additional background to the SN spectrum. Although SNe II predominantly occur in late-type galaxies, we use an early-type galaxy template as the host background as an approximation that is sufficient to assess the impact of spectral background variation. 

\begin{figure*}[hbt!]
\centering
\subfigure[]{\label{spec_back_04}\includegraphics[width=0.49\textwidth]{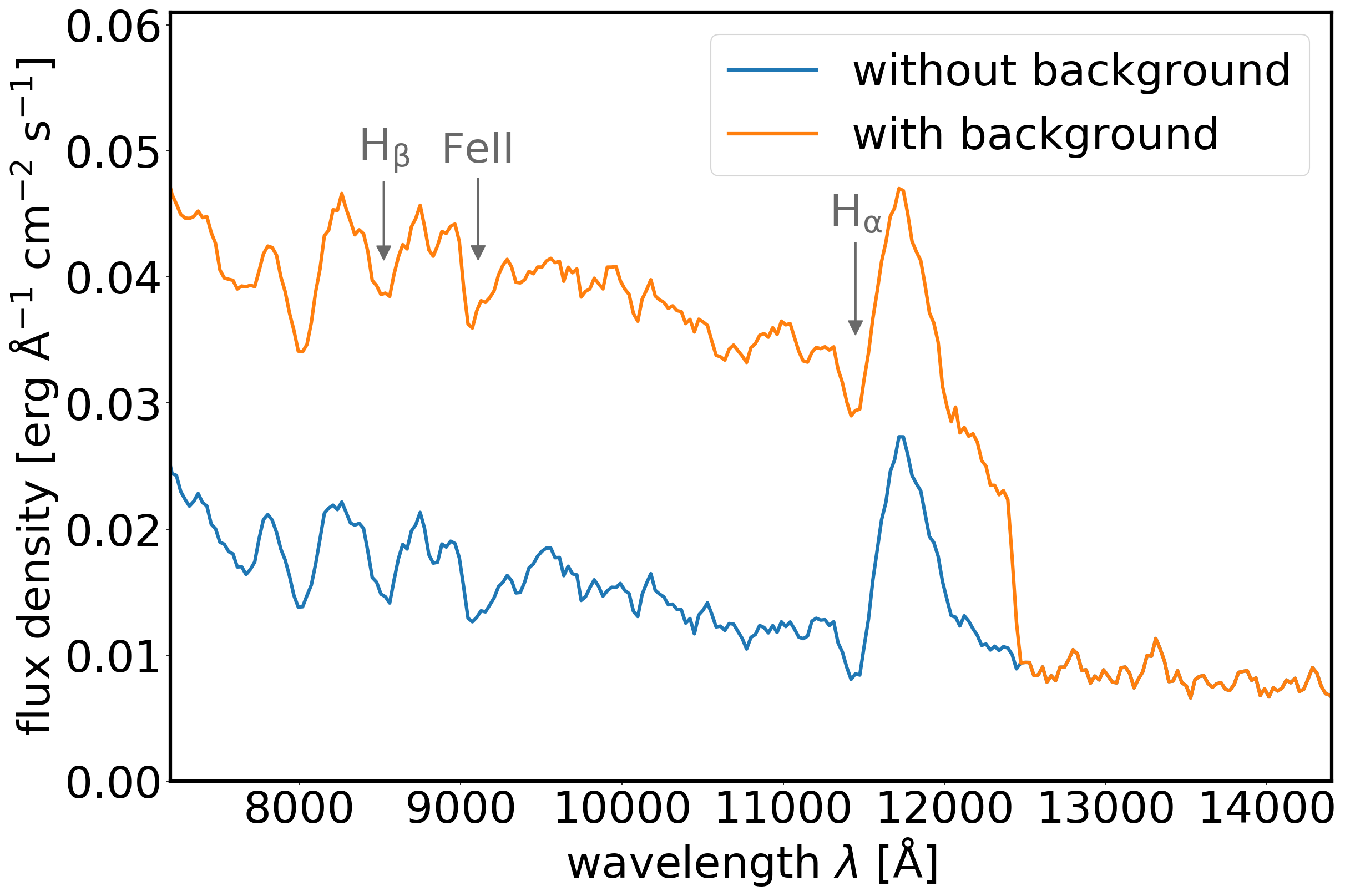}}\hfill
\subfigure[]{\label{spec_back_08}\includegraphics[width=0.49\textwidth]{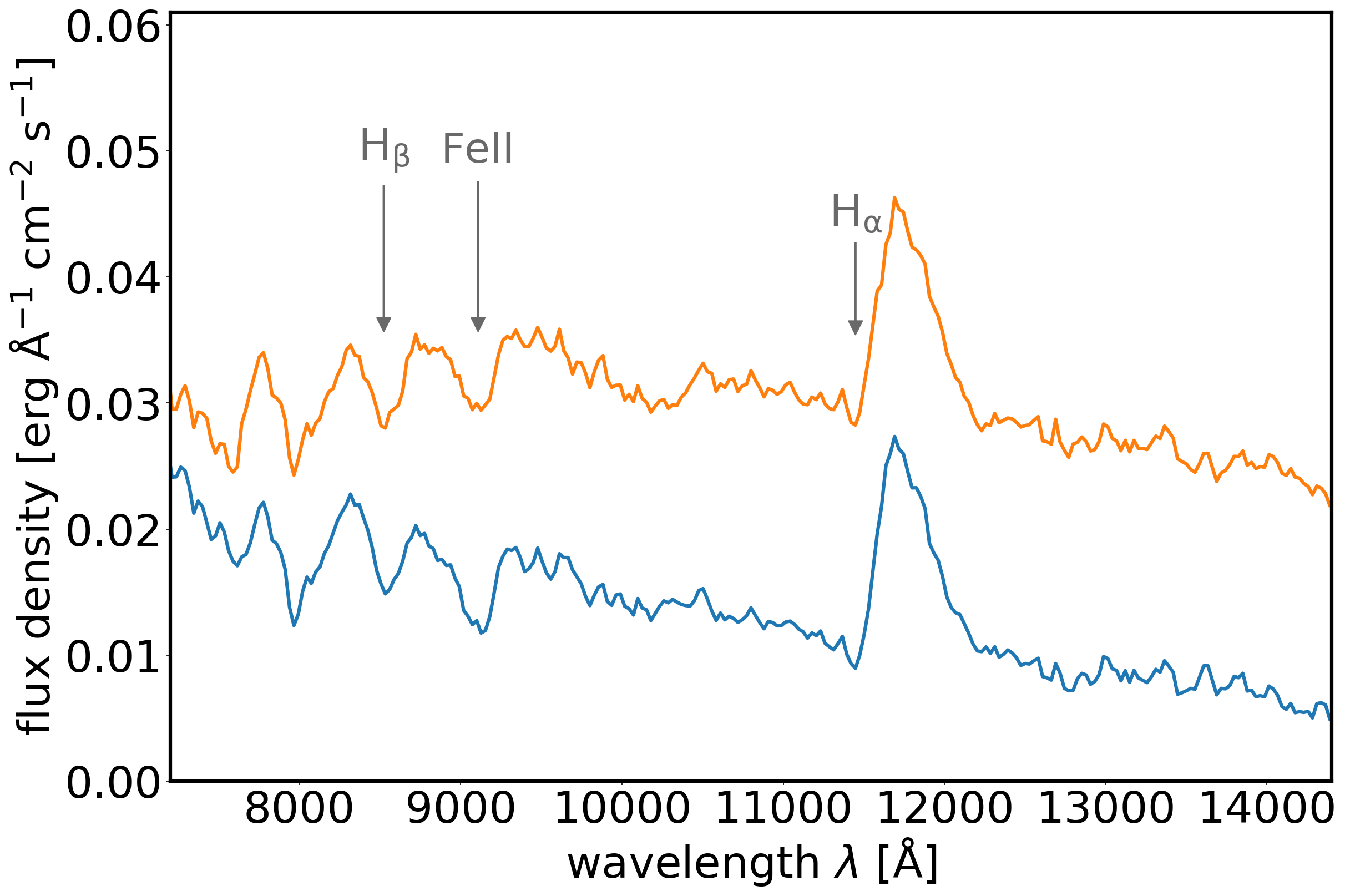}}\
\caption{\label{spec_back}
Model spectra of SN 1999em at redshift \zs\ = 0.8 with resolution $R = 150$ in blue and with added background of an early-type galaxy spectrum in orange for the epoch of 22 days. The absorption lines considered in the phase inference are labeled. Panel (a) shows the inclusion of a spectral background due to lens galaxy light at \zd\ = 0.4. The sharp drop around 12500 \AA \ in the spectrum with background (orange) stems from the truncated template galaxy spectrum. Panel (b) shows the inclusion of a spectral background due to host galaxy light at \zs\ = 0.8. The two model spectra of SN 1999em in panels (a) and (b) are created with different noise realizations and therefore show different shapes and depths of the individual lines.}
\end{figure*}

An example of the superposition for each assumed background scenario is shown in Fig. \ref{spec_back} for $R = 150$ spectra at a phase of 22.0 days after explosion. For the case of the spectral background at the lens redshift \zd\ $= 0.4$, the sharp drop in the spectrum around 12500 \AA \ stems from the galaxy template spectrum being truncated at a rest-frame wavelength of about 9000 \AA. However, this is not a problem, since the absorption lines used for phase retrieval are covered by the galaxy template in both cases.

A visual inspection of spectra influenced by the additional background shows that the absorption lines appear slightly flatter and broader than in the SN-only spectrum, potentially reducing the precision of determining the absorption line minimum.
In the investigated cases, no strong absorption or emission feature in the template spectrum coincides with any of the phase-retrieval absorption lines.
In future applications of this method, if spectral features of the SN coincide with absorption or emission features from the host or lens galaxy, the affected spectral lines can be excluded from the analysis as the time-delay measurement remains feasible with a reduced set of spectral features, although with slightly reduced precision and accuracy \citep{Bayer2021}.
Nebular emission lines associated with star-forming regions, which are likely present in typical host galaxies of SNe II, are arguably the strongest source of localized contamination. However, for the usual set of H\,\textsc{ii}-region emission lines (Balmer lines, [O\,\textsc{iii}], [N\,\textsc{ii}], and [S\,\textsc{ii}]), the effect is fully predictable, since the SN and its host galaxy are at the same redshift. The impact turns out to be limited: only the blueshifted Fe\,\textsc{ii} absorption of the SN might be contaminated (by [O\,\textsc{iii}]) in this case, whereas the H$\alpha$ and H$\beta$ absorptions remain clean. Therefore, a phase retrieval based on H$\alpha$ and H$\beta$ is still feasible \citep{Bayer2021}.

Our simulations further do not incorporate dust extinction, since we anticipate its effect to be minimal on the phase retrieval. Dust extinction could either occur at the SN host itself or at the foreground lens galaxy.  In the former case, the extinction would affect all the multiple images equally and would not affect phase retrieval.  In the latter case, differential extinction between the SN images could redden the spectra of SN images differently, potentially affecting the phase retrieval.  This is unlikely to occur since most of the time-delay lenses useful for cosmology with sufficiently long time delays are lensed by massive early-type galaxies with low amounts of dust; in fact, most of the lensed quasar systems analyzed by \citet{Eliasdottir2006} have extinction coefficient in the V-band, $A_{\rm V}$ values $\lesssim0.3$.  Furthermore, one could use dust extinction laws to correct for spectral distortions caused by dust extinction.  Given these considerations, we do not expect dust extinction to significantly affect our phase retrieval.

\begin{figure*}[hbt!]
\centering
\subfigure[]{\label{spec_back_04}\includegraphics[width=0.49\textwidth]{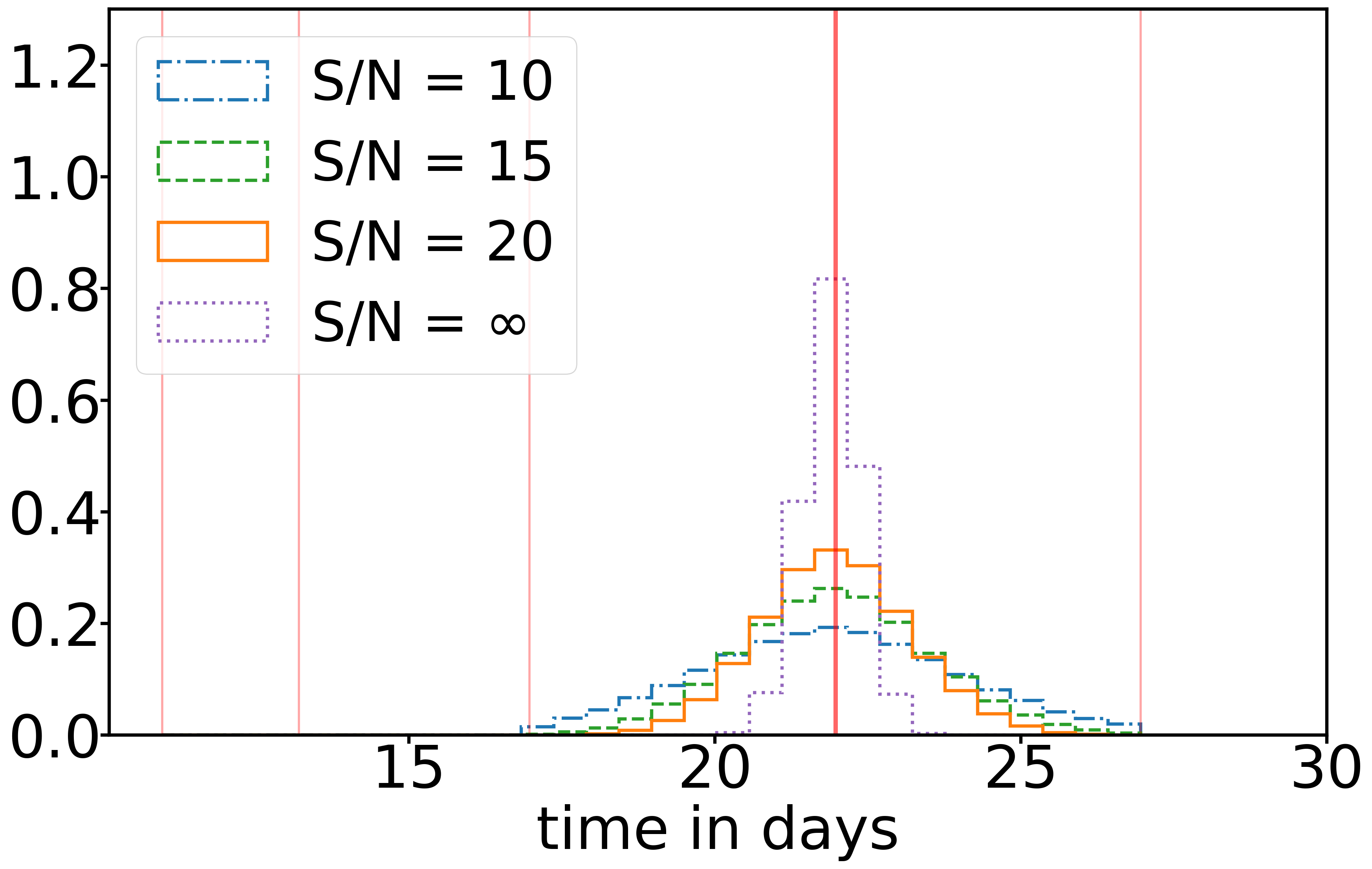}}\hfill
\subfigure[]{\label{spec_back_08}\includegraphics[width=0.49\textwidth]{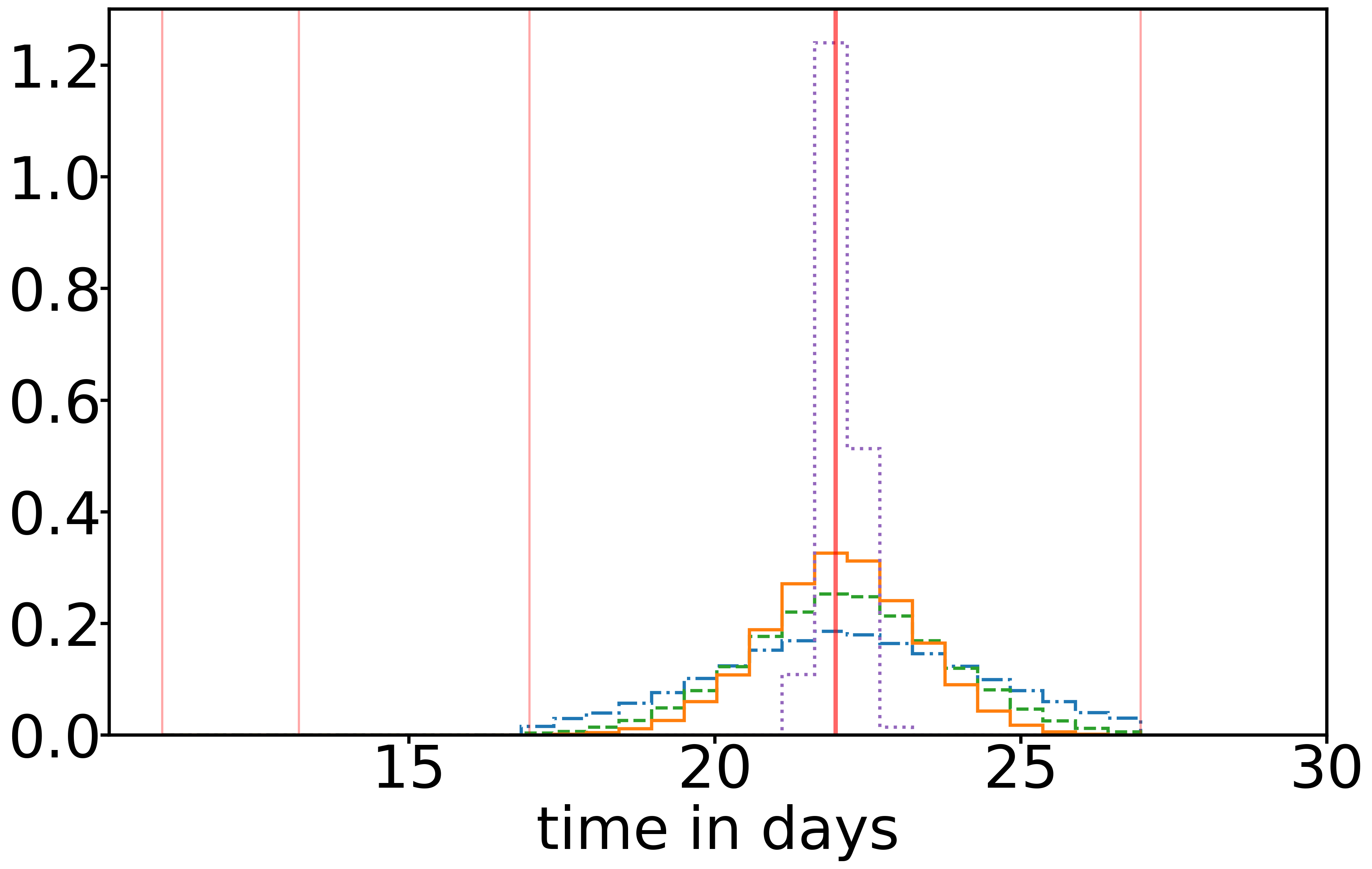}}\
\caption{\label{phase_back}
Histograms of the retrieved phases of the
  second SN image combining the absorption lines Fe\,\textsc{ii},
  H$\mathrm{\alpha}$, and H$\mathrm{\beta}$ for the two cases with added background spectrum. The phase inference
  is done for $S/N$ = 10, 15, 20, and noiseless spectra. The thin red vertical lines in the histograms indicate the epochs of the 5
  available spectra of the first SN image. The thick vertical line
  marks the input value. The phase of the second SN image is correctly
  recovered. Panel (a): Spectral background from lens galaxy at \zd\ = 0.4. Panel (b): Spectral background from host galaxy at \zs\ = 0.8.}
\end{figure*}

The phase retrieval histograms of the combined absorption lines are shown in Fig. \ref{phase_back}, and the corresponding values with 1$\sigma$ uncertainties are shown in Table \ref{phase_uncertainties_background}.
\begin{table*}[hbt!]
\caption{Retrieved phases and 1$\sigma$ uncertainties of the phase retrievals with background from the lens or the host galaxy.}
\label{phase_uncertainties_background}
\centering
\begin{tabular}{l *{4}{c}}
\hline
\hline
\backslashbox{background}{$S/N$}
&\makebox[3em]{10}&\makebox[3em]{15}&\makebox[3em]{20}
&\makebox[3em]{$\infty$} \\
\hline  
\zd\ $ = 0.4$ & 21.9 $\pm$ 2.0 days & 22.0 $\pm$ 1.6 days & 22.1 $\pm$ 1.3 days & 21.9 $\pm$ 0.5 days \\ 
\hline 
\zs\ $ = 0.8$ & 22.1 $\pm$ 2.1 days & 22.1 $\pm$ 1.6 days & 22.1 $\pm$ 1.3 days & 22.0 $\pm$ 0.3 days \\ 
\hline 
\end{tabular}
\tablefoot{
The values are computed combining the three absorption lines Fe\,\textsc{ii},
  H$\mathrm{\alpha}$, and H$\mathrm{\beta}$ for a spectral resolution of $R = 150$. The results concerning the background spectrum considered at the lens galaxy distance are indicated with \zd\ $= 0.4$. For the host galaxy background, we indicate \zs\ $ = 0.8$. We assume the noise and the microlensing in these lines are
  uncorrelated.}
\end{table*}
For both considered cases with additional background, the resulting precision and accuracy of the retrieved phase are comparable to the values retrieved without background in Sect. \ref{sec: SN phase inference from spectra}. 
Some precision values are even higher than in the case without background, which results from statistical fluctuations, and the H$\mathrm{\beta}$ being skewed to appear more symmetrical (see Fig. \ref{spec_back}), which the Gaussian fit matches better than a highly skewed line.
Overall, this additional investigation into the potential negative impact of a background spectrum from the lens or host shows that our phase retrieval method still achieves a precision better than 10\%, which is necessary for high-precision cosmology. 

\end{document}